\documentclass[final,onefignum,onetabnum]{siamart190516}

\usepackage{amsfonts}
\usepackage{graphicx}
\usepackage{epstopdf}

\usepackage[export]{adjustbox}
\usepackage{algorithm}
\usepackage{algorithmicx}
\usepackage{algpseudocode}
\usepackage{amsmath}
\usepackage{amssymb}
\usepackage{appendix}
\usepackage{caption}
\usepackage{enumerate}
\usepackage{enumitem}
\usepackage{filecontents}
\usepackage{float}
\usepackage{mathtools}
\usepackage{multirow}
\usepackage{physics}
\usepackage{stackengine} 
\usepackage{setspace}
\usepackage{subfig}
\usepackage{tcolorbox}
\tcbuselibrary{skins}

\usepackage{ifthen}

\ifpdf
  \DeclareGraphicsExtensions{.eps,.pdf,.png,.jpg}
\else
  \DeclareGraphicsExtensions{.eps}
\fi

\newsiamremark{remark}{Remark}
\newsiamremark{hypothesis}{Hypothesis}
\crefname{hypothesis}{Hypothesis}{Hypotheses}
\newsiamthm{claim}{Claim}

\headers{Coarse-graining multi-agent stochastic systems}{D. Stepanova, H. M. Byrne, P. K. Maini\and T. Alarc\'on}
\title{A method to coarse-grain multi-agent stochastic systems with regions of multistability\thanks{\today.
\funding{This work is supported by a grant of the Obra Social La Caixa Foundation on Collaborative Mathematics awarded to the Centre de Recerca Matemàtica through a scholarship awarded to D.S. D.S. and T.A. have been partially funded by the CERCA Programme of the Generalitat de Catalunya. They also acknowledge MINECO (\url{https://www.ciencia.gob.es/}) for funding under grants MTM2015-71509-C2-1-R and RTI2018-098322-B-I00. D.S. and T.A. participate in project 2017SGR01735 which was awarded by AGAUR (\url{https://agaur.gencat.cat/en/inici/index.html}) but with no actual funding. The funders had no role in study design, data collection and analysis, decision to publish, or preparation of the manuscript. H.M.B. and P.K.M. received no specific funding for this work.}}}

\author{Daria Stepanova\footnotemark[2]\ \footnotemark[3]\ \footnotemark[6] 
\and Helen M. Byrne\footnotemark[4]
\and Philip K. Maini\footnotemark[4]
\and Tom\'as Alarc\'on\footnotemark[5]\ \footnotemark[2]\ \footnotemark[3]\
}

\usepackage{amsopn}

\newcommand{\of}[1]{\left(#1\right)}
\newcommand{\af}[1]{\left[#1\right]}
\newcommand{\uf}[1]{\left\{#1\right\}}

\DeclareMathOperator{\prob}{P}

\makeatletter
\newcounter{savesection}
\newcounter{apdxsection}

\newcommand\unappendix{\par
  \setcounter{apdxsection}{\value{section}}%
  \setcounter{section}{\value{savesection}}%
  \setcounter{subsection}{0}%
  \gdef\thesection{\@arabic\c@section}}
\makeatother

\ifpdf
\hypersetup{
  pdftitle={CG paper},
  pdfauthor={Daria Stepanova}
}
\fi

\begin{document}

\maketitle
\makeatletter
\renewcommand{\thefootnote}{\fnsymbol{footnote}}
\makeatother
\footnotetext[2]{Centre de Recerca Matem\`atica, Bellaterra (Barcelona) 08193, Spain}
\footnotetext[3]{Departament de Matem\`atiques, Universitat Aut\`onoma de Barcelona, Bellaterra (Barcelona) 08193, Spain}
\footnotetext[4]{Wolfson Centre for Mathematical Biology, Mathematical Institute, University of Oxford, Oxford OX2 6GG, UK}
\footnotetext[5]{Instituci\'o Catalana de Recerca i Estudis Avan\c{c}ats (ICREA), Barcelona 08010, Spain}
\footnotetext[6]{\email{dstepanova@crm.cat}}

\renewcommand{\thefootnote}{\arabic{footnote}}


\begin{abstract}
Hybrid multiscale modelling has emerged as a useful framework for modelling complex biological phenomena. However, when accounting for stochasticity in the internal dynamics of agents, these models frequently become computationally expensive. Traditional techniques to reduce the computational intensity of such models can lead to a reduction in the richness of the dynamics observed, compared to the original system. Here we use large deviation theory to decrease the computational cost of a spatially-extended multi-agent stochastic system with a region of multi-stability by coarse-graining it to a continuous time Markov chain on the state space of stable steady states of the original system. Our technique preserves the original description of the stable steady states of the system and accounts for noise-induced transitions between them. We apply the method to a bistable system modelling phenotype specification of cells driven by a lateral inhibition mechanism. For this system, we demonstrate how the method may be used to explore different pattern configurations and unveil robust patterns emerging on longer timescales. We then compare the full stochastic, coarse-grained and mean-field descriptions via pattern quantification metrics and in terms of the numerical cost of each method. Our results show that the coarse-grained system exhibits the lowest computational cost while preserving the rich dynamics of the stochastic system. The method has the potential to reduce the computational complexity of hybrid multiscale models, making them more tractable for analysis, simulation and hypothesis testing.
\end{abstract}

\begin{keywords}
Large deviation theory, coarse-graining, phenotype pattern formation, multiscale modelling, hybrid modelling
\end{keywords}

\begin{AMS}
60F10 (Large deviations), 92C15 (Developmental biology, pattern formation), 92C42 (Systems biology, networks), 92B05 (General biology and biomathematics), 92-08 (Computational methods for problems pertaining to biology)
\end{AMS}


\section{Introduction}

When modelling a biological process, one has to make choices on how detailed the model should be in order to capture the characteristic features of the system. At the same time, the model should be as simple as possible in order to facilitate its analysis and numerical simulations. The evolution of systems with large numbers of agents (e.g. molecules, cells, species) can be described by the average behaviour of their agents, or their mean-field limits using (ordinary or partial) differential equations (\cite{boareto2015jagged,byrne1995mathematical,macklin2009multiscale}). Dynamical systems theory provides methods and techniques for the analysis and numerical simulations of such systems. This description might become insufficient when the system comprises agents with internal variables that change in time, thus altering the agents' behaviour, or when the system is not `large enough' to be described accurately by the mean-field equations. For these systems, stochastic descriptions are employed \cite{perez2016intrinsic} (for example, continuous time Markov chains, CTMCs, or stochastic differential equations, SDEs). In biological systems, the number of agents is finite and some level of noise is always present which can affect the system dynamics \cite{perez2016intrinsic}. While exhibiting richer dynamics than deterministic systems, stochastic models are more computationally intensive. 

Furthermore, in order to formulate a theoretical model of a biological phenomenon, it is often necessary to account for dynamics that act on different temporal and/or spatial scales \cite{bardini2017multi,deutsch2020multi}. This has led to the development of hybrid multiscale models, in which different modelling techniques may be applied at each scale and then efficient coupling algorithms are used to integrate these models (see, e.g., \cite{buske2011comprehensive,deisboeck2011multiscale,osborne2010hybrid} and references therein). In many of these models, individual entities (cells, species, etc.) are considered as discrete agents which are, themselves, equipped with models for their internal states determining the behaviour (e.g. subcellular signalling, cell cycle, response to extracellular stimuli). Such models have great potential for generating insights into the behaviour of a system (e.g., endothelial cell rearrangements \cite{bentley2014role}, cell differentiation and tissue organization in intestinal crypts \cite{buske2011comprehensive}, and multiscale cancer modelling \cite{chaplain2020multiscale}). However, they frequently become numerically intractable because of their complexity (e.g. the internal dynamics of agents) \cite{bardini2017multi}. This limits possible applications of these models. 

\renewcommand{\thesubfigure}{\alph{subfigure}}
\captionsetup{singlelinecheck=off}
\begin{figure}[htbp]
\centering 
\subfloat[][\label{PhenotypeSwitch_Motivation_Config}]{\includegraphics[height = 4.3cm,valign=b]{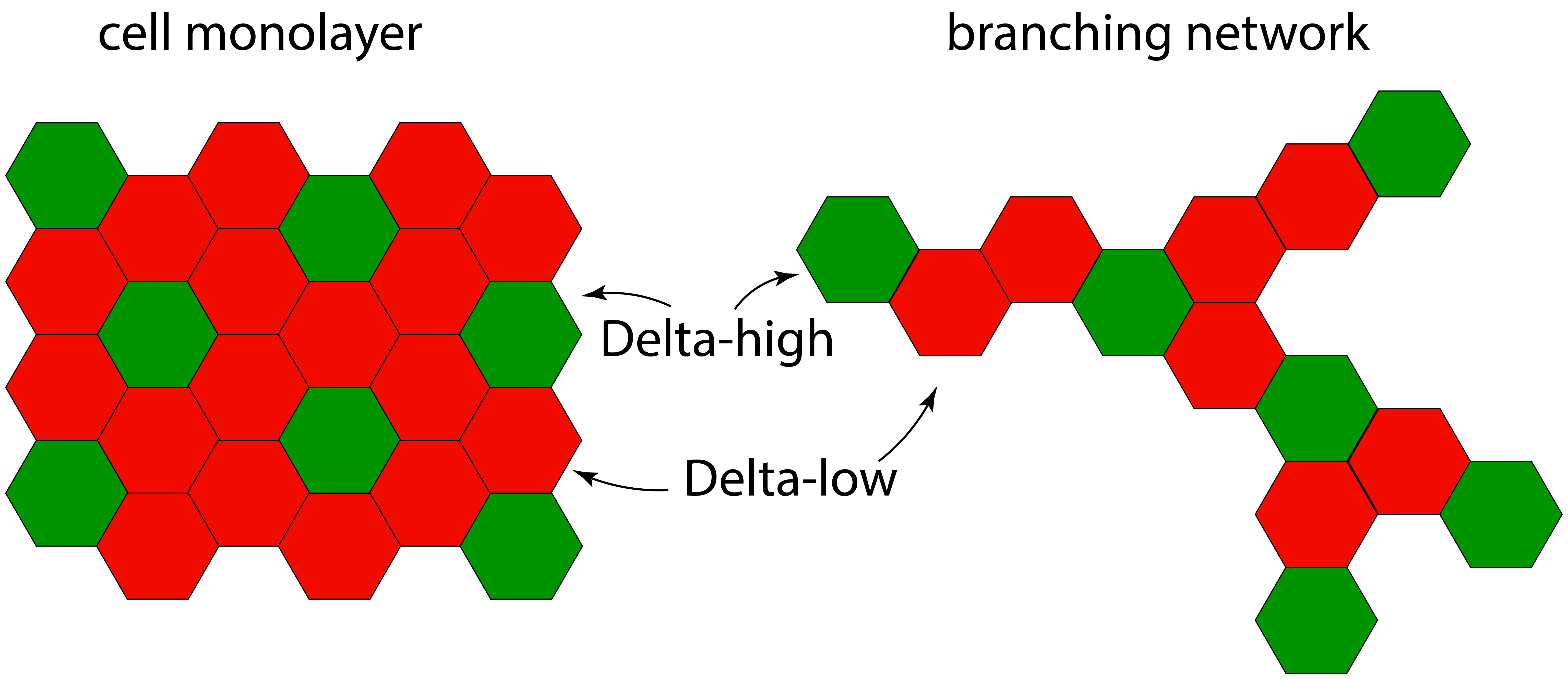}}

\subfloat[][\label{PhenotypeSwitch_Motivation_Trajectory}]{\includegraphics[height = 2.5cm,valign=b]{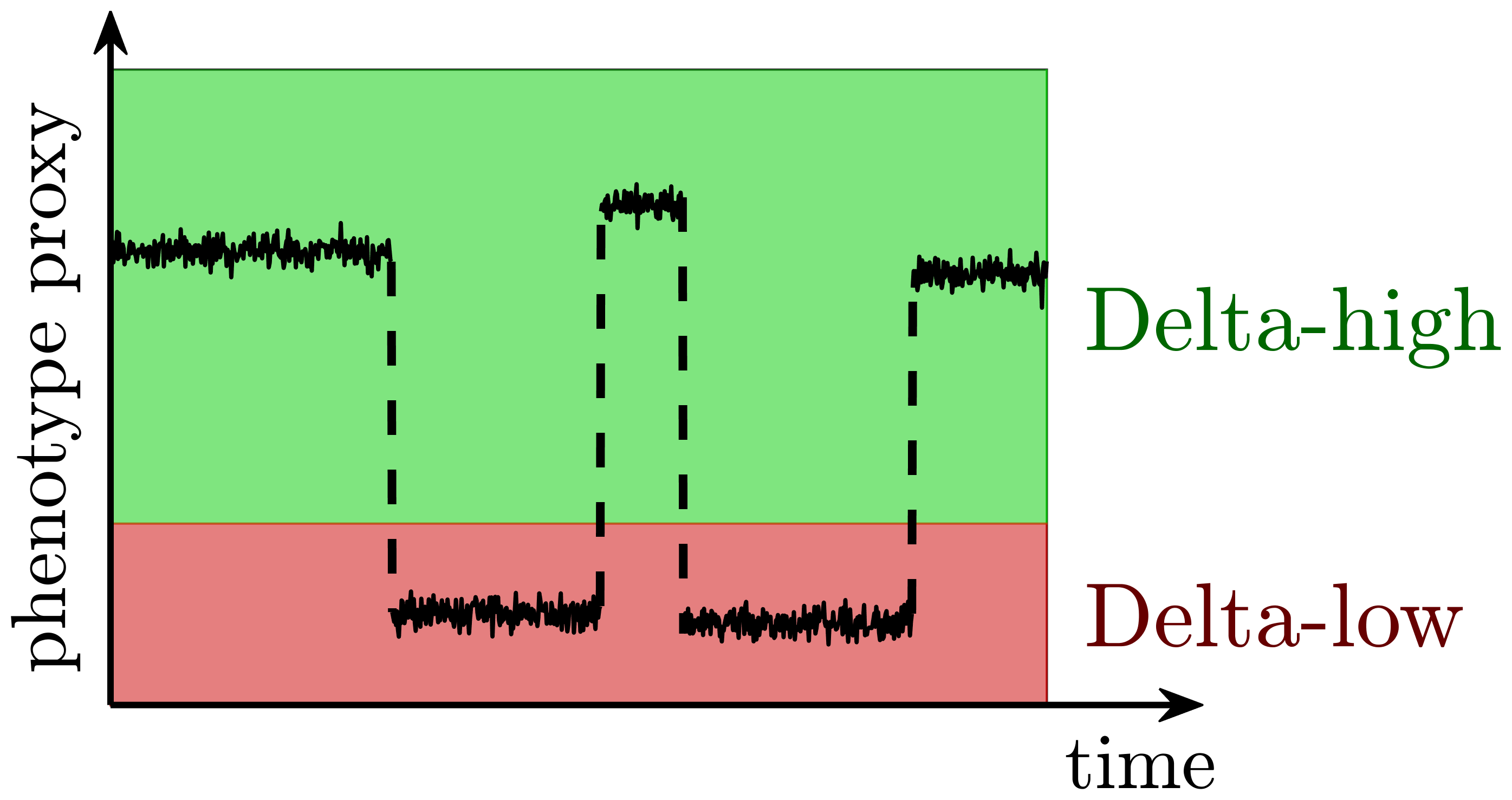}}
\hspace{0.5cm}
\subfloat[][\label{PhenotypeSwitch_Motivation_Switch}]{\includegraphics[height = 2.5cm,valign=b]{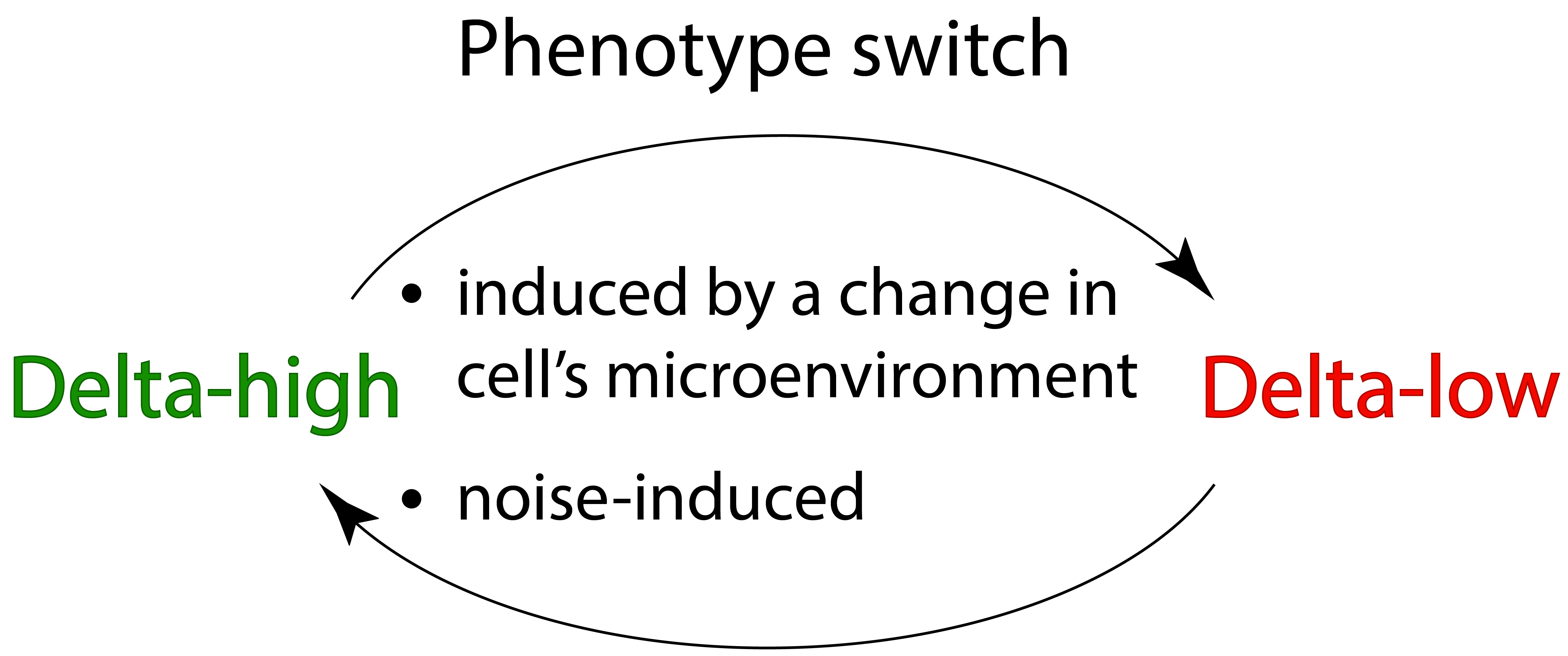}}
\caption{\textbf{Cell phenotype specification.} (a) Phenotype (Delta-high and Delta-low cells) patterning of cells induced by a mechanism of lateral inhibition in two different domains: a cell monolayer and a branching network. (b) Dynamic time evolution of phenotype adaptation of an individual cell. Using a phenotype proxy, e.g. level of Delta, allows for identification of a continuous cell phenotype. (c) Phenotype switches, as in (b) (dashed vertical lines), occur due to either a change in a cell's microenvironment or naturally present noise in intracellular signalling.} \label{PhenotypeSwitch_Motivation}
\end{figure}

In this work, we explain how to reduce the computational complexity of a hybrid model by coarse-graining the internal dynamics of its agents when these are described by a stochastic system with multiple steady states. The method involves applying large deviation theory (LDT) to reduce the dynamics of the stochastic system to a continuous time Markov chain (CTMC) on the state space of its stable steady states. LDT provides a theoretical framework with which to quantify how small time-dependent fluctuations can lead to significant deviations from the mean-field behaviour (\textit{rare events}) such as transitions between stable steady states which cannot occur in deterministic systems \cite{freidlin1998random}. This approach has previously been used to study rare, noise-induced events in individual stochastic systems \cite{de2017coarse,de2018minimum,dykman1994large,perez2016intrinsic,poppe2018physical,roma2005optimal}, but to our knowledge, this is its first application to a multi-agent model. 

In previous work, we developed a multiscale model of angiogenesis \cite{stepanova2021multiscale}, the process of growth of new blood vessels from pre-existing ones \cite{heck2015computational}, which accounts for gene expression patterns (phenotypes) of endothelial cells (ECs) at the subcellular scale. For prescribed levels of extracellular stimuli, the system is either monostable (i.e. only one cell phenotype exists) or bistable (i.e. two stable steady states, cell phenotypes, coexist). Cell phenotype is specified via contact-dependent cross-talk with neighbouring ECs via the VEGF-Delta-Notch signalling pathway \cite{blanco2013vegf,gerhardt2003vegf}. VEGF, or vascular endothelial growth factor, is the activating external stimulus; Delta and Notch are transmembrane ligands and receptors, respectively, which can trans-bind, (i.e. a ligand on one cell can bind to a receptor on another cell, thus allowing the two cells to `communicate'). Cells adjust their gene expression in order to maintain a pattern of two distinct phenotypes, Delta-high and Delta-low cells (see \Cref{PhenotypeSwitch_Motivation_Config,PhenotypeSwitch_Motivation_Trajectory}). We use the internal level of Delta as a proxy to distinguish between the phenotypes. In angiogenesis, the Delta-high (Delta-low) cells are referred to as tip (stalk) cells \cite{blanco2013vegf}. The number of transmembrane proteins in this signalling pathway is on the order of thousands for each cell \cite{boareto2015jagged}. Therefore, in order to formulate a mathematical model, it is tempting to use deterministic mean-field equations to describe the kinetic reactions of this signalling pathway. However, deterministic descriptions cannot account for noise-induced transitions between stable steady states or, in the case of this signalling pathway, phenotypic switches, which can occur in regions of bistability (see \Cref{PhenotypeSwitch_Motivation_Trajectory,PhenotypeSwitch_Motivation_Switch}). Since branching patterns of vascular networks are affected by the distribution of cells with different phenotypes, such phenotype transitions are potentially significant. Therefore, we modelled the subcellular signalling pathways stochastically, which increased the computational cost of the model. This example illustrates a general problem associated with computational and, in particular, hybrid models: in order to preserve emergent features of the system, such as continuous cell phenotypes and noise-induced phenotype switches, the model becomes computationally intractable for large lattice simulations.

We illustrate the coarse-graining method by reference to the subcellular model of the VEGF-Delta-Notch signalling pathway that defines cell phenotype. The core Delta-Notch signalling pathway plays a key role in phenotype adaptation in cell types which can form cell monolayers, such as epithelial sheets \cite{monk2001spatiotemporal,sprinzak2010cis}, bristle patterning in \textit{Drosophila} \cite{cohen2010dynamic,hunter2016coordinated,corson2017self}, and neural precursor cells \cite{formosa2013lateral}. In all of these biological processes, the lateral inhibition mechanism of Delta-Notch signalling generates spatial patterns of cells with alternating fates (phenotypes). In the VEGF-Delta-Notch model, the stationary distribution of VEGF serves as an activating extracellular stimulus for the particular case of endothelial cells. In other cell types, which use the lateral inhibition mechanism to communicate, the external stimulus may differ from VEGF. In this paper we perform our simulations for two spatial geometries: a cell monolayer and a branching network (\Cref{PhenotypeSwitch_Motivation_Config}). For our model of multicellular VEGF-Delta-Notch signalling, we show typical simulation results of the coarse-grained system which allows us to explore different configurations of spatial patterns in a single realisation of the model (due to phenotypic switches). We then demonstrate how this dynamic exploration of possible patterns may be used to uncover robust patterns emerging at long timescales. We finally compare the spatio-temporal dynamics and computational cost of the full stochastic CTMC, the coarse-grained and the deterministic mean-field descriptions. Our results show that the coarse-grained model, while preserving the continuous description of cell phenotype and rare events of phenotype switching, is more computationally efficient than the other two systems. Thus, it significantly reduces the computational complexity of the model without sacrificing the rich dynamics of the original stochastic system. 

The remainder of the paper is organised as follows. In \cref{sec:TheoreticalBackground}, we review the hybrid (multiscale) modelling approach (\cref{subsec:HybridModels}) and summarise large deviation theory (\cref{subsec:LargeDeviationTheory}). This provides us with the information needed to formulate the coarse-grained model in \cref{sec:CoarseGraining}. In \cref{subsec:IndividualCellSystem}, we start by coarse-graining the individual agent system and checking the accuracy of the method. We then extend the technique to a multi-agent system in \cref{subsec:MulticellularSystem} where we outline a general algorithm for formulating and simulating the coarse-grained model. In \Cref{sec:Results}, we present typical simulation results for the model of the VEGF-Delta-Notch signalling pathway (\cref{subsec:PatterningCG}) and compare the full stochastic, coarse-grained and mean-field systems via metrics which quantify the spatial patterns formed by the two cell phenotypes and we also compare computational cost of simulations (\cref{subsec:ModelComparison}). The paper concludes in \cref{sec:Discussion} with a summary of our findings and suggestions for future research directions.

\section{Theoretical background}
\label{sec:TheoreticalBackground}
\subsection{Hybrid models}
\label{subsec:HybridModels}

\begin{figure}[htbp]
\centering
\subfloat{\includegraphics[width=11cm]{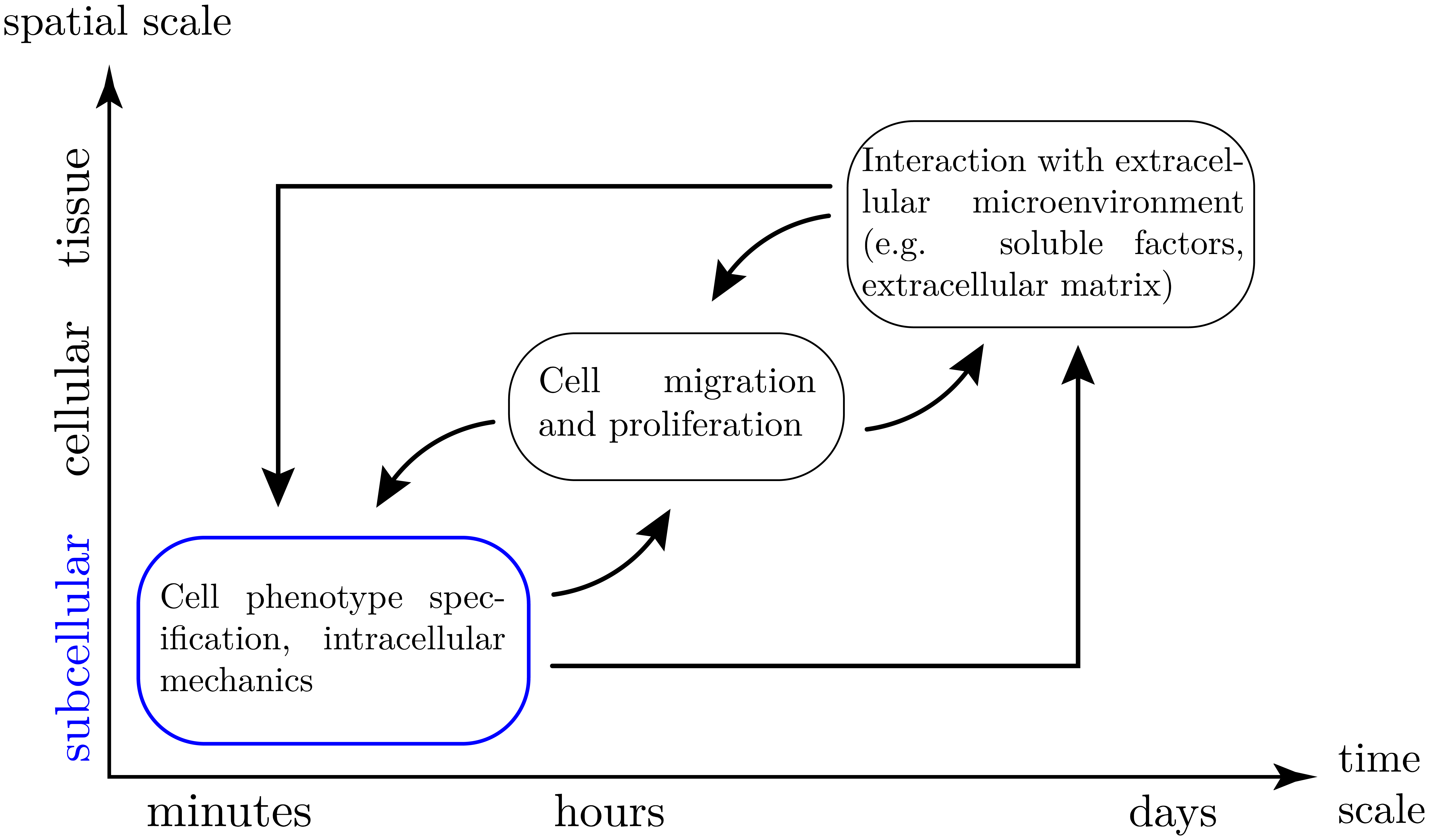}}
\caption{\textbf{A schematic diagram illustrating characteristic spatial and temporal scales of a typical biological process and coupling between them.} The VEGF-Delta-Notch signalling pathway, which serves as an illustrative example for application of the CG method, acts at the subcellular scale (highlighted in blue) on a timescale shorter than other processes (e.g. cell migration, cell-extracellular matrix interaction at the tissue scale) involved in the multiscale model of angiogenesis \cite{stepanova2021multiscale}. As a result, we may use LDT theory to coarse-grain its dynamics.}
\label{SchematicScales}
\end{figure}

Biological systems are often highly complex, involving processes that may interact across multiple spatial and temporal scales (see \Cref{SchematicScales}). From a general perspective, the subcellular scale is characterised by intracellular chemistry (e.g. gene expression, signal transduction and receptor/ligand dynamics). Subcellular processes determine behaviour at the cellular scale and may generate emergent properties at the tissue scale. In addition to this upward coupling across spatial scales, there is downward coupling whereby extracellular chemicals and biomechanical cues influence the  subcellular chemistry/mechanics within a cell. In this way, dynamic interactions, encompassing all the scales, can occur (\Cref{SchematicScales}). 

From the theoretical perspective, models which consider only processes at a single spatial/temporal scale do not allow for investigation of emergent features which manifest at other scales (for example, collective migration or phenotype patterning which arise from individual cell dynamics and govern tissue scale organisation). Equally, difficulties associated with the physical interpretation of parameters in phenomenological models, i.e. large scale models which capture the overall evolution of a biological process, make it challenging to fit the model to biological data. In particular, this abstract parameter construct hinders model calibration/validation and limits potential applications of the models. Multiscale models, which couple processes at different spatial and/or temporal scales, have the potential to address these issues \cite{bernard2013build}.

A challenge in formulating a multiscale model relates to the number of entities (protein, cells, extracellular components, etc.) that should be included at each scale of interest. Using the same mathematical formalism to model processes involving entities which vary in number by several orders of magnitude may lead to the omission of essential features or make the model computationally intractable. Hybrid approaches are increasingly being recognised as suitable tools for trying to overcome problems of this type and have become a key part of multiscale modelling \cite{deisboeck2011multiscale,deutsch2020multi}. The central idea is to employ the modelling framework most suitable to each subprocess and then to couple them. For example, the extracellular environment and signalling cues are usually modelled deterministically due to the large number of proteins involved. On the other hand, cells may be treated as individual entities, equipped with a subcellular model which determines their behaviour (e.g. proliferation, cell polarity and migration). This framework has been used to develop multiscale models of cancer (see reviews \cite{deisboeck2011multiscale,rejniak2012state} and references therein), angiogenesis \cite{heck2015computational}, collective cell migration \cite{deutsch2020multi}, among other examples \cite{bardini2017multi}.

Hybrid modelling allows for efficient parameter estimation and model visualisation, forging interdisciplinary collaboration between researchers in theoretical modelling and experimental biology \cite{bardini2017multi,osborne2010hybrid}. There is also the potential of using high-throughput experimental data to develop more detailed multiscale models. As an example, one of the aspects of biological systems that has received little attention in theoretical modelling is the effect of stochasticity in the response of individual entities to external stimuli \cite{deutsch2020multi}. Hybrid modelling allows investigation of this effect on the collective, emergent behaviour. However, increasing computational complexity makes these models intractable for large-scale simulations \cite{deisboeck2011multiscale}.

This challenge motivated us to develop a technique which reduces the computational complexity of a model while preserving its stochasticity. The method is applicable to systems characterised by stochastic processes which exhibit multistability and which evolve on timescales shorter than those associated with other system processes. The example that we study in this paper is of this type: the subcellular dynamics of cell fate determination via lateral inhibition (a bistable, stochastic system) act on a shorter timescale than those associated with, for example, cell migration, and tissue scale processes such as the dynamics of extracellular soluble factors (e.g. diffusion, secretion by cells, degradation) \cite{heck2015computational} (\Cref{SchematicScales}). This observation motivates us to use large deviation theory to coarse-grain the dynamics associated with intracellular signalling to produce a jump process (i.e. a Markov chain) on the stable state space of the steady states of the original system which describes the VEGF-Delta-Notch pathway. 

\subsection{Large deviation theory (LDT)}
\label{subsec:LargeDeviationTheory}

In the presence of noise, small fluctuations can drive significant deviations from mean-field behaviour such as, for example, transitions from one stable steady state to another. These transitions are usually referred to as \textit{rare events} since their likelihood is small. LDT is predicated on the assumption that when rare events occur, the system follows the least unlikely paths. Deviations from these paths occur with very small probability (i.e. smaller than the probability of a rare event). Specifically, Freidlin-Wentzell's theory of large deviations predicts that the deviations are exponentially suppressed \cite{freidlin1998random}, making such transitions `predictable'. LDT provides the means to analyse the frequency of rare events and to identify the maximum likelihood path (minimum action path, MAP) along which these transitions can occur.

A stochastic differential equation (SDE) of a diffusion process, $x^{\epsilon} \in \mathbb{R}^n$, has the following form

\begin{equation} \label{SDEgeneral}
\mathrm{d} x^{\epsilon} (t) = b(x^{\epsilon}) \mathrm{d} t + \sqrt[]{\epsilon} \sigma (x^{\epsilon}) \mathrm{d} W,
\end{equation}
where $b: \mathbb{R}^n \rightarrow \mathbb{R}^n$ is a drift vector, $a(x^{\epsilon}) = (\sigma \sigma^T)(x^{\epsilon})$ is a diffusion tensor ($\sigma: \mathbb{R}^n \rightarrow \mathbb{R}^n \times \mathbb{R}^m$, $m$ corresponds to the number of kinetic reactions in the system), $W$ is a Wiener process in $\mathbb{R}^m$ and $\epsilon = \Omega^{-1}$ is noise amplitude.
 
The mean-field limit of \Cref{SDEgeneral}, $x(t) \in \mathbb{R}^n$, solves the following differential equation:
 
 \begin{equation} \label{ODEgeneral}
 \dv{x}{t} = b(x).
 \end{equation}

Assume that \Cref{ODEgeneral} has two stable steady states, $x_1,~ x_2 \in \mathbb{R}^n$, whose basins of attraction form a complete partition of $\mathbb{R}^n$. We are interested in transitions from $x_1 \rightarrow x_2$ (and $x_2 \rightarrow x_1$) which cannot be accounted for unless noise is present in the system.
 
A key player in LDT is the action functional
 
 \begin{equation*} 
 S_T(\psi) = \begin{cases}
 \displaystyle \int_0^T L(\psi, \dot{\psi}) \mathop{dt}, & \text{if $\psi \in C(0,T)$ is absolutely continuous and} \\[-7pt]
 & \text{the integral converges,} \\[5pt]
 + \infty, & \text{otherwise,}
 \end{cases}
\end{equation*}
which is computed for a transition path $\psi : [0,T] \rightarrow \mathbb{R}^n$ from $x_1$ to $x_2$ ($\psi(0) = x_1$ and $\psi(T) = x_2$, $T$ is the transition time). Here  $L(x, y) = \displaystyle \sup_{\theta \in \mathbb{R}^n } \of{\langle y, \theta \rangle - H(x,\theta)} $ is the large deviation Lagrangian, with $\langle \cdot, \cdot \rangle$ being the Euclidean scalar product in $\mathbb{R}^n$ and $H(x,\theta)$ being the Hamiltonian associated with $L(x,y)$. The particular form of the Hamiltonian depends on the dynamical system under consideration (in \cref{supp-appendix:gMAM}, we explain how to define the Hamiltonian for an SDE such as \Cref{SDEgeneral} and a general birth-death CTMC).

\begin{figure}[htbp]
\centering
\subfloat{\includegraphics[width=7cm]{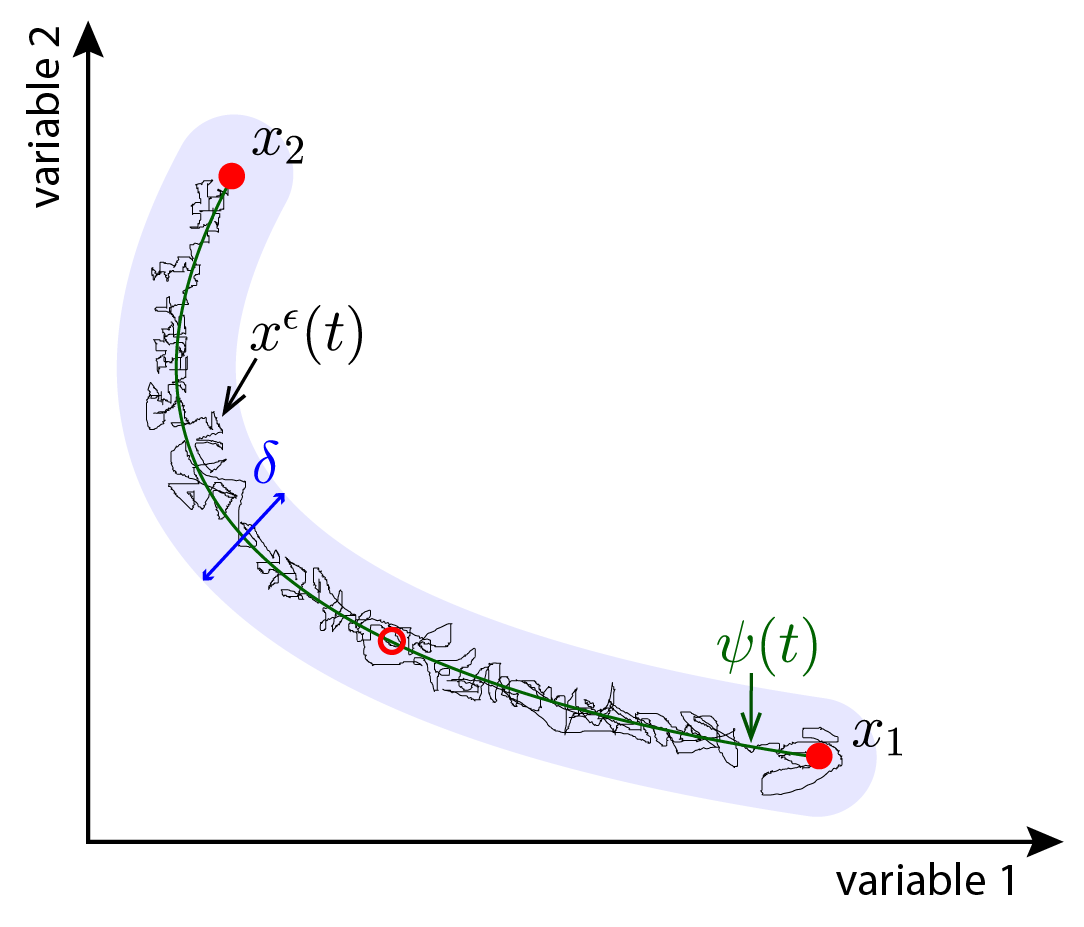}}
\caption{\textbf{An illustration of a transition path between two stable steady states of an arbitrary bistable system.} The two stable steady states, $x_1$ and $x_2$, are marked by filled red circles; an unstable saddle point is marked by an unfilled red circle. The transition path, $\psi(t)$, from $x_1$ to $x_2$ is shown by a thick green line, whereas a single stochastic trajectory, $x^{\epsilon}(t)$, is indicated by a thin black path. The shaded blue region indicates a $\delta$-neighbourhood around $\psi(t)$ ($\delta$ as defined in \Cref{Probability}).}
\label{MapIllustration}
\end{figure}

The action functional is used to estimate the probability that a trajectory $x^{\epsilon}(t)$ lies in a narrow neighbourhood, of width $\delta >0$, of a given path $\psi \in C(0,T)$ (see \Cref{MapIllustration} for an illustration):

 \begin{equation} \label{Probability}
 \prob \uf{ \sup_{0 \leq t \leq T} \mid x^{\epsilon}(t) - \psi(t) \mid < \delta~\middle\vert ~ x^{\epsilon}(0) = x_1} \approx \exp{- \epsilon^{-1} S_T(\psi)}.
\end{equation}

Since the probability function in \Cref{Probability} decreases as the action functional, $S_T(\psi)$, increases, the maximum likelihood path, $\psi^*$, is the minimiser of $S_T(\cdot)$. This leads naturally to the idea of the quasipotential:
 
\begin{equation} \label{Quasipotential}
V(x_1,x_2) = \displaystyle \inf_{T>0} ~ \inf_{\psi \in \overline{C}_{x_1}^{x_2}(0,T)} S_T(\psi).
\end{equation}
Here $\overline{C}_{x_1}^{x_2}(0,T)$ is the space of absolutely continuous functions $f: [0,T] \rightarrow \mathbb{R}^n$ such that $f(0) = x_1$ and $f(T) = x_2$. Roughly speaking, the quasipotential gives an estimate of how `difficult' it is to move from $x_1$ to $x_2$. 

On timescales which are much longer than those associated with relaxation to a stable steady state, the dynamics of \Cref{SDEgeneral} can be reduced, or \textit{coarse-grained}, to that of a CTMC on the state space of the two stable steady states, $\uf{x_1,x_2}$, with transition rates

\begin{equation} \label{CG_rates}
k_{x_1 \rightarrow x_2} \asymp \exp \of{ -\epsilon^{-1} V(x_1,x_2) }, \qquad k_{x_2 \rightarrow x_1} \asymp \exp \of{ -\epsilon^{-1} V(x_2,x_1) }. 
\end{equation}
\noindent Here $\asymp$ denotes log-asymptotic equivalence so that $f(\epsilon) \asymp g(\epsilon)$ if and only if \\ ${\lim_{\epsilon \rightarrow  0} \frac{\log f(\epsilon)}{\log g(\epsilon)}=1}$.

In practice, most double minimisation problems, such as \Cref{Quasipotential}, do not have a solution for finite $T>0$. Furthermore, closed-form Lagrangians exist for SDEs of the type defined by \Cref{SDEgeneral} but not for general birth-death CTMCs. \Cref{Quasipotential} can be reformulated in terms of a Hamiltonian system of the form

\begin{equation*}  \label{HamiltonianEquations}
\dv{\phi}{t} = \pdv{H(\phi,\theta)}{\theta}, \qquad \dv{\theta}{t} = - \pdv{H(\phi, \theta)}{\phi}.
\end{equation*}
\noindent This problem must be solved as a boundary-value problem, i.e. $\phi(0) = x_1$ and $\phi (T) = x_2$, on an infinite time interval, $T \rightarrow \infty$, \cite{grafke2017long} which makes it a non-trivial numerical problem. Thus the traditional LDT methods are inapplicable in most cases.

One way to resolve these problems is to reformulate the minimisation problem defined by \Cref{Quasipotential} on the space of curves (i.e. transition paths from one stable steady state to another). In \cite{heymann2008geometric}, Heymann and Vanden-Eijnden proved that the minimisation problem defined by \Cref{Quasipotential}, is equivalent to 

\begin{equation} \label{QuasipotentialGeometrical}
V(x_1,x_2) = \displaystyle \inf_{\phi} \widehat{S}(\phi), ~~\text{with}  ~~ \widehat{S}(\phi) = \displaystyle \sup_{\substack{\hat{\theta}:[0,1] \rightarrow \mathbb{R}^n \\ H(\phi,\hat{\theta}) =0}} \displaystyle \int_0^1 \langle \phi',\hat{\theta} \rangle \mathop{d\alpha},
\end{equation}
where $\phi: [0,1] \rightarrow \mathbb{R}^n$ is a curve from $x_1$ to $x_2$ parametrised by standard arc length.

The geometric reformulation, \Cref{QuasipotentialGeometrical}, resolves analytically the issue of the infinite time, $T$, in the original minimisation problem. Furthermore, only the Hamiltonian is needed. In this respect, the method is more general as it can be applied to SDEs, CTMCs and other systems for which the Hamiltonian is known (see \cref{supp-appendix:gMAM} in \hyperlink{SuppMaterial}{Supplementary Material}). 

In \cite{heymann2008geometric}, an algorithm was developed to efficiently compute $V(x_1,x_2) $ and the corresponding minimiser, $\phi^*$, from the geometric reformulation. The algorithm is known as the geometric minimum action method (gMAM) and the minimiser, $\phi^*$, of the action functional is referred to as the minimum action path (MAP) (for more details see \cref{supp-appendix:gMAM}).

Once the quasipotential has been computed, the coarse-grained system is given by a CTMC, with rates defined by \Cref{CG_rates}.

\section{Coarse-graining (CG)}
\label{sec:CoarseGraining} 

We now illustrate how the theory described in the previous section can be used to coarse-grain a specific hybrid multiscale model, one for which the internal dynamics of the agents are described by multistable stochastic systems. This property is characteristic of, for example, systems driving cell fate (phenotype) determination. We begin by using LDT to formulate a CG model for a system comprising a single agent (here a cell). The subcellular signalling pathway, which we use to illustrate the method, is the VEGF-Delta-Notch pathway (see \cref{supp-appendix:VEGFDeltaNotch} in \hyperlink{SuppMaterial}{Supplementary Material} and \cite{stepanova2021multiscale} for details). This pathway regulates phenotypic adaptation via lateral inhibition \cite{collier1996,monk2001spatiotemporal}. This system meets the requirements for application of the CG technique: \textit{(a)} it is bistable; its stable steady states are associated with cellular phenotypes (Delta-high and Delta-low cells); \textit{(b)} we are interested in its evolution on timescales longer than the typical time for relaxation to an equilibrium since other processes (e.g. cell migration and dynamics of extracellular matrix) act on longer timescales (see \Cref{SchematicScales}). 

We then extend the method to the general case of multi-agent systems. Here the dynamics of each entity is coarse-grained to a CTMC on the state space of its stable states, and coupling between the internal dynamics of individual agents is achieved via the external variables whose dynamics depend on the states of neighbouring agents and/or the time evolution of these variables. We outline below how we apply this method to a monolayer of cells (motivated by phenotype patterning via the core Delta-Notch pathway in cell monolayers \cite{monk2001spatiotemporal}) and a branching network (angiogenesis-motivated application \cite{stepanova2021multiscale}) that interact via VEGF-Delta-Notch signalling. 

\subsection{Individual agent system} 
\label{subsec:IndividualCellSystem}

\begin{figure}[htbp]
\centering
\subfloat{\includegraphics[width=13cm]{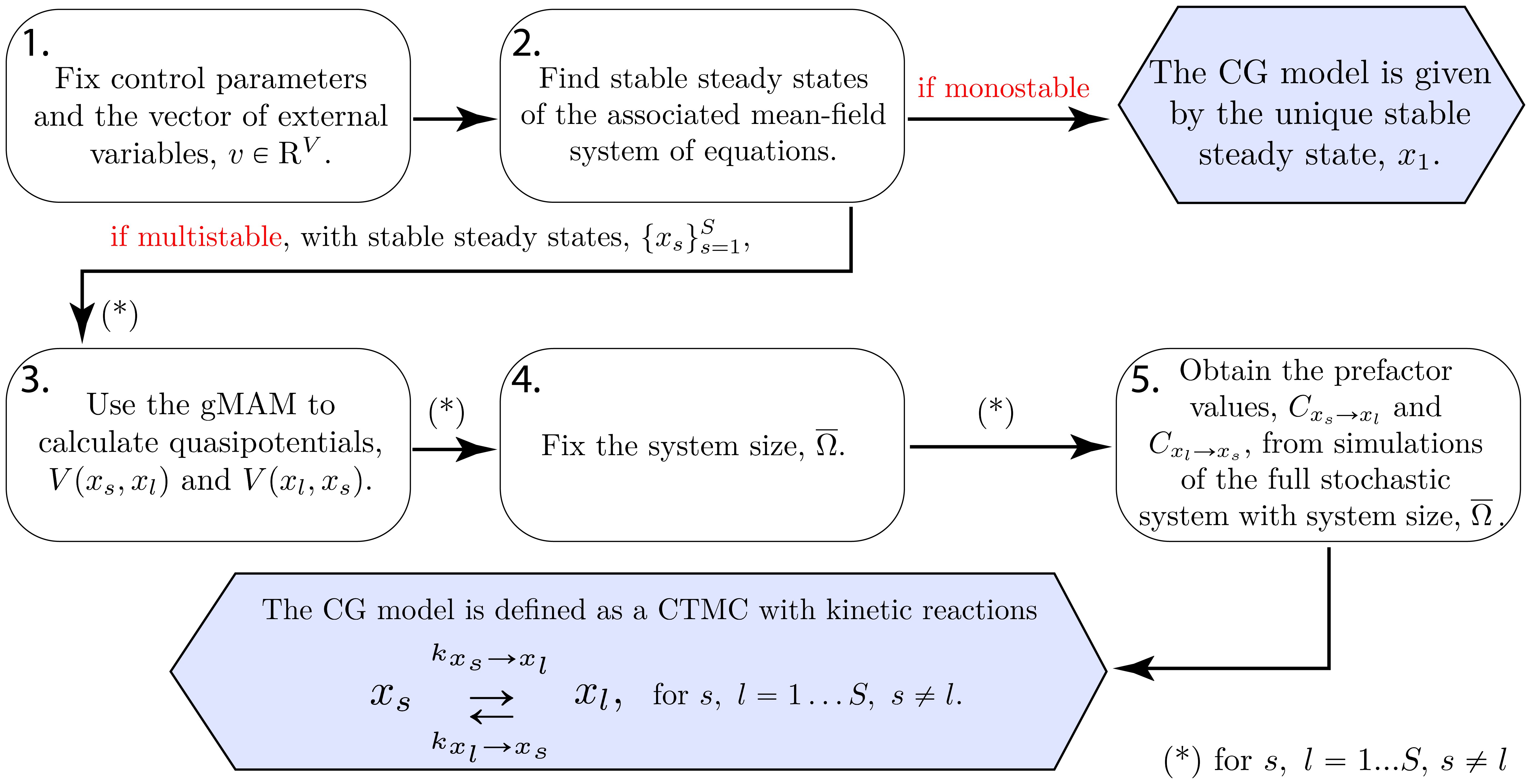}}
\caption{\textbf{A flowchart of the procedure used to coarse-grain a multistable stochastic system for an individual entity.} The steady state solutions, quasipotential and prefactor depend on the model parameters and external variables, $v \in \mathrm{R}^V$ ($V$ indicates the dimension of the vector of external variables). Here the transition rates, $k_{x_s \rightarrow x_l}$, are defined by \Cref{CG_rates_DN}, the prefactor, $C_{x_s \rightarrow x_l}$, is determined from \Cref{PrefactorFromData}, and $\overline{\Omega}$ is given by \Cref{OmegaEstimate}.}
\label{FlowchartIndividualCell}
\end{figure} 

Our algorithm for coarse-graining a stochastic system with a region of multistability involving a single entity is illustrated in \Cref{FlowchartIndividualCell}. For the particular case of VEGF-Delta-Notch signalling, a cell's internal state (phenotype) depends on two model parameters (inputs) corresponding to the extracellular levels of Delta and Notch, $v=\of{d_{ext}, n_{ext}} \in \mathrm{R}^2$ (see \cref{supp-appendix:VEGFDeltaNotch}). We fix the values of the model parameters and the external variables, $v$ (see \Cref{supp-Params}). We then use the mean-field system defined by \Cref{supp-eq:DN_single_nondimensional} to compute the steady state solutions. For this example, the values of the external variables, $v$, are chosen so that the system is bistable; the two stable steady states correspond to Delta-high and Delta-low cell phenotypes, $\uf{x_1,x_2} = \uf{\text{Delta-high},~\text{Delta-low}}$, and the unstable steady state is an unstable saddle. Our goal is to compute the transition rates of the CG system which we approximate as follows:

\begin{equation} \label{CG_rates_DN}
k_{x_s \rightarrow x_l} \approx C_{x_s \rightarrow x_l}\exp \of{ -\Omega V(x_s, x_l) }, \quad s,~l \in \uf{1,2},~s \neq l.
\end{equation} 
We note that the prefactor, $C_{x_s \rightarrow x_l}$, arises from the asymptotic equivalence relation defined by \Cref{CG_rates}. The system size is given by $\Omega = \epsilon^{-1}$, where $\epsilon$ is the noise level.

\begin{figure}[htbp]
\subfloat[][system size, $\Omega=70$ (noise level, $\epsilon=1/\Omega \approx 0.014$) \label{MAP_Path_s2t}]{\includegraphics[height = 8cm,valign=b]{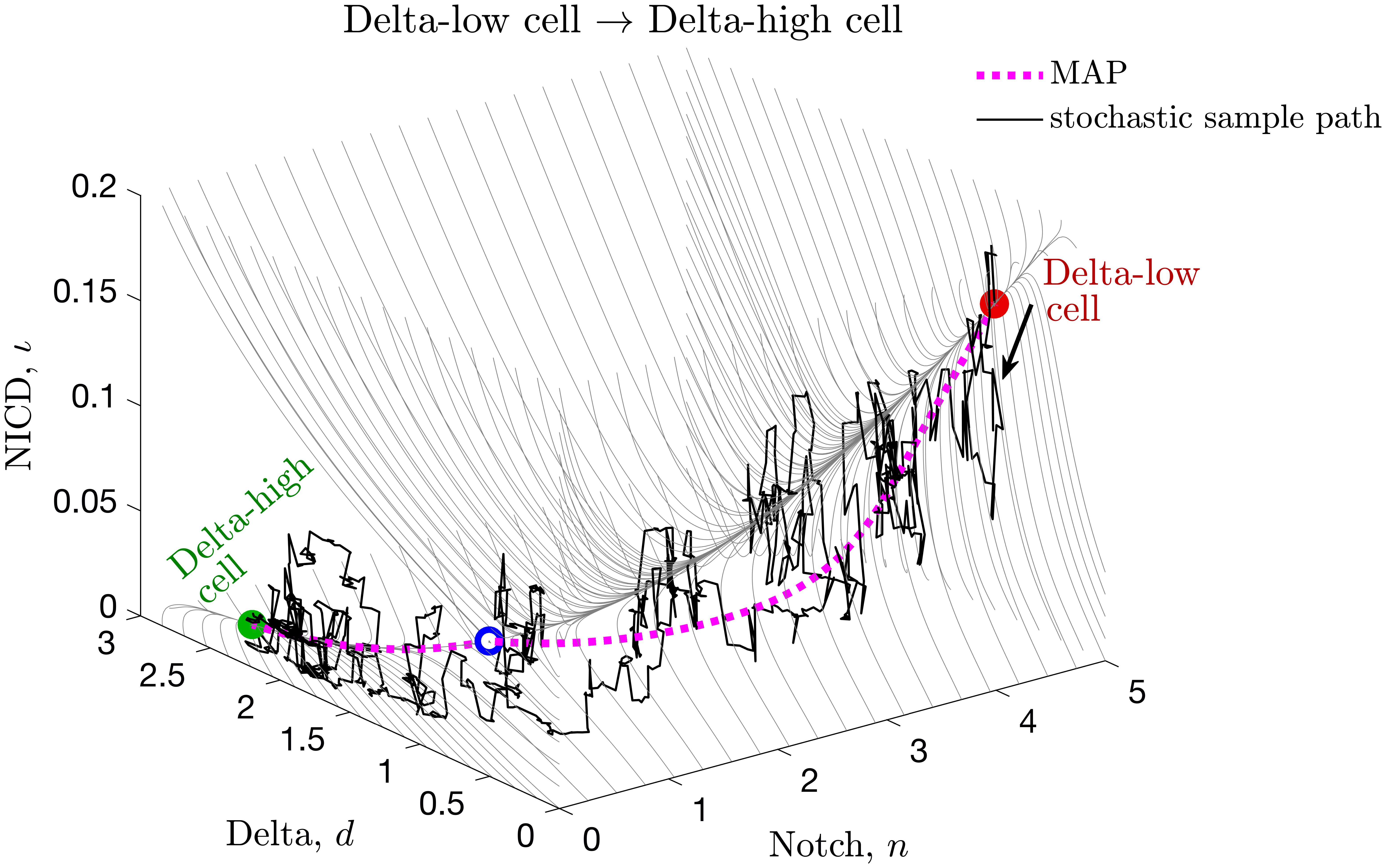}}
\hfill
\subfloat[][system size, $\Omega=450$ (noise level, $\epsilon=1/\Omega \approx 0.002$) \label{MAP_Path_t2s}]{\includegraphics[height = 8cm,valign=b]{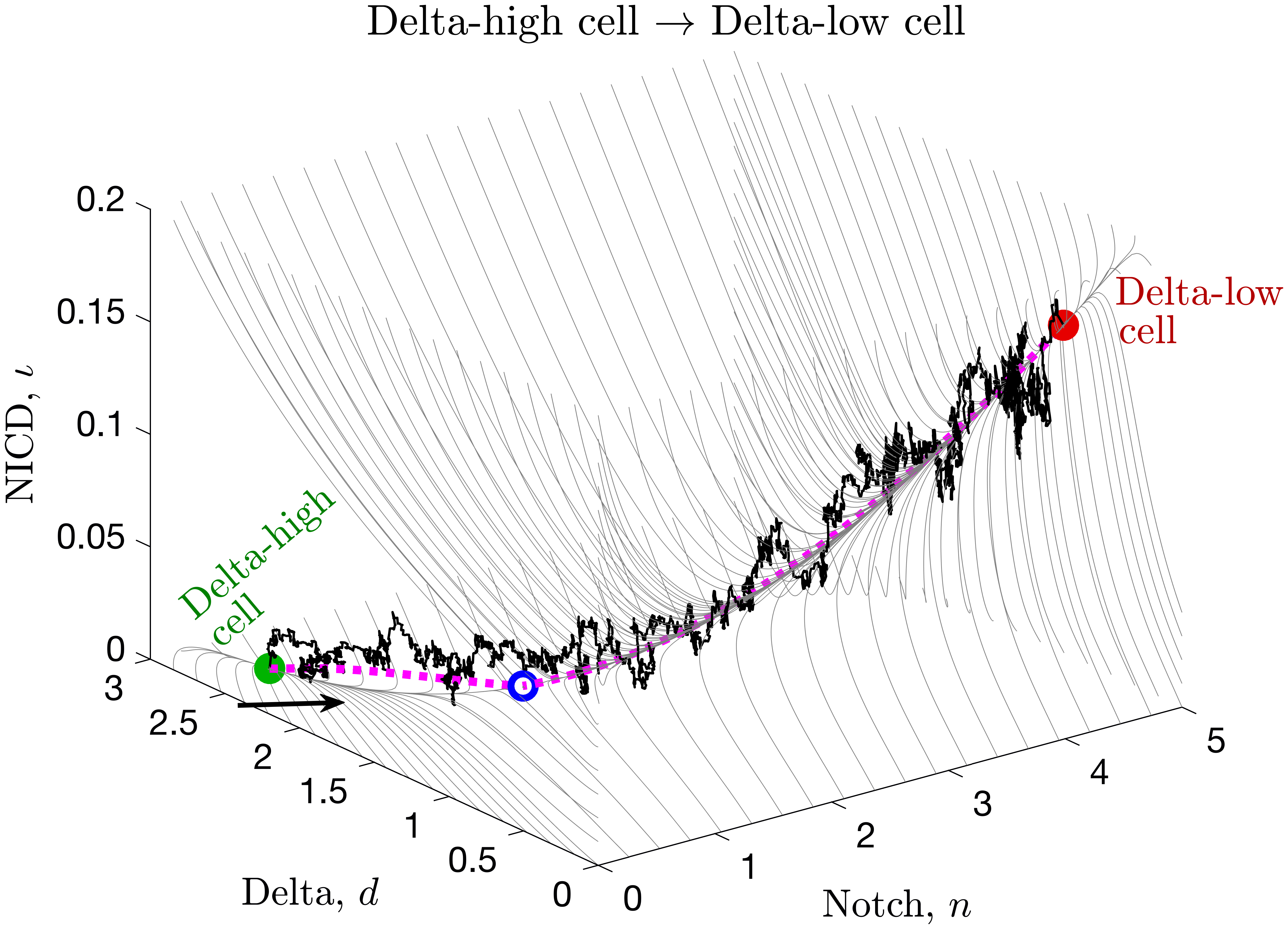}}
\caption{\textbf{An illustration of the minimum action paths (MAPs) and stochastic sample paths for transitions between the Delta-high and Delta-low cell phenotypes.} We computed the MAPs (indicated by the dotted magenta lines) for the subcellular VEGF-Delta-Notch system in an individual cell using the gMAM for transitions from (a) Delta-low to Delta-high cell and (b) Delta-high to Delta-low cell. The stochastic sample paths obtained by simulating the full stochastic CTMC model (\Cref{supp-FullStochasticSystem}) with the system sizes (a)  $\Omega=70$,  (b)  $\Omega=450$, are plotted in black. The thin grey lines indicate streamlines of the corresponding mean-field system (\Cref{supp-eq:DN_single_nondimensional}). The Delta-high (Delta-low) cell stable steady state is indicated by a green (red) filled circle; the unstable saddle by a blue unfilled circle. The plots represent three-dimensional projections of the full five-dimensional system as defined by \Cref{supp-eq:DN_single_nondimensional}. Parameter values are fixed as indicated in \Cref{supp-Params}.}
\label{Map_with_SamplePath}
\end{figure}

We use the gMAM to compute the quasipotential values and corresponding paths (MAPs) for transitions between the Delta-high and Delta-low phenotypes (for more details, see \cref{supp-appendix:MAP} in \hyperlink{SuppMaterial}{Supplementary Material}). An illustrative example is shown in \Cref{Map_with_SamplePath}, where we compare the MAPs and sample paths of the full stochastic CTMC for an individual cell (see also \Cref{supp-FullStochasticSystem} in \cref{supp-appendix:VEGFDeltaNotch}). Several characteristic features of the phenotype transitions are noteworthy. First, the dynamics of the MAP can be split into two parts: the transition from the steady state of origin to the saddle point (for example, from the Delta-low phenotype to the saddle point, indicated by the blue circle in \Cref{MAP_Path_s2t}) which is possible due to the presence of noise. The main contribution to the quasipotential comes from this transition. The MAP from the unstable saddle point to the stable steady state of destination (from the saddle point indicated by the blue circle to the Delta-high phenotype in \Cref{MAP_Path_s2t}) follows the fastest route given by the deterministic heteroclinic orbit connecting the steady states (i.e. the unstable saddle and the stable Delta-high cell state). The second noteworthy feature of the phenotype transitions is that, as the level of noise, $\epsilon$, decreases, the stochastic sample path follows the MAP more closely (compare \Cref{MAP_Path_s2t,MAP_Path_t2s} for which $\Omega = \epsilon^{-1} = 70$ and $\Omega = \epsilon^{-1}= 450$, respectively).

\renewcommand{\thesubfigure}{\alph{subfigure}}
\captionsetup{singlelinecheck=off}

To fully determine the CG transition rates, the prefactor value, $C_{x_s \rightarrow x_l}$, must be estimated. From \Cref{CG_rates_DN}, for $s,~l \in \uf{1,2},~s \neq l$, we have

\begin{subequations}
\begin{align}
\log \langle T^{\Omega}_{x_s \rightarrow x_l} \rangle &\approx \Omega V(x_s,x_l) - \log C_{x_s \rightarrow x_l}~, \label{PrefactorMeanPassageTime} \\ 
\log C_{x_s \rightarrow x_l}  &\approx \Omega V(x_s,x_l) -\log \langle T^{\Omega}_{x_s \rightarrow x_l} \rangle~, \label{PrefactorFromData}
\end{align} \label{Prefactor}
\end{subequations}
where $\langle T^{\Omega}_{x_s \rightarrow x_l} \rangle = 1/k_{x_s \rightarrow x_l}$ is the mean passage time between the stable steady states, $x_s$ and $x_l$ (Delta-high and Delta-low phenotypes), for a fixed value of the system size, $\Omega$. $\langle T^{\Omega}_{x_s \rightarrow x_l} \rangle$ can be determined from direct simulation of the full stochastic model using the reaction kinetics given in \Cref{supp-FullStochasticSystem}. 

\renewcommand{\thesubfigure}{\alph{subfigure}}
\captionsetup{singlelinecheck=off}
\begin{figure}[htbp]
\centering 
\subfloat[][\label{OmegaConvergence_s2t}]{\includegraphics[height = 5.5cm,valign=b]{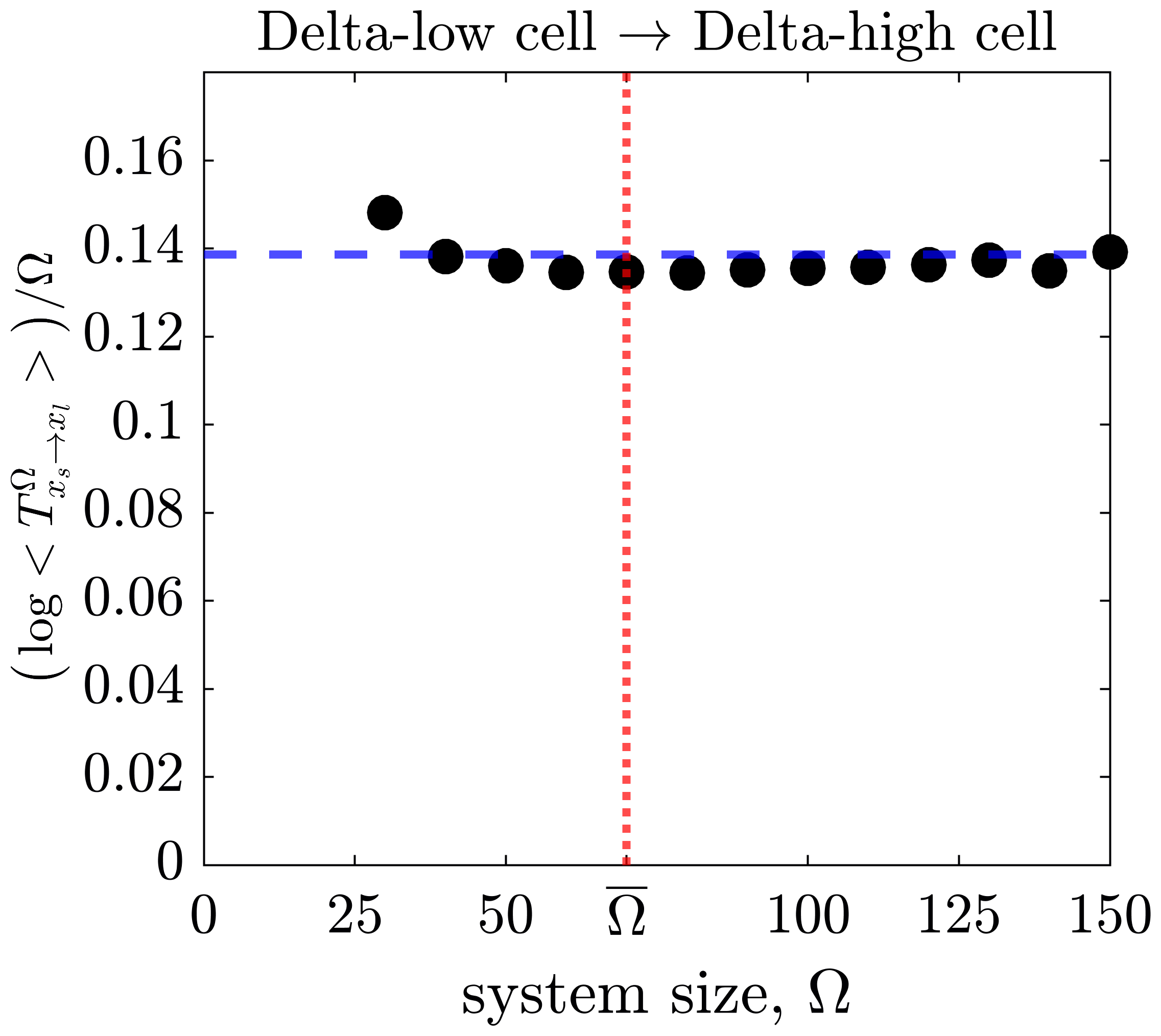}}
\hfill
\subfloat[][\label{OmegaConvergence_t2s}]{\includegraphics[height = 5.5cm,valign=b]{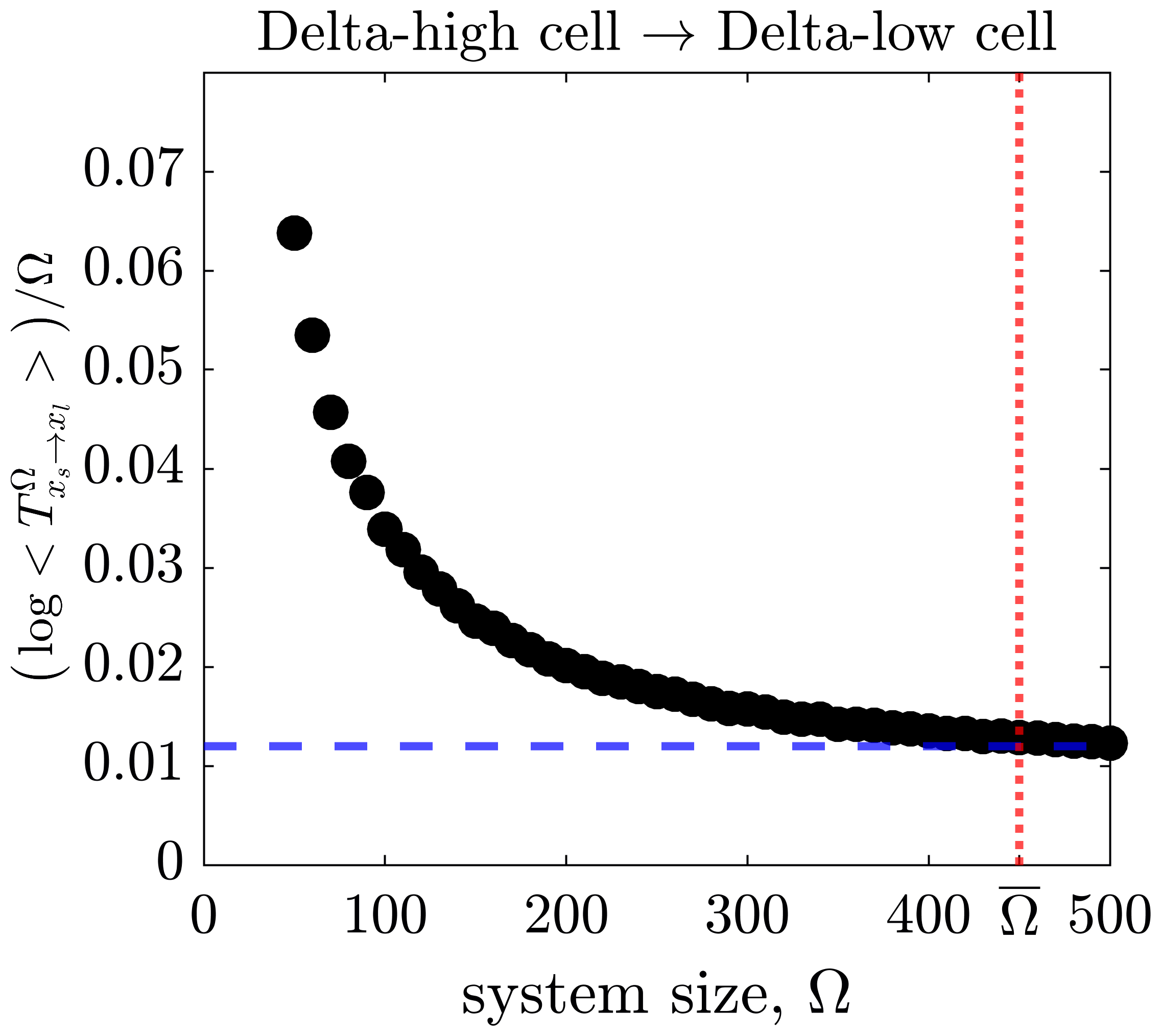}}
\vspace*{0.3cm}
\caption{\textbf{Convergence of the quasipotential, $V(x_s,x_l)$, as the system size, $\Omega$, increases.} We ran 1000 realisations of the stochastic VEGF-Delta-Notch model for an individual cell (see \Cref{supp-FullStochasticSystem}) for fixed values of $d_{ext}=0.2$, $n_{ext}=0.5$ and increasing system size, $\Omega$. We plotted the convergence to the quasipotential value (a) $V(\text{Delta-low}, \text{Delta-high})$ and (b) $V(\text{Delta-high},\text{Delta-low})$ as a function of $\Omega$ (black circle markers). For these parameter values, transitions from the Delta-low to Delta-high phenotype are less likely to occur (higher noise levels, $\epsilon = \Omega^{-1}$, and/or longer transition times are needed) than transitions from the Delta-high to Delta-low phenotype (see \Cref{PrefactorMeanPassageTime}). Therefore, the perturbations of this random event are smaller and convergence is reached for higher values of noise. This is why lower values of $\Omega$  in (a) suffice to accurately determine the prefactor value from \Cref{Prefactor}. The blue dashed lines indicate the value of the corresponding quasipotential computed via the gMAM; the red dotted lines indicate $\overline{\Omega}$ from \Cref{OmegaEstimate}. All other parameter values are fixed as indicated in \Cref{supp-Params}.}
\label{OmegaConvergence}
\end{figure}

An accurate estimate of the quasipotential (as obtained via the gMAM) allows us to obtain the prefactor given the mean passage time, $\langle T^{\Omega}_{x_s ~\rightarrow ~x_l} \rangle$, for a single value of the system size, $\Omega$. However,  the approximate relation in \Cref{Prefactor} is valid in the limit $\Omega \rightarrow \infty$ (see \Cref{OmegaConvergence}). Thus, $\Omega$ should be chosen sufficiently large to achieve convergence in \Cref{Prefactor} and, at the same time, not too large in order to ensure that transitions between the phenotypes occur in a computationally feasible time, since the waiting times for transitions between stable steady states increase exponentially as $\Omega$ grows. Specifically, we fix a maximum simulation time, $T_{max}$, and an average prefactor value, $\bar{C}$ (determined computationally from simulations), and approximate the corresponding system size, $\overline{\Omega}$, as:

\begin{equation} \label{OmegaEstimate}
\overline{\Omega} \approx \frac{\log T_{max} + \log \bar{C} }{V(x_s, x_l)}.
\end{equation}
Then the prefactor, $C_{x_s \rightarrow x_l}$, can be approximated using \Cref{PrefactorFromData} with $\Omega = \overline{\Omega}$. 

\begin{figure}[htbp]
\centering 
\subfloat[][ $\text{Delta-low} \rightarrow \text{Delta-high} $ \label{PrefactorFit_s2t}]{\includegraphics[height = 4.8cm,valign=b]{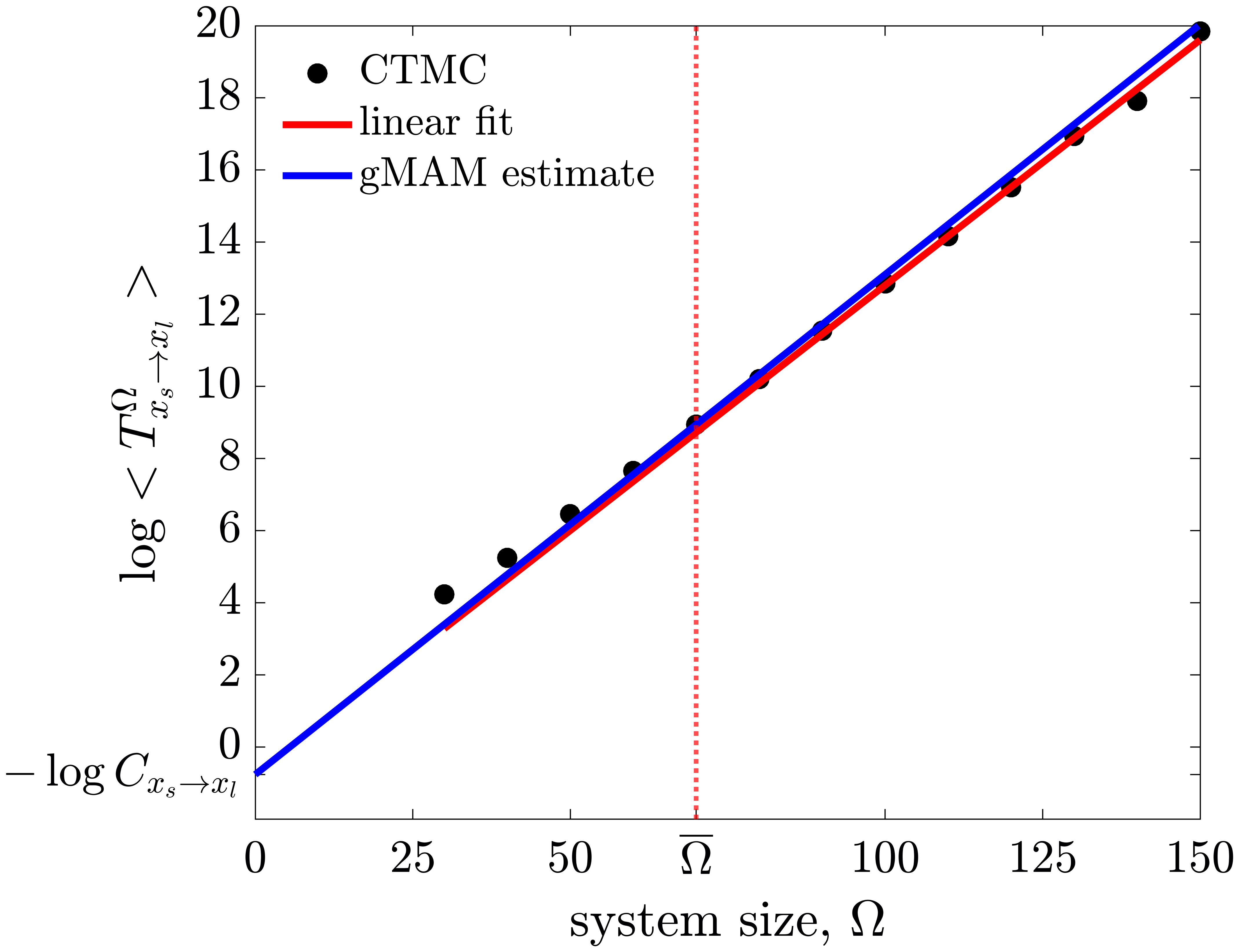}}
\hfill
\subfloat[][ $\text{Delta-high} \rightarrow \text{Delta-low}$ \label{PrefactorFit_t2s}]{\includegraphics[height = 4.8cm,valign=b]{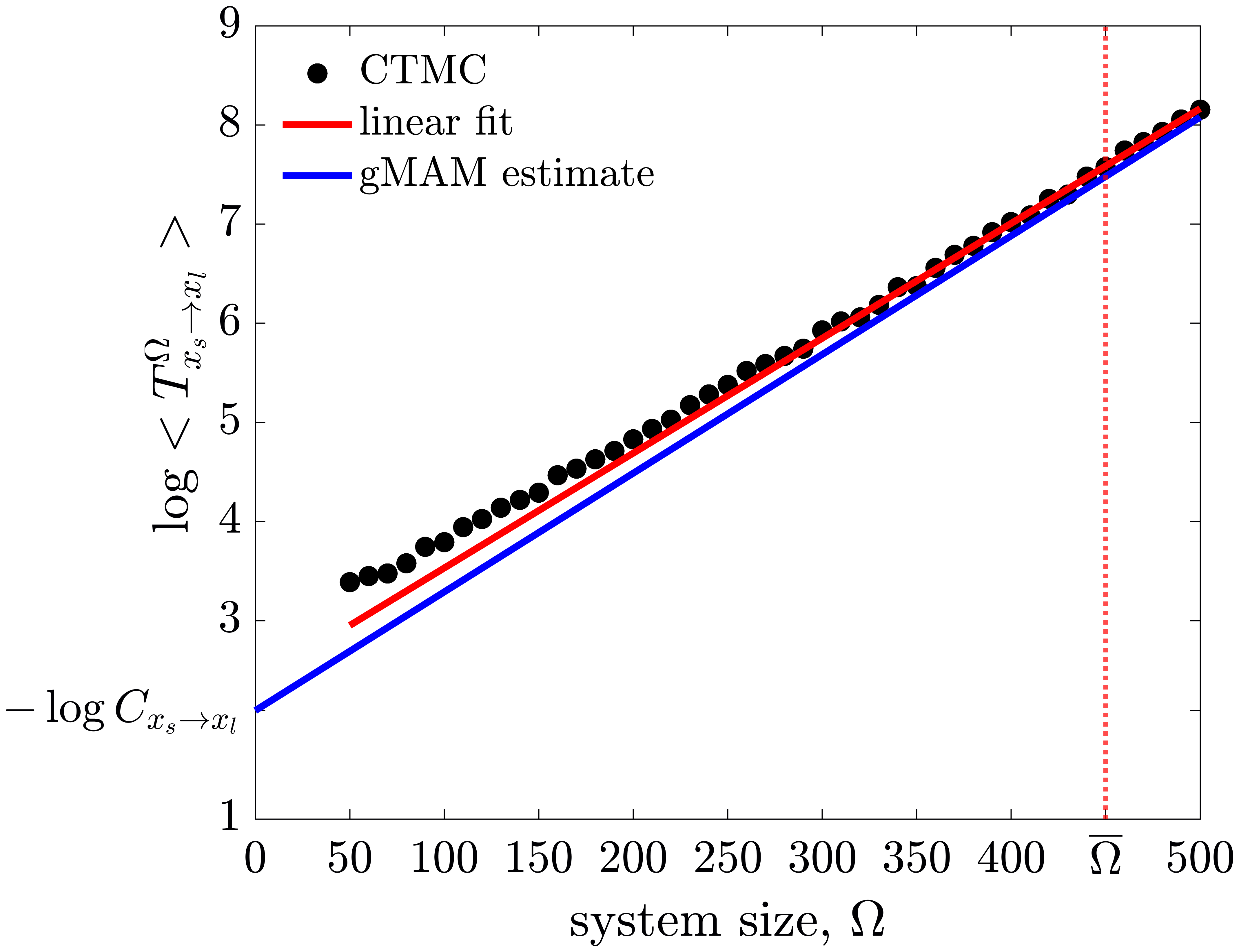}}
\caption{\textbf{Prefactor estimation.} Comparison of prefactor estimates obtained from simulations of the full stochastic CTMC model (black circles) and estimates obtained using the gMAM-quasipotential and mean passage times for a single value of the system size, $\overline{\Omega}$ (blue line), see \Cref{PrefactorMeanPassageTime}. The linear fit of the full stochastic data (red line) was performed for values of $\Omega$ such that the corresponding sample $\uf{T^{\Omega}_{x_s ~\rightarrow ~x_l}}$ is exponentially distributed (high levels of noise might affect the distribution of these transitions). Panel (a) corresponds to the transition from Delta-low to Delta-high phenotype; panel (b) corresponds to the transition from Delta-high to Delta-low phenotype. The red dotted lines indicate $\overline{\Omega}$ from \Cref{OmegaEstimate}. All other parameter values are fixed as indicated in \Cref{supp-Params}.}
\label{PrefactorFit}
\end{figure}

From \Cref{PrefactorMeanPassageTime}, we know that $\log \langle T^{\Omega}_{x_s ~\rightarrow ~x_l} \rangle$ is a linear function of $\Omega$ whose slope and intercept are given by the quasipotential, $V(x_s, x_l)$, and $\of{- \log C_{x_s ~ \rightarrow ~ x_l}}$, respectively. Thus, in order to check the accuracy of our estimate for the system size, $\overline{\Omega}$ (\Cref{OmegaEstimate}), we compared linear fitting of data obtained from the full stochastic CTMC model for increasing $\Omega$, with the estimate obtained from the gMAM quasipotential and the prefactor extracted from simulations with system size, $\overline{\Omega}$. The results presented in \Cref{PrefactorFit} show that the estimates converge as $\Omega$ increases, confirming the accuracy of the two methods.

\renewcommand{\thesubfigure}{\alph{subfigure}}
\captionsetup{singlelinecheck=off}

To summarise, we coarse-grain the stochastic VEGF-Delta-Notch dynamics as follows (see \Cref{FlowchartIndividualCell}):

\vspace*{0.3cm}
\begin{enumerate}[label=\Roman*]
\item Fix the model parameter values and the vector of external variables, $v$, which, for this system, is given by the extracellular levels of Delta and Notch, $v=\of{d_{ext},n_{ext}}$. 
\item Compute the steady states of the corresponding mean-field system (\Cref{supp-eq:DN_single_nondimensional}).
\item Formulate the CG model:
\begin{enumerate}[label=\roman*]
\item If, for the given $v=\of{d_{ext},n_{ext}}$, the system is monostable (either Delta-high or Delta-low cell steady state exists), then the quasipotential value to arrive at this state is 0. The value of the other quasipotential can be assumed infinite (since the system is monostable, this transition is impossible). For example, if the only stable steady state is the Delta-high cell, then $V(\text{Delta-low}, \text{Delta-high})=0$ and $V(\text{Delta-high}, \text{Delta-low}) = \infty$. The CG model is defined by its unique stable steady state. 
\item If the system is within the bistable regime (both Delta-high and Delta-low steady states are stable), then the CG model is defined as a CTMC on the state space of $\{x_s,x_l\}=\{\text{Delta-high},$ $\text{Delta-low}\}$. The transition rates are given by \Cref{CG_rates_DN}. The quasipotential, $V(x_s,x_l)$, is approximated using the gMAM; the prefactor value, $C_{x_s ~ \rightarrow ~ x_l}$, is obtained via \Cref{PrefactorFromData} from stochastic simulations of the full VEGF-Delta-Notch model for a fixed value of the system size, $\overline{\Omega}$, defined by \Cref{OmegaEstimate}.
\end{enumerate}
\item The CG model can be simulated using any variant of the SSA, such as, for example, the classical Gillespie algorithm \cite{gillespie1976general}.
\end{enumerate}

\vspace*{10pt}
The above method generalises naturally for systems with an arbitrary number of stable steady states (see \Cref{FlowchartIndividualCell}). In this case, the quasipotential and the corresponding prefactor must be approximated for each pair of stable steady states. The method can also be applied to systems which possess other attractors, e.g. limit cycles \cite{de2018minimum,freidlin1998random}.

\subsection{Multi-agent system}
\label{subsec:MulticellularSystem}

\begin{figure}[htbp]
\centering
\subfloat{\includegraphics[width=13cm]{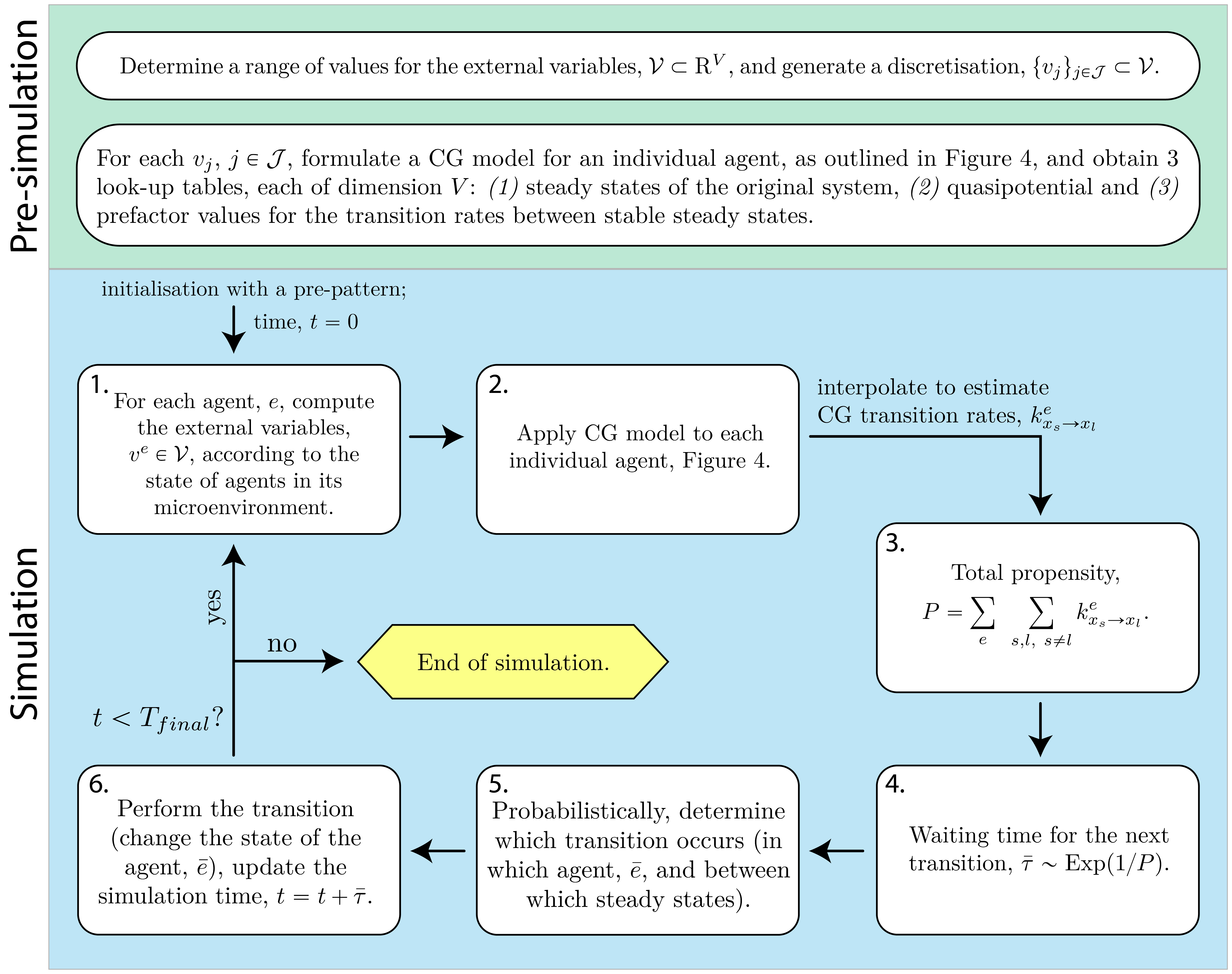}}
\caption{\textbf{A flowchart of the procedure to coarse-grain a multi-agent stochastic system with a region of multistability.} A pseudocode of the simulation algorithm for the multi-agent CG model is presented in \cref{supp-appendix:SimulationAlgorithm}. The \textit{simulation} part of the diagram illustrates an iteration of the Gillespie algorithm for simulation of multi-agent CG systems. Here $T_{final}$ stands for the final simulation time; $\mathrm{Exp}(\lambda)$ is an exponential distribution of intensity, $\lambda$.}
\label{FlowchartMulticellular}
\end{figure} 

In this section we show how the CG method can be applied to multi-agent systems with a region of multistability. In this case, the dynamics of \textit{each} agent is coarse-grained to that of a CTMC between its stable steady states for given values of the external variables, $v$, which establish the coupling between the internal dynamics of individual agents ($v$ depends on the state of agents in the local environment of the focal agent and/or time, and defines its internal state, e.g. phenotype). If the dynamics of an individual agent are independent of its neighbours and time (i.e. the values of the external variables are constant) then we use the CG method described in \cref{subsec:IndividualCellSystem} (see also \Cref{FlowchartIndividualCell}). A suitable range of values for the external variables, $v \in \mathcal{V}$, where $\mathcal{V} \subset \mathrm{R}^V$, can be determined by simulating the original multiscale model. Here $V$ indicates the dimension of the vector of external variables, $v$. In order to reduce the computational cost in the multi-agent CG system, it is convenient to calculate \textit{a priori} look-up tables for the steady states, quasipotential and prefactor values for a discretisation, $\uf{v_j}_{j \in \mathcal{J}} \subset \mathcal{V}$ (here, $j$ indexes entries in the generated discretisation; $\mathcal{J}$ is the size of the discretisation). Interpolation routines can then be used to establish an input-output relationship between an arbitrary $v \in \mathcal{V}$ and the values of the corresponding steady states and the transition rates between them. Therefore, we split the general CG method for multi-agent systems into two steps (see \Cref{FlowchartMulticellular}):

\vspace*{10pt}
\begin{enumerate}[label=(\roman*)]
\item \textbf{Pre-simulation:} calculate look-up tables for the system steady states, quasi-potential and prefactor values for each entry in a discretisation, $\uf{v_j}_{j \in \mathcal{J}}$, for a range of values of the external variables, $\mathcal{V} \subset \mathrm{R}^V$.
\item \textbf{Simulation:} the CG model is simulated (via, e.g., the Gillespie algorithm) as a CTMC on a state space defined by the steady states of all of its entities, with the coupling maintained via the external variables, $v$, updated at each simulation step according to entities' local environments and/or time. 
\end{enumerate}

\vspace*{10pt}
We now provide more details on the pre-simulation and simulation steps. 

\subsubsection{Pre-simulation: look-up tables}
\label{subsubsec:LookUpTables}

Pre-computed look-up tables of system steady states, quasipotential and prefactor values are used to interpolate the values of the system steady states and the CG transition rates between them for an arbitrary set of values of the external variables, $v$, without calculating them explicitly at each step during simulations of the CG model. In a general setting, the dimension of each table is equal to $V$, the dimension of the vector of external variables.

The steady states must be computed numerically for each entry $v_j$ in the discretisation, $\uf{v_j}_{j \in \mathcal{J}}$, using the mean-field limit for an individual entity (as described in \cref{subsec:IndividualCellSystem}). For values of $v_j$ that fall within the multistability region, the quasipotential is computed via the gMAM in a pair-wise manner, for each pair of stable steady states, $\uf{x_s}_{s=1}^{\mathcal{S}}$. The last look-up table corresponds to the prefactor, $C_{x_s \rightarrow x_l}$, $x_s,~x_l \in 1 \ldots \mathcal{S}$, which must be approximated for each $v_j$ within the multistability region. The prefactor values are obtained from \Cref{PrefactorFromData} as before, using the mean passage times, $\langle T^{\Omega}_{x_s ~\rightarrow ~x_l} \rangle$, which are determined by simulating the full stochastic model with the system size, $\Omega = \overline{\Omega}$, defined by \Cref{OmegaEstimate}.

\subsubsection{Simulation algorithm}
\label{subsubsec:SimulationAlgorithm}

Once all the look-up tables have been computed, the multi-agent CG system can be simulated as a standard Gillespie algorithm (or one of its variants, e.g., Next Subvolume method \cite{elf2004spontaneous}) in which the total propensity, $P$, at each time step is computed as a sum of transitions, $k^e_{x_s \rightarrow x_l}$, for each entity, $e$, to switch its (stable) state (see \Cref{FlowchartMulticellular}). The steady states corresponding to each entity (and the transition rates between them) for the exact value of the external variables, $v^e \in \mathcal{V}$, ($v^e$ has to be computed for each entity, $e$, according to its microenvironment) are interpolated via appropriate numerical routines. We present pseudocode for the simulation procedure in \cref{supp-appendix:SimulationAlgorithm}. 

Note that our CG method does not account for the initial, relatively short (compared to the LDT timescale), relaxation time during which the system relaxes onto the timescale on which the CG approximation is valid. Thus, it is necessary to obtain an initial stable steady state configuration, i.e to \textit{pre-pattern} the system, using either the full stochastic CTMC or the mean-field model (see \Cref{FlowchartMulticellular} and line 5 in \Cref{supp-SimulationAlgorithm}). The final simulation time for the pre-patterning should be large enough to ensure that the system relaxes to an equilibrium. Since this procedure is performed only once, it does not affect the computational complexity of the CG simulations. We have chosen to use the mean-field system to pre-pattern our simulations since it is less time-consuming and the stochasticity (i.e. transitions between phenotypes) is preserved later in the CG simulation loop. 

\section{Results}
\label{sec:Results}

For illustrative purposes, we consider the specific example of spatial phenotype patterning via the Delta-Notch lateral inhibition mechanism in response to an external signalling cue (VEGF). First, we provide more details about our implementation of the CG model and present typical simulation results and the robust patterns that emerge at long times. We then discuss the relative merits of the CG method, using a variety of metrics to compare its performance with the original stochastic and mean-field systems. We used the Next Subvolume method \cite{elf2004spontaneous} for simulations of the full stochastic CTMC and the Euler-Lagrange method (explicit scheme) for the numerical integration of the mean-field equations. 

\subsection{CG model of spatial cell phenotype patterning} \label{subsec:CGModel}

\begin{figure}[htbp]
\centering
\subfloat{\includegraphics[height=5cm]{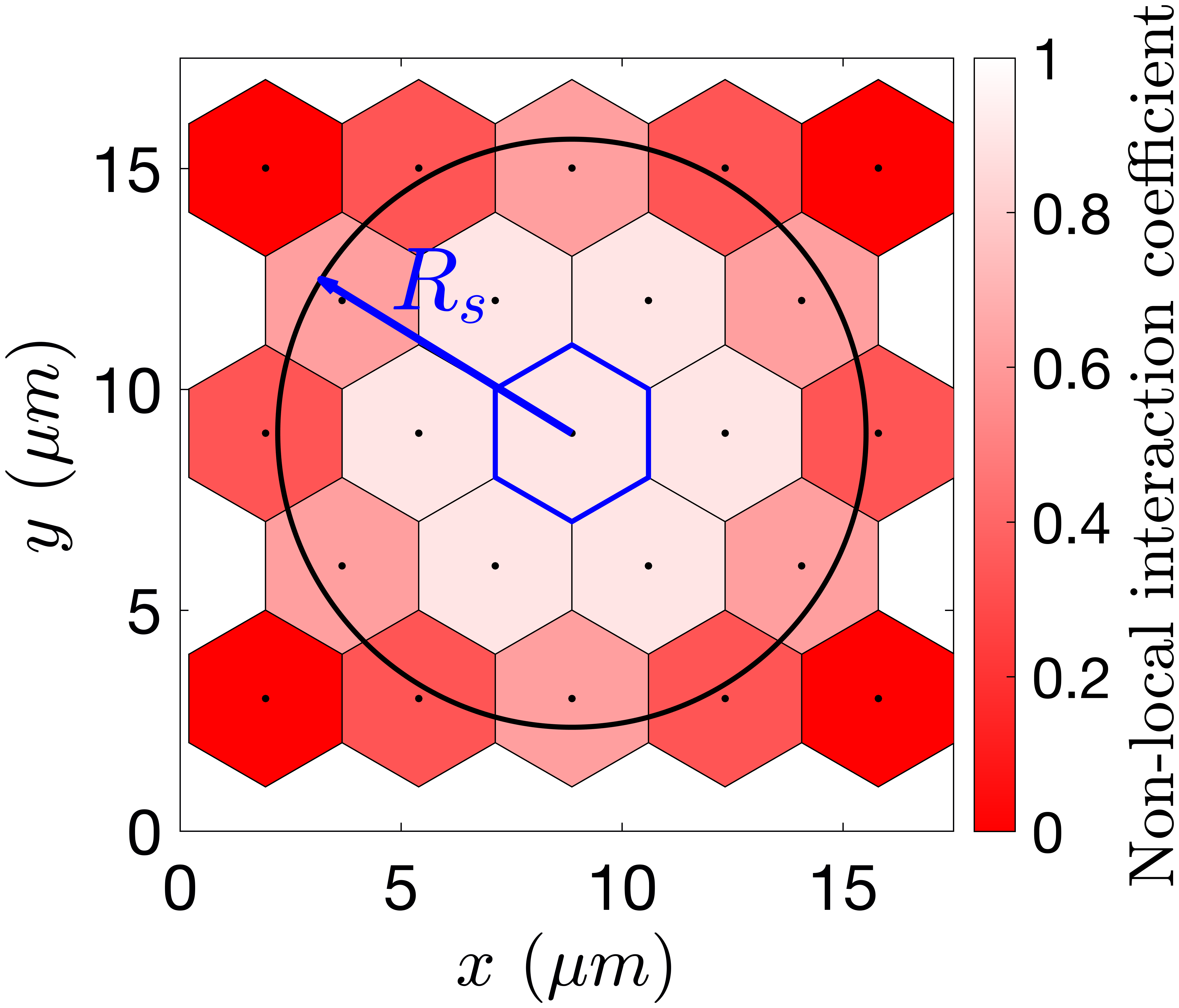}}
\caption{\textbf{A schematic diagram showing the non-local interactions in the multicellular VEGF-Delta-Notch model.} Cell-to-cell interactions may be non-local (i.e. beyond immediate neighbours on a given lattice) provided they lie within an interaction radius, $R_s$. The diagram illustrates the weights of interactions between the focal cell (highlighted in blue) and cells in its neighbourhood, for a regular hexagonal lattice (the weights are defined as a normalised area of the overlap between a neighbouring voxel and the circular neighbourhood of the focal cell, see \Cref{supp-ExtDN} in \cref{supp-appendix:VEGFDeltaNotch}).}
\label{InteractionRadius}
\end{figure}

The multicellular VEGF-Delta-Notch (i.e. the Delta-Notch signalling pathway coupled with external VEGF stimulation) model is bistable (see \cref{supp-appendix:VEGFDeltaNotch} in \hyperlink{SuppMaterial}{Supplementary Material}). When simulated in a two-dimensional geometry, it produces `salt-and-pepper' patterns in which the phenotypes of neighbouring cells alternate between Delta-high and Delta-low states \cite{stepanova2021multiscale}. For this model, cross-talk between individual cells is achieved via external variables, $d_{ext}$ and $n_{ext}$, which represent the levels of Delta and Notch, respectively, summed over cells in a circular neighbourhood with a fixed interaction radius, $R_s$ (see \Cref{InteractionRadius} and \cref{supp-appendix:VEGFDeltaNotch}). Hence, for this system, $v=\of{d_{ext},n_{ext}}$ defines a cell's internal state (phenotype) and the dimension of the pre-computed look-up tables is 2 (see \cref{subsubsec:LookUpTables}). We determined a suitable range, $\mathcal{V}=\af{0, d_{ext}^{max}} \times \af{0, n_{ext}^{max}} \subset \mathrm{R}^2$, for these variables by running 100 realisations of the multiscale model of angiogenesis (the number of realisations depends on the model of interest).

We then generated a regular discretisation of $\mathcal{V}$, $\uf{v_j}_{j \in \mathcal{J}}$, with a grid $100 \times 100 $. For each $v_j$ in this grid, we computed the steady states for the mean-field limit defined by \Cref{supp-eq:DN_single_nondimensional} using non-linear solvers from the C++ GNU Scientific Library (GSL). We note that, once the steady states of the full system have been computed, the subcellular variables $\iota$, $r_2$ and $r_2^*$, corresponding to the Notch intracellular domain, VEGF receptor 2 (VEGFR2) and VEGF-VEGFR2 complexes, respectively, (see definitions in \cref{supp-appendix:VEGFDeltaNotch} in \hyperlink{SuppMaterial}{Supplementary Material}) are redundant; it is not necessary to track these variables because the input-output relationship between $v=\of{d_{ext}, n_{ext}}$, and the steady states completely defines the configuration of the system.

\renewcommand{\thesubfigure}{\alph{subfigure}}
\captionsetup{singlelinecheck=off}
\begin{figure}[htbp]
\centering 
\subfloat[][\label{QasipotentialIllustration_trajectory_D}]{\includegraphics[height = 4cm,valign=b]{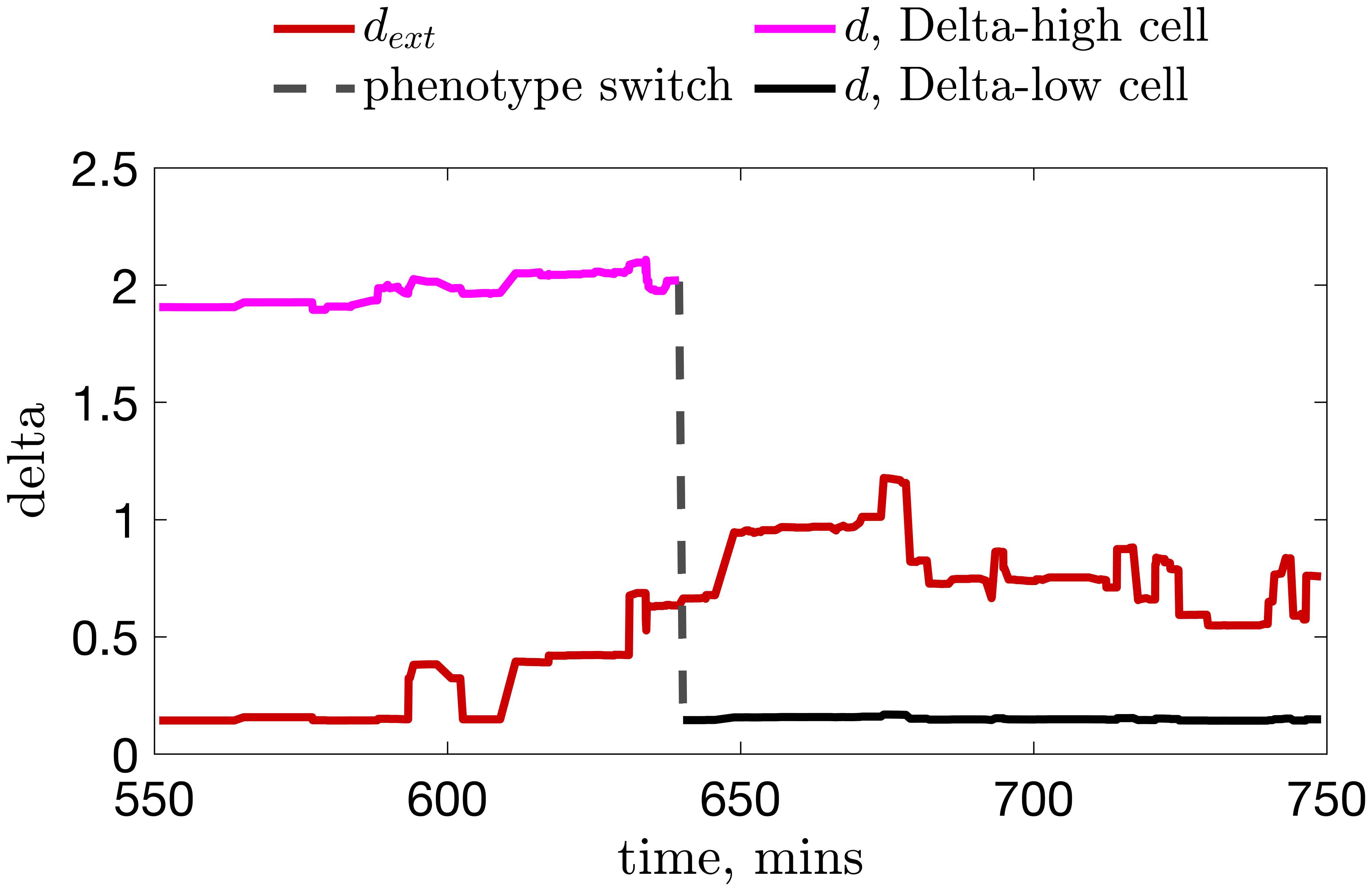}}
\hfill
\subfloat[][\label{QasipotentialIllustration_trajectory_N}]{\includegraphics[height = 4cm,valign=b]{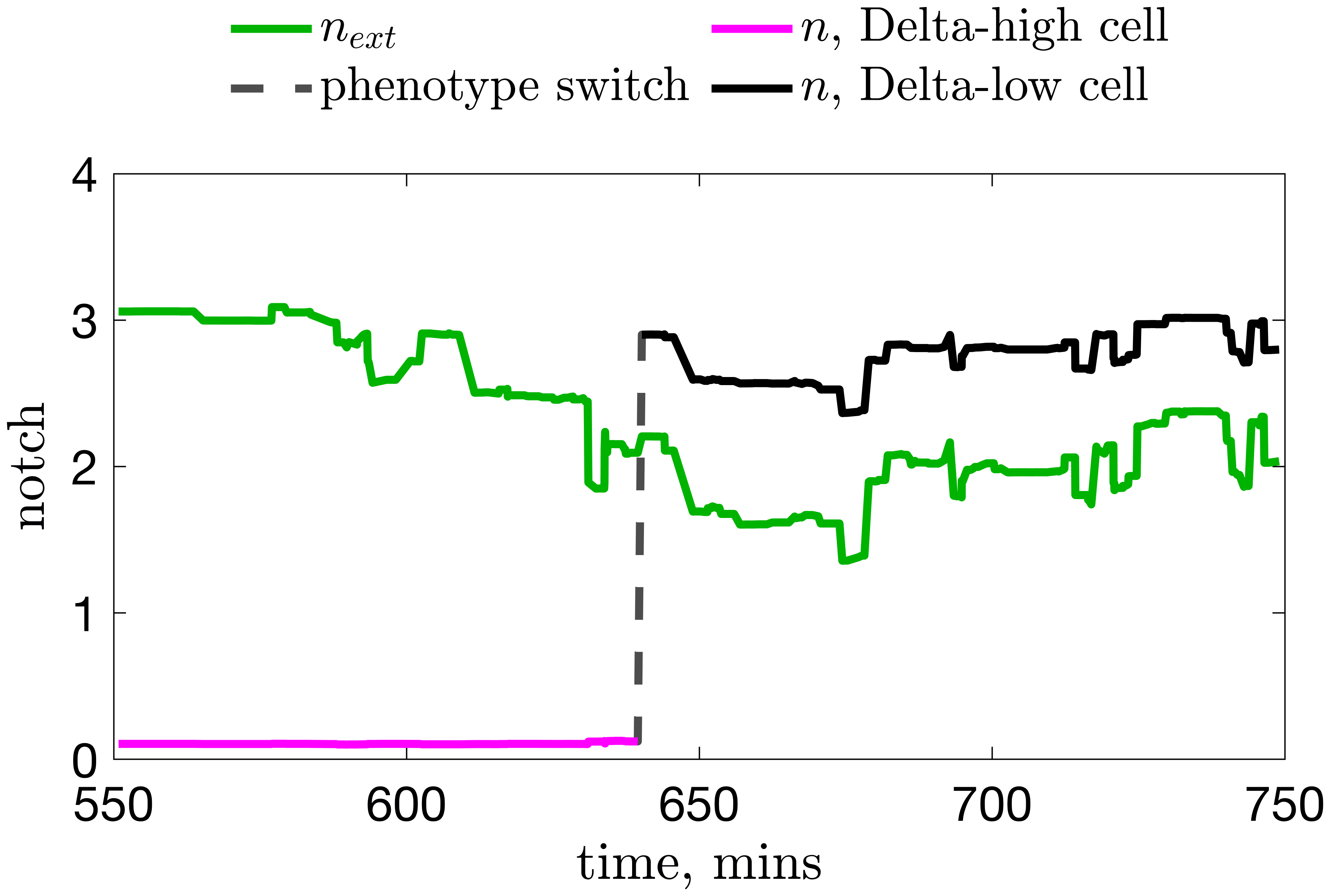}}

\subfloat[][ \label{QasipotentialIllustration_s2t}]{\includegraphics[height = 5cm,valign=b]{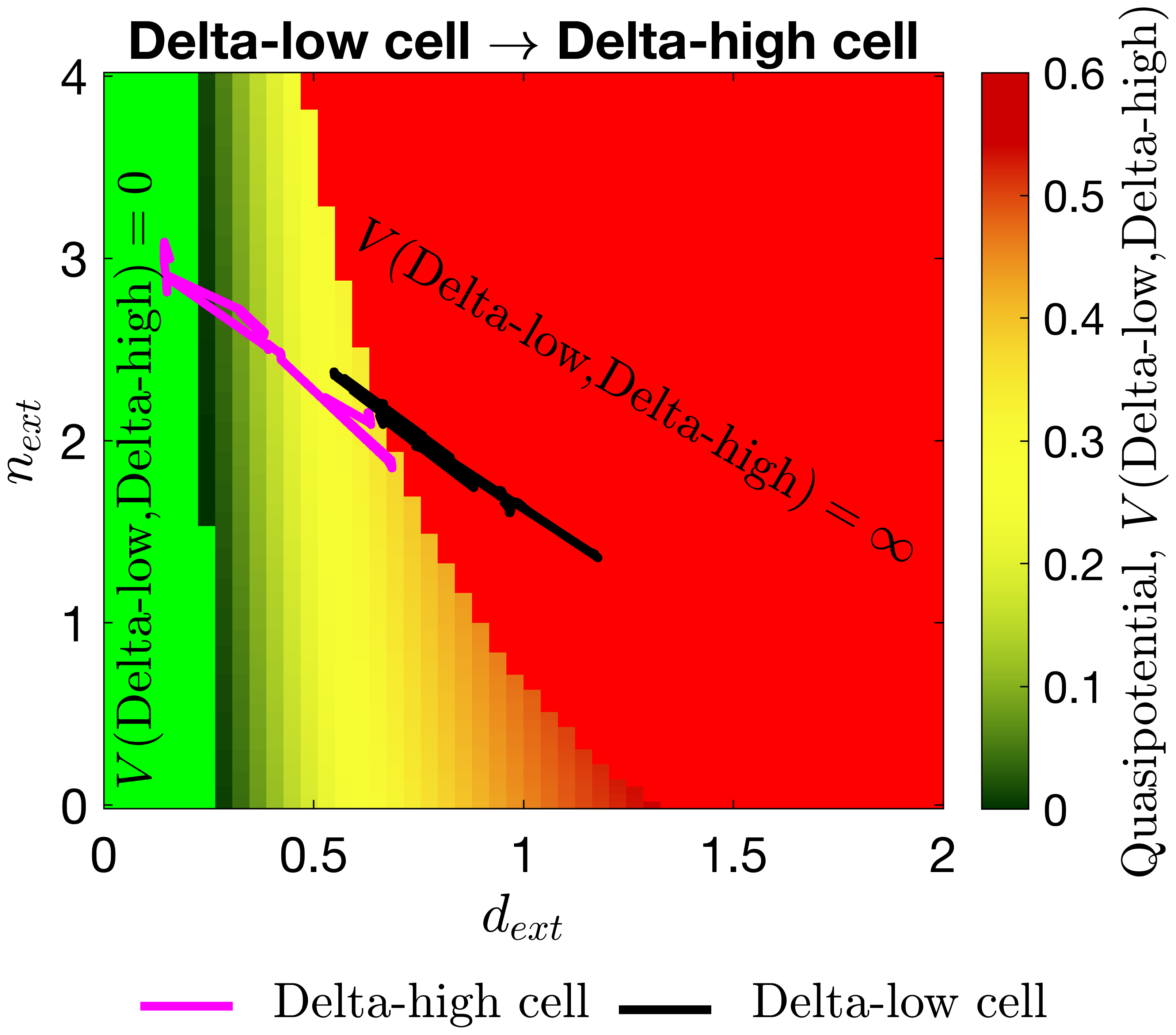}}
\hfill
\subfloat[][ \label{QasipotentialIllustration_t2s}]{\includegraphics[height = 5cm,valign=b]{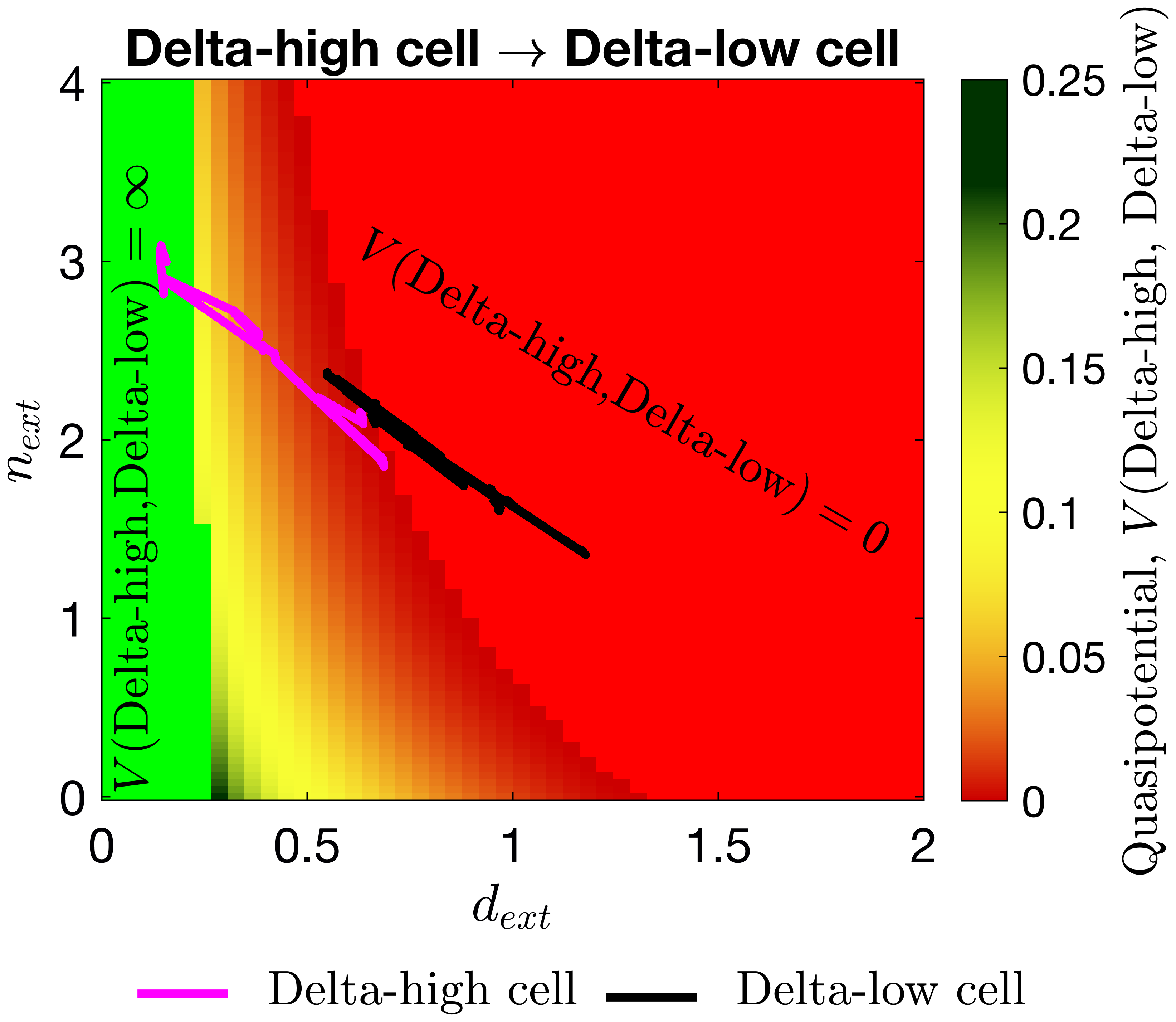}}
\caption{\textbf{An illustration of the quasipotential surfaces.}  \underline{Upper panels}: a noise-induced transition from Delta-high (in magenta) to Delta-low (in black) phenotype of a single cell during a simulation of the angiogenesis model \cite{stepanova2021multiscale} plotted as a function of the focal cell's (a) Delta and (b) Notch levels. The external Delta, $d_{ext}$, (Notch, $n_{ext}$) for the focal cell is computed as a weighted sum of the Delta (Notch) levels of its neighbours as defined by \Cref{supp-MulticellularMeanField}. \underline{Lower panels}: 2D projections of the quasipotential surfaces (c) $V(\text{Delta-low},\text{Delta-high})$ and (d) $V(\text{Delta-high},\text{Delta-low})$ as functions of $d_{ext}$ and $n_{ext}$. The monostability region in which the unique stable steady state corresponds to a Delta-high (Delta-low) cell is coloured green (red). The colour bar indicates the value of the corresponding quasipotential. The trajectory (as in panels (a) and (b)) plotted on the quasipotential surfaces (in (c) and (d)), illustrates that phenotype switches are more likely to occur for lower values of the quasipotential. Parameter values are fixed as indicated in \Cref{supp-Params}.} 
\label{QasipotentialIllustration}
\end{figure}

For values of $v_j$ that fall within the bistability region, we computed the quasipotential values of the transitions between phenotypes (see \Cref{QasipotentialIllustration}), using the gMAM (see \cref{supp-appendix:gMAM} in \hyperlink{SuppMaterial}{Supplementary Material}). We also used the full stochastic system to check those values of the quasipotential for which a phenotype switch is more likey to occur. As expected, most phenotype transitions occur close to the boundary of the bistability region, where values of the quasipotential are lower. For example, \Cref{QasipotentialIllustration_trajectory_D,QasipotentialIllustration_trajectory_N} show a sample path of the full stochastic system for an individual cell during a simulation of the multi-agent model \cite{stepanova2021multiscale}. The cell undergoes a noise-induced switch from a Delta-high to a Delta-low phenotype. \Cref{QasipotentialIllustration_s2t,QasipotentialIllustration_t2s} show the same sample path projected onto the quasipotential surfaces. These plots show that phenotypic switches are more likely to occur when the values of external Delta and Notch, $\of{d_{ext},n_{ext}}$, are such that the quasipotential, $V(x_1,x_2)=V(\text{Delta-high},\text{Delta-low})$, is small.

We constructed a look-up table of prefactor values, $C_{x_s \rightarrow x_l}$, $x_s,~x_l \in \{\text{Delta-high},$ $\text{Delta-low}\}$, by approximating the mean passage times, $\langle T^{\overline{\Omega}}_{x_s \rightarrow x_l} \rangle$, (sample size of 1000 realisations) for an individual cell to switch its phenotype from simulations of the full stochastic CTMC (\Cref{supp-FullStochasticSystem}) with the system size, $\overline{\Omega}$, given by \Cref{OmegaEstimate}.

We then implemented the CG model in C++ using \Cref{supp-SimulationAlgorithm}. In order to establish an input-output relationship between an arbitrary $v=\of{d_{ext},n_{ext}}$ and the corresponding cell phenotypes and transition rates, we used bilinear interpolation routines from the C++ GNU Scientific Library (GSL) (\textit{gsl\_interp2d} routines). The model was then simulated using the standard Gillespie algorithm. We used no-flux boundary conditions to compute for each cell the extracellular levels of Delta and Notch in all our simulations.

\subsection{Spatial patterning in the CG model}
\label{subsec:PatterningCG}

\begin{figure}[htbp]
\centering 
\subfloat[][ \label{PatternConfigurations_S1}]{\includegraphics[height = 6cm,valign=b]{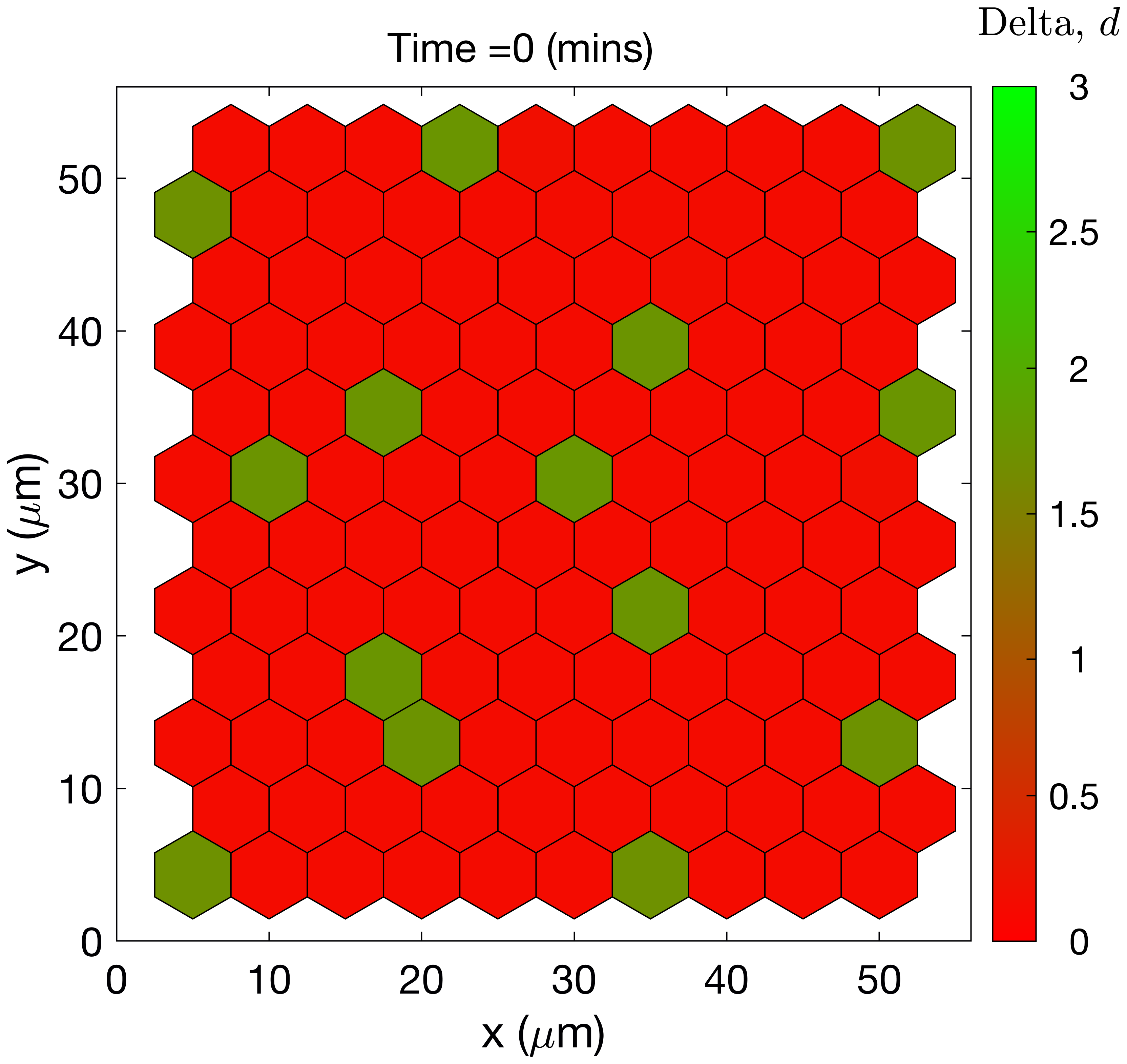}}
\hfill
\subfloat[][ \label{PatternConfigurations_S2}]{\includegraphics[height = 6cm,valign=b]{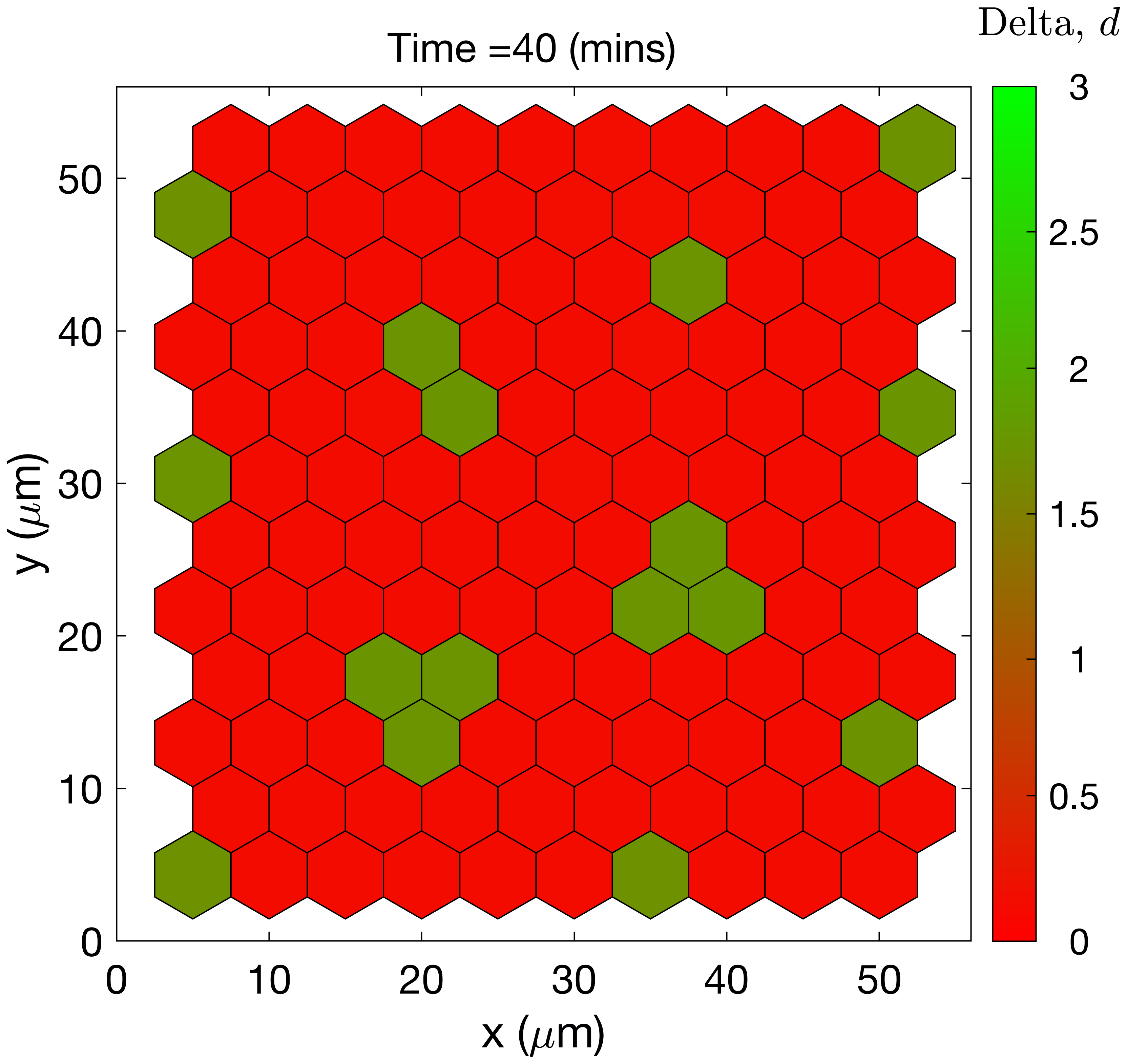}}

\subfloat[][ \label{PatternConfigurations_S3}]{\includegraphics[height = 6cm,valign=b]{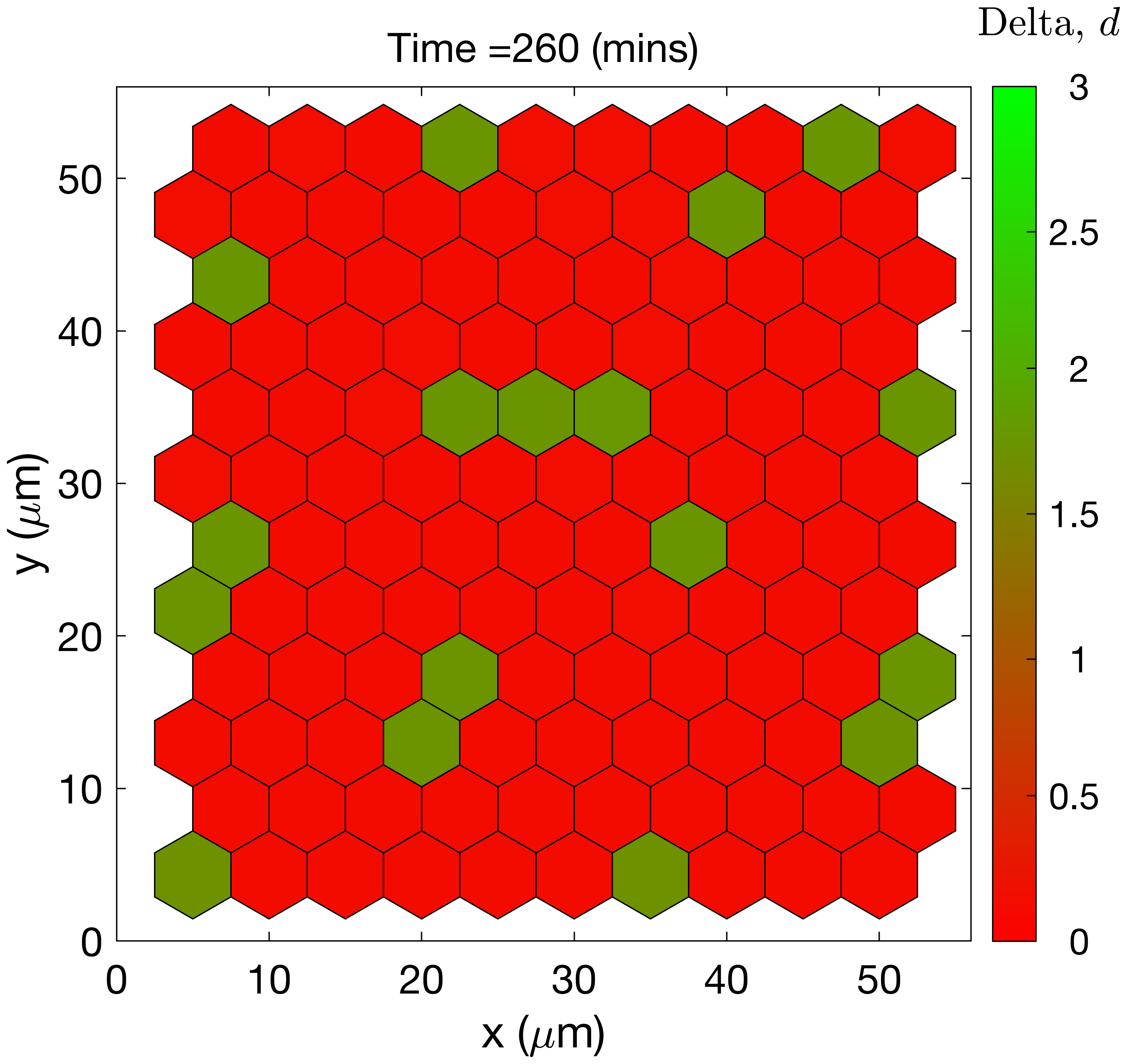}}
\hfill
\subfloat[][ \label{PatternConfigurations_S4}]{\includegraphics[height = 6cm,valign=b]{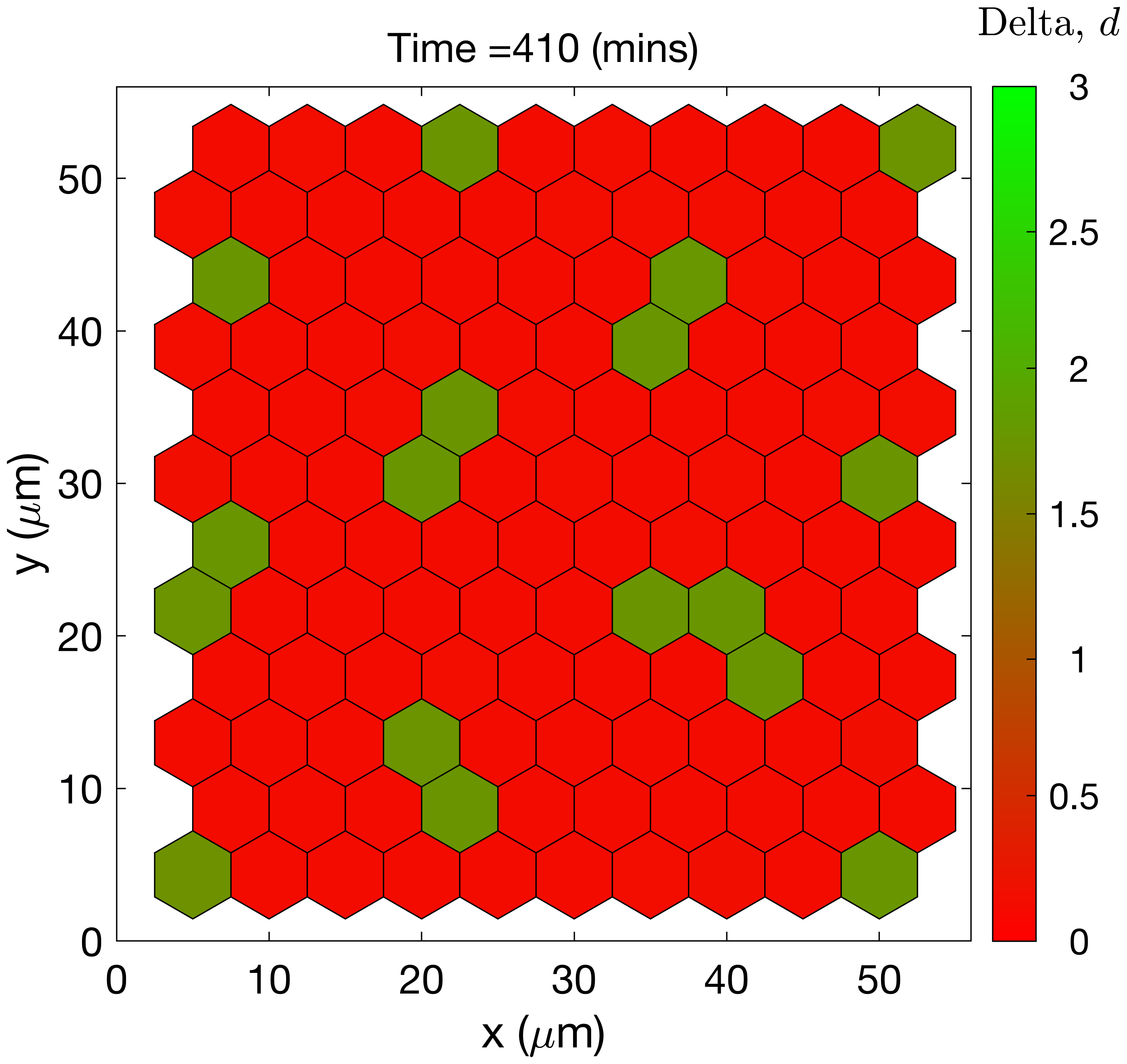}}

\subfloat[][ \label{PatternConfigurations_TipProportion}]{\includegraphics[height = 4cm,valign=b]{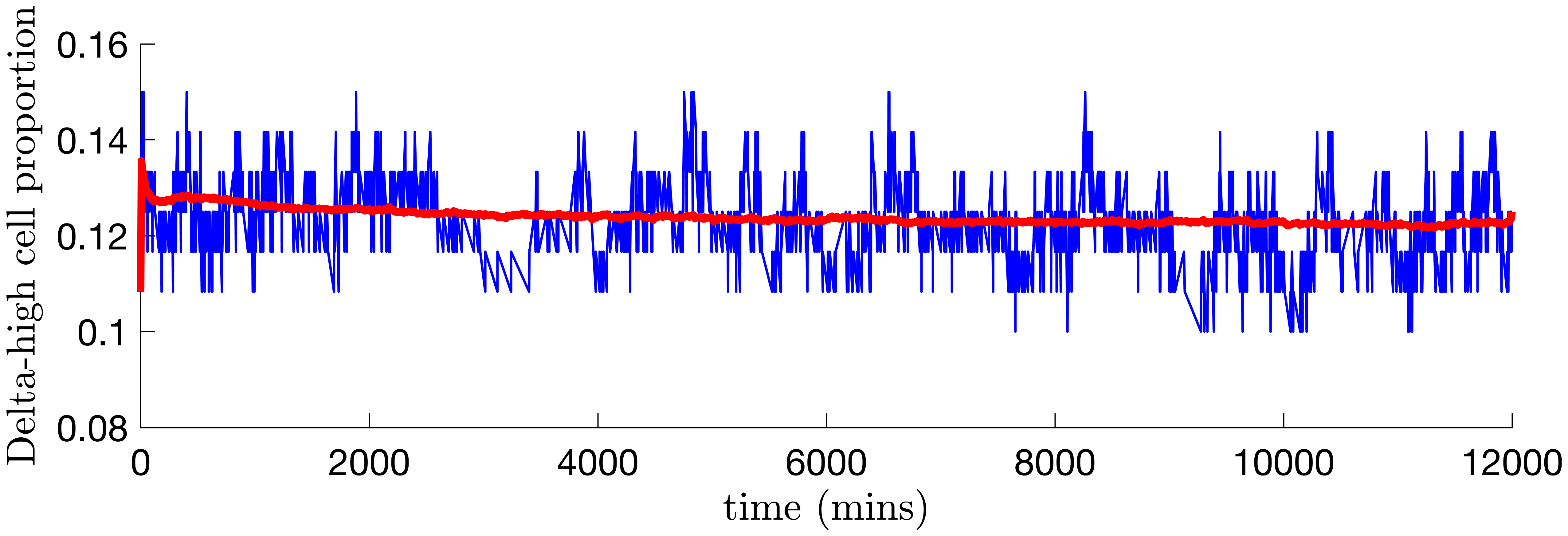}}
\caption{\textbf{Different pattern configurations explored by the CG model.} (a)-(d) Series of plots showing how the distribution of cell phenotypes changes over time during a single simulation of the CG model. The colour bar indicates the level of Delta. (a) $t=0$; (b) $t=40$; (c) $t=260$; (d) $t=410$ minutes. (e) Time evolution of the Delta-high cell proportion (defined as a ratio of cells with the Delta-high phenotype to the total cell number) for a single simulation of the CG model (blue line) and averaged over $1000$ realisations (red line). For these simulations, the interaction radius and system size were fixed at $R_s = 15 \mu m$ and $\Omega=100$, respectively; the values of the remaining parameters were fixed as indicated in \Cref{supp-Params}.} 
\label{PatternConfigurations}
\end{figure}

In order to illustrate the CG model, we first ran numerical simulations on a small cell monolayer ($10 \times 12$ voxels). The results presented in \Cref{PatternConfigurations_S1,PatternConfigurations_S2,PatternConfigurations_S3,PatternConfigurations_S4} show how the distribution of Delta-high and Delta-low cells changes over time during a typical CG realisation (see also \hyperlink{MovieS1}{Movie S1}). Starting from an initial pre-pattern (\Cref{PatternConfigurations_S1}), noise-induced phenotype transitions enable the system to explore different pattern configurations for the given geometry, while the proportion of Delta-high cells remains on average constant (see \Cref{PatternConfigurations_TipProportion}). 

\renewcommand{\thesubfigure}{\alph{subfigure}}
\captionsetup{singlelinecheck=off}

The mean proportion of Delta-high cells (and, thus, the spatial pattern) during simulations of the CG system depends on the interaction radius, $R_s$. For values of $R_s$ corresponding to nearest-neighbours interaction ($R_s \leq 1.5h$, where $h$ is the voxel width), we observe classical patterns of alternating Delta-high and Delta-low cells (i.e. the so-called salt-and-pepper pattern \cite{collier1996}; see \Cref{supp-PatternVaryingR10}). As $R_s$ increases, the number of Delta-low cells that may be inhibited by a focal Delta-high cell increases, causing the proportion of Delta-high cells in the spatial patterns to decrease \cite{stepanova2021multiscale}. Thus, for larger values of $R_s$ ($R_s> 1.5h$), Delta-high cells are separated by larger distances (see \Cref{supp-PatternVaryingR20,supp-PatternVaryingR30,supp-PatternVaryingR40}). These results for CG simulations are consistent with those obtained for the full multicellular stochastic model of the VEGF-Delta-Notch signalling pathway \cite{stepanova2021multiscale}. The ability of the CG system to explore different spatial patterns increases as the size of the interaction radius, $R_s$, grows, and the corresponding emerging patterns are more diverse (see \Cref{PatternConfigurations_S1,PatternConfigurations_S2,PatternConfigurations_S3,PatternConfigurations_S4,supp-PatternVaryingR20,supp-PatternVaryingR30,supp-PatternVaryingR40}).

It is noteworthy that spatial patterns explored in simulations of the CG model differ in their robustness to noise. In particular, the mean passage time for a phenotype switch, and, thus, a change in the pattern, to occur, which is equal to the inverse of the total propensity, $P$, depends on the values of the quasipotential, $V(x_s,x_l)$, for all entities in the system. Here, the total propensity, $P$, for a phenotype switch event is defined as a sum of transition rates, $k^e_{x_s \rightarrow x_l}$, for each cell with index, $e$, to change its state from $x_s$ to $x_l$, see \Cref{FlowchartMulticellular}. When, via random exploration, the system finds a configuration for which the values of $V(x_s,x_l)$ are larger, the waiting time for a phenotype switch increases and the configuration is more resilient to further changes.

This feature of the CG method facilitates exploration of new robust spatial patterns which cannot practically be achieved using other numerical frameworks: \textit{(i)} simulations of the full stochastic model are too computationally intensive, which makes the exploration of these patterns infeasible because of the longer timescales needed; \textit{(ii)} the deterministic framework does not allow for transitions between stable steady states, which makes this exploration impossible; \textit{(iii)} the complexity of analytic methods needed to verify the stability of a pattern of a system with non-local interactions does not permit exploration of complex pattern configurations \cite{o2012continuum}.

\renewcommand{\thesubfigure}{\alph{subfigure}}
\captionsetup{singlelinecheck=off}
\begin{figure}[htbp]
\centering 
\subfloat[][ \label{OptimalPattern_CellHighlighted}]{\includegraphics[height = 6.5cm,valign=b]{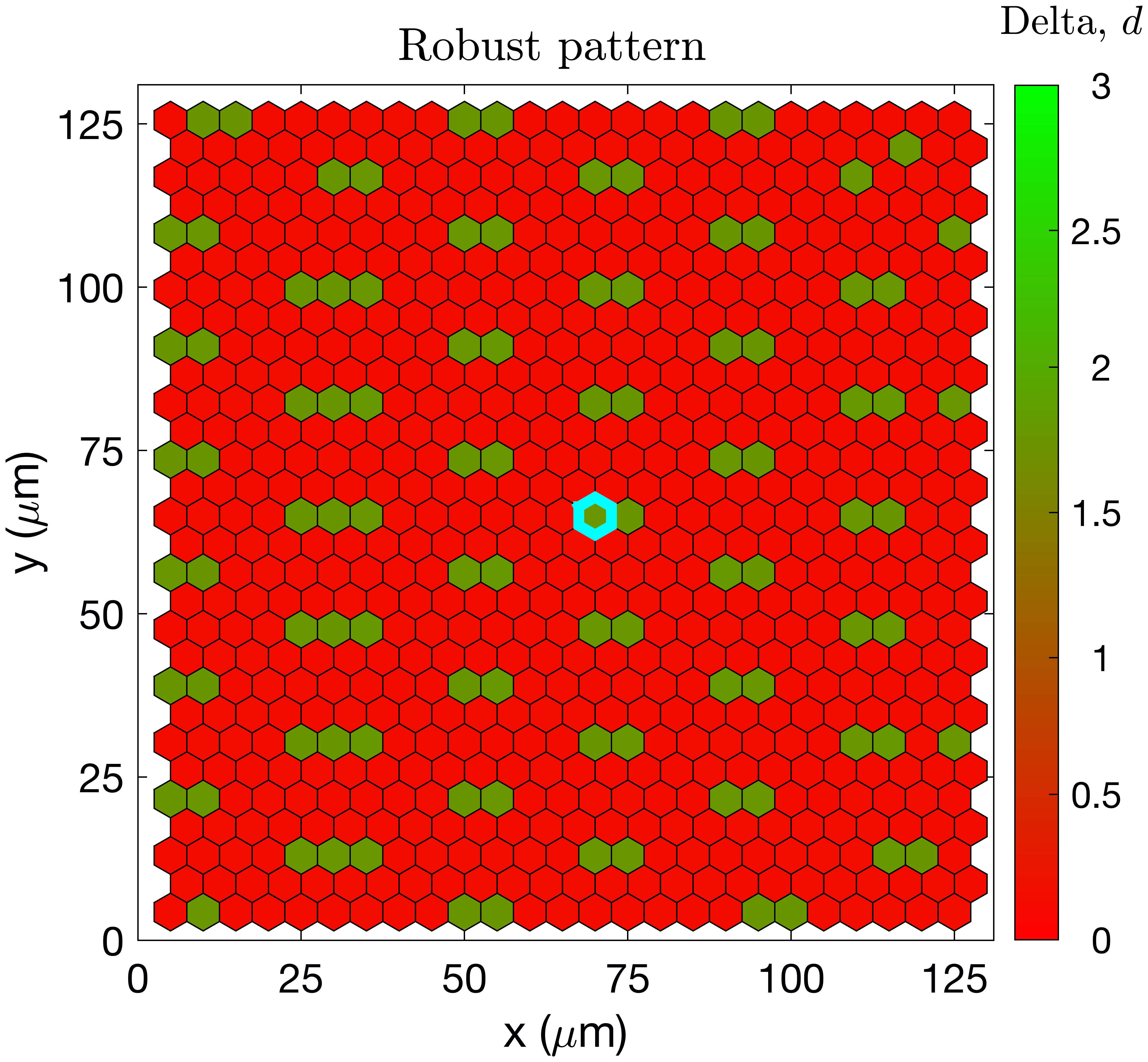}}
\hfill
\subfloat[][ \label{OptimalPattern_TotalPropensity}]{\includegraphics[height = 6.2cm,valign=b]{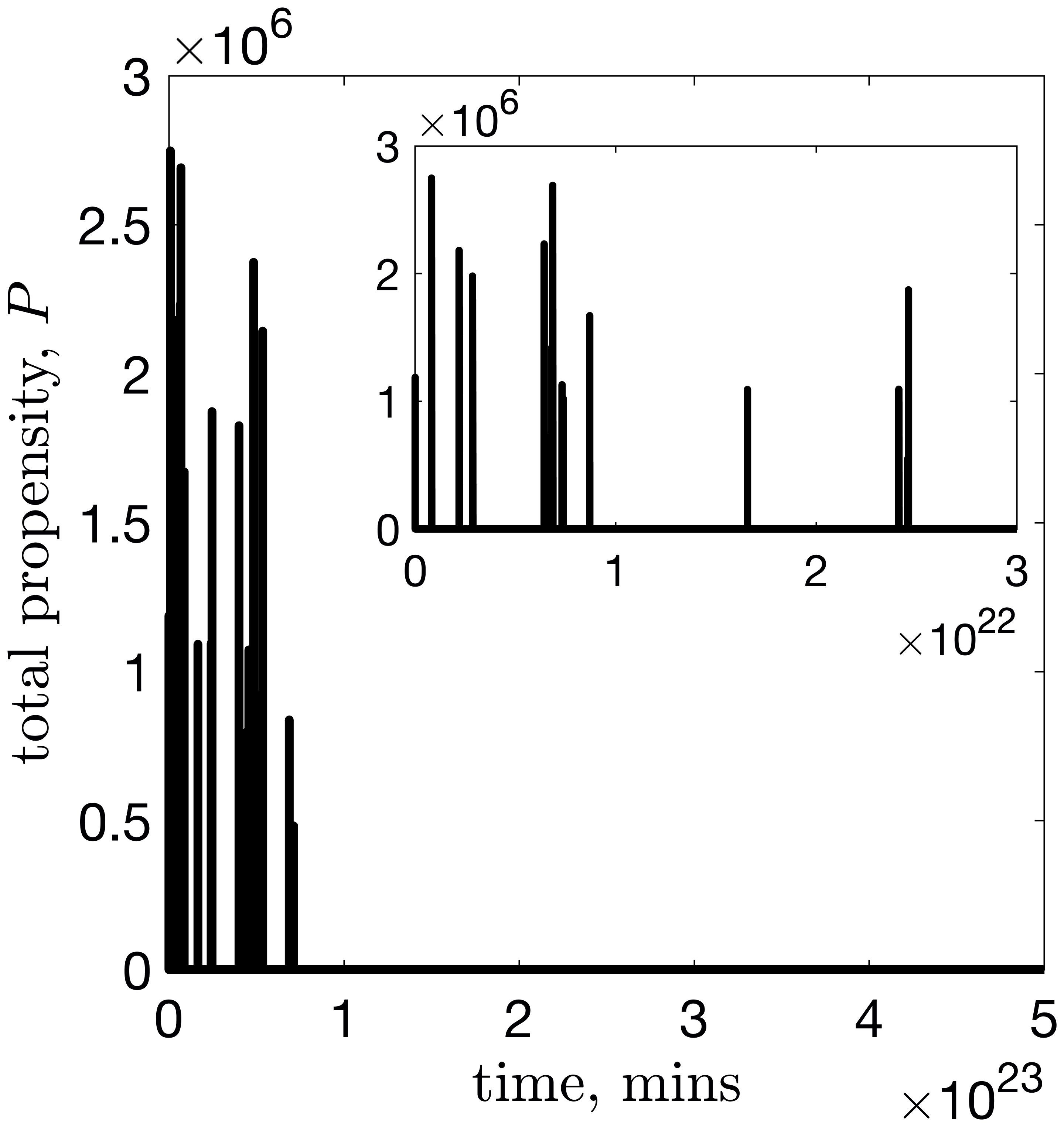}}

\subfloat[][ \label{OptimalPattern_DeltaLevel}]{\includegraphics[height = 5.5cm,valign=b]{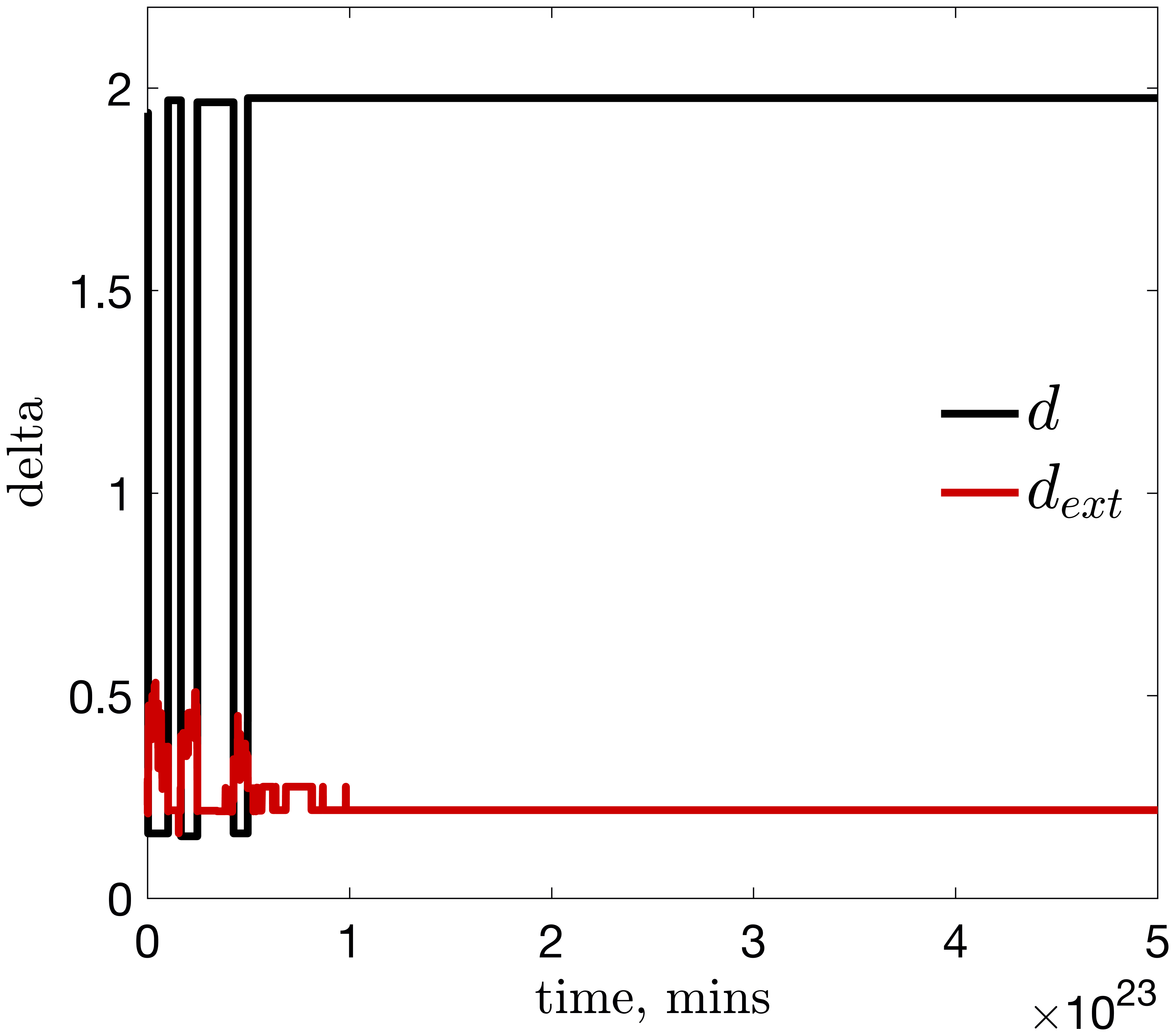}}
\hfill
\subfloat[][ \label{OptimalPattern_TransitionRate}]{\includegraphics[height = 5.8cm,valign=b]{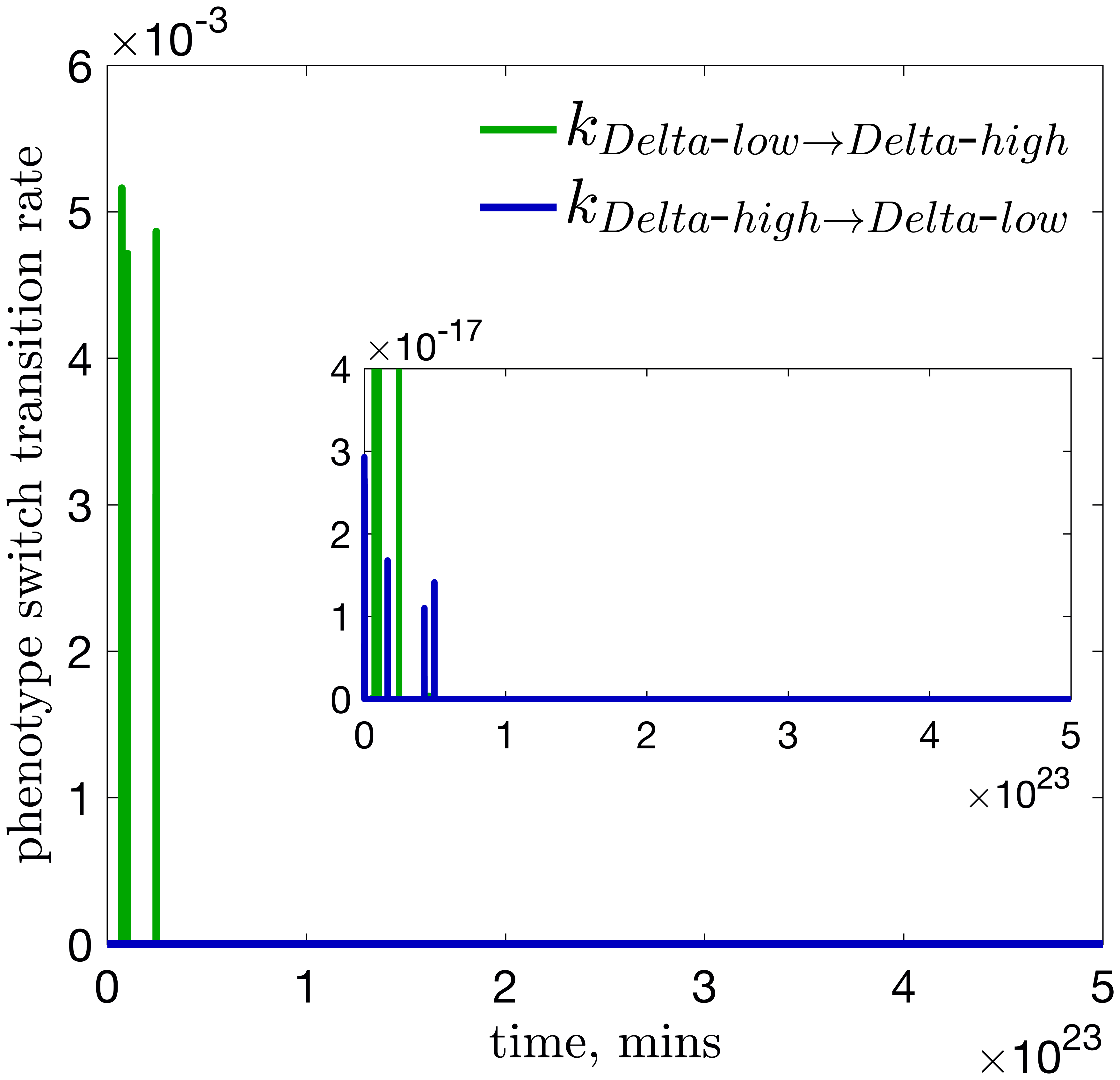}}
\caption{\textbf{Emergence of robust pattern configurations in simulations of the CG model.} At long times, via exploration of different pattern configurations, the dynamics of the CG system evolve to a robust pattern in which any further phenotype switches are unlikely. (a) A typical emergent pattern for a single realisation of the CG model (the colour bar indicates the level of Delta, $d$, for each cell). (b) The time evolution of the total propensity, $P$, for a phenotype switch to occur. Cells in the border rim (three-cell width) are excluded from $P$ since, due to the model geometry, they do not possess a `robust' configuration of neighbours. As $P$ decreases to $0$, the waiting time for a phenotype switch to occur approaches infinity, and the pattern becomes more robust to change. (c)-(d) The dynamics of an individual cell (outlined in cyan in (a)) during this simulation. (c) Temporal evolution of the internal level of Delta, $d$, (defining cell phenotype: high (low) values of $d$ correspond to Delta-high (Delta-low) phenotype) and that in its microenvironment, $d_{ext}$. (d) Temporal evolution of transition rates for a phenotype switch for this cell. We note that the large difference in the order of values for transition rates for the total propensity, $P$, of the lattice ($O(10^6)$), plot (b), and for an individual cell ($O(10^{-17})-O(10^{-3})$), plot (d), comes from the contribution to $P$ of transition rates for cells which are, for the given values of the external variables, on the border of the bistability region (see \Cref{QasipotentialIllustration}). For these simulations, the interaction radius and system size were fixed at $R_s = 15 \mu m$ and $\Omega=1000$, respectively; the values of all remaining parameters were fixed as indicated in \Cref{supp-Params}.} 
\label{OptimalPattern}
\end{figure}

We now present simulation results which illustrate the ability of the CG method to uncover new spatial patterns for the VEGF-Delta-Notch system at long times. We fixed the interaction radius at $R_s = 3.0 h = 15 \mu m $ ($h= 5 \mu m$ is the voxel width), so that interactions occur between cells that are first and second order neighbours in the lattice; the noise amplitude was fixed at $\epsilon = \Omega^{-1} = 0.001$. We ran a CG simulation on a medium size monolayer of cells (see \Cref{OptimalPattern_CellHighlighted} and \hyperlink{MovieS2}{Movie S2}). Starting from the initial pre-pattern, the CG model explores various patterns until it eventually settles on a more robust configuration (shown in \Cref{OptimalPattern_CellHighlighted}). In order to confirm our prediction regarding pattern robustness, we plotted the temporal evolution of the total propensity of the lattice, $P$, in \Cref{OptimalPattern_TotalPropensity}. As its value decreases, $P \rightarrow 0$, the mean waiting time for a change in the spatial pattern becomes infinite, which accounts for the robustness of the emerging pattern. We also considered the dynamics of an individual cell (its position in the monolayer is highlighted by a cyan line in \Cref{OptimalPattern_CellHighlighted}). \Cref{OptimalPattern_DeltaLevel} shows how the phenotype of this cell changes over time: at early times, the cell switches between Delta-high and Delta-low phenotypes (low (high) values of subcellular Delta, $d$, correspond to Delta-low (Delta-high) phenotype). As the spatial pattern settles to a robust configuration, the cell's environment, i.e. the levels of Delta of its neighbours, $d_{ext}$, stop changing and the cell acquires a Delta-high phenotype that remains unchanged for the rest of the simulation. The transition rates for phenotype switches for this cell (\Cref{OptimalPattern_TransitionRate}) exhibit similar dynamics to the total propensity, $P$, of the whole lattice (\Cref{OptimalPattern_TotalPropensity}).

Our CG simulation results show that this robust pattern configuration is not unique. However, we note that the spatial patterns tend to have a regular structure; for example, Delta-high cells may be organised in similar clusters comprising two or three cells as in the pattern shown in \Cref{OptimalPattern_CellHighlighted}. These configurations have lower values of the total propensity, $P$. Cells on the border of the lattice undergo phenotype switches (see \hyperlink{MovieS2}{Movie S2}), since they cannot attain this `more robust' combination of neighbours for the given geometry (since we use no-flux boundary conditions in our simulations).

\subsection{Comparison of the full stochastic, coarse-grained and mean-field frameworks}
\label{subsec:ModelComparison}

We compared the dynamics of the multicellular VEGF-Delta-Notch model using three frameworks: \textit{(i)} full stochastic CTMC, \textit{(ii)} CG, and \textit{(iii)} mean-field descriptions. Simulated (using any of these frameworks) on a 2D domain, the model produces a characteristic pattern of ECs with two cell phenotypes (see, for example, \Cref{PatternConfigurations,supp-PatternVaryingR}). Since the CG approximation describes the long-term behaviour of the system, when its evolution is dominated by the timescale associated with phenotypic switches, it does not account for the initial relaxation onto a quasi-steady state pattern. Thus, the three frameworks cannot be compared with respect to their behaviour at early evolution times. Instead, we quantified the final pattern and the computational cost of simulations. The final simulation time, $t=T_{final}$, was chosen sufficiently large to ensure that a steady state pattern had been established for the mean-field simulations (since stochastic systems do not have a steady state pattern in a classical sense). In order to systematically compare the three frameworks, we used the same final simulation time, $t=T_{final}$, for the other two systems.

We used the following set of metrics to compare the dynamics of the three mathematical descriptions (\hyperlink{SuppMaterial}{Supplementary Material}):

\begin{itemize}
\item \textbf{Delta-high cell proportion}, which is defined as the ratio of the number of cells with Delta-high phenotype to the total number of cells in the system;
\item \textbf{distribution of Delta-high cell clusters}, which provides a breakdown of sizes of Delta-high cell clusters (adjacent cells with Delta-high phenotype, e.g., a single Delta-high cell, two adjacent Delta-high cells, etc.) in a steady pattern configuration;
\item \textbf{computational cost}, which is defined as the average CPU time (in seconds) to perform a single realisation of model simulation.
\end{itemize}

Since the pre-calculated look-up tables for the CG simulations (\Cref{subsubsec:LookUpTables}) were computed for a fixed set of model parameters (see \Cref{supp-Params}), we held them fixed for all simulations. However, the cell-to-cell interaction radius, $R_s$, which is used in the multicellular simulations to determine for each cell, $e$, the vector of extracellular variables, $v^e=\of{d^e_{ext},n^e_{ext}}$, may vary. In our simulations, we used $R_s \in \uf{5,~7.5,~10,~12.5,~15}~ \mu m$ which correspond to experimental observations of the distance over which cell-to-cell interaction can occur in endothelial cells \cite{du2016three} (which corresponds to up to three cells in the interaction circle). Nonetheless, from a theoretical point of view, this quantity can take any value greater than the half-width of a voxel, $R_s> 0.5 h$, where $h$ is the voxel width (we fix $h=5 \mu m$ in our simulations). In addition, for the full stochastic CTMC and CG descriptions, we vary the noise amplitude, $\epsilon = 1/ \Omega$, by changing the system size parameter, $\Omega$. We used $\Omega \in \uf{50,~100,~200,~500,~1000}$. The larger the value of $\Omega$, the closer will be the dynamics of a stochastic system to its mean-field description. For each numerical setup ($R_s$ and $\Omega$), we ran 100 realisations.

We considered two simulation geometries: a 2D cell monolayer and a branching network.

\paragraph*{\textbf{Setup 1: a cell monolayer}} We first ran numerical simulations on a cell monolayer (see \Cref{supp-InitialSetup_M}). This spatial geometry was motivated by the biological process of cell fate specification induced by lateral inhibition via Delta-Notch signalling in flat domains. Examples of such cell fate specification include bristle patterning in \textit{Drosophila} notum \cite{cohen2010dynamic,hunter2016coordinated,corson2017self}, and differentiation of neural precursors in neurogenesis \cite{formosa2013lateral} (see \cite{bocci2020understanding,monk2001spatiotemporal} and references therein for other examples). The fixed stationary distribution of the VEGF serves as an external stimulus which enhances lateral inhibition via Delta-Notch signalling. We chose VEGF as an illustrative example, although, depending on the specific system, other extracellular signals will provide cell stimulus.

\renewcommand{\thesubfigure}{\alph{subfigure}}
\captionsetup{singlelinecheck=off}
\begin{figure}[htbp]
\centering 
\subfloat[][\label{ResultsMonolayer_TipProportion}]{\includegraphics[height = 4.3cm,valign=b]{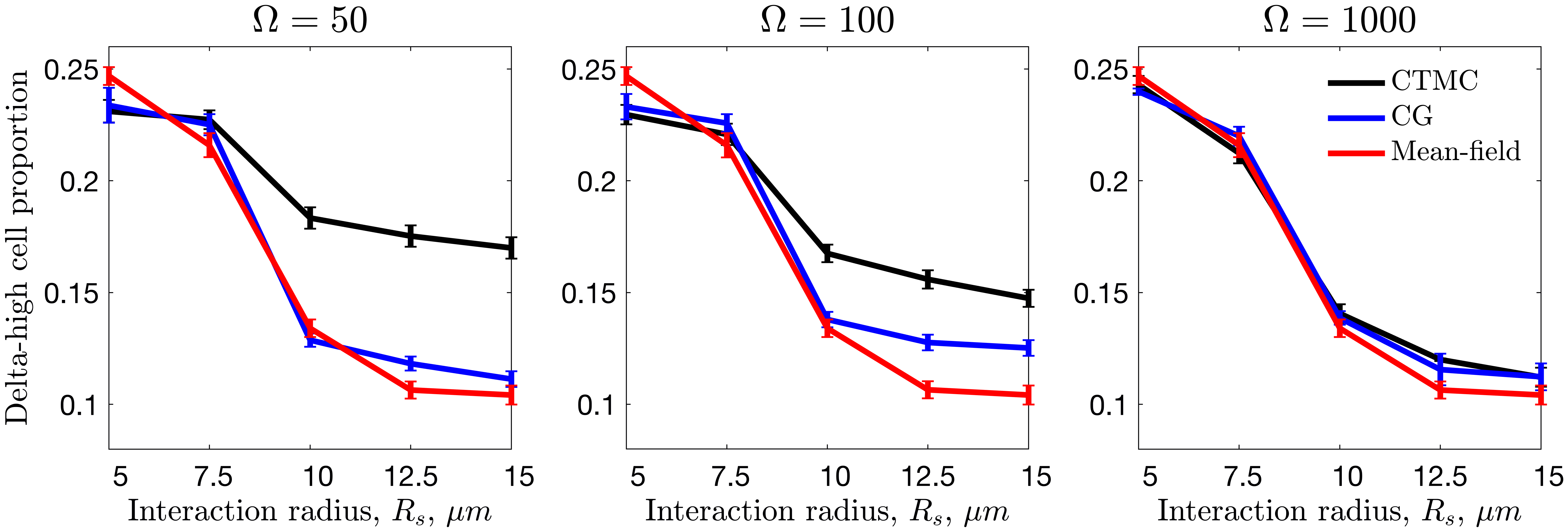}}

\subfloat[][\label{ResultsMonolayer_Clusters}]{\includegraphics[height = 4.2cm,valign=b]{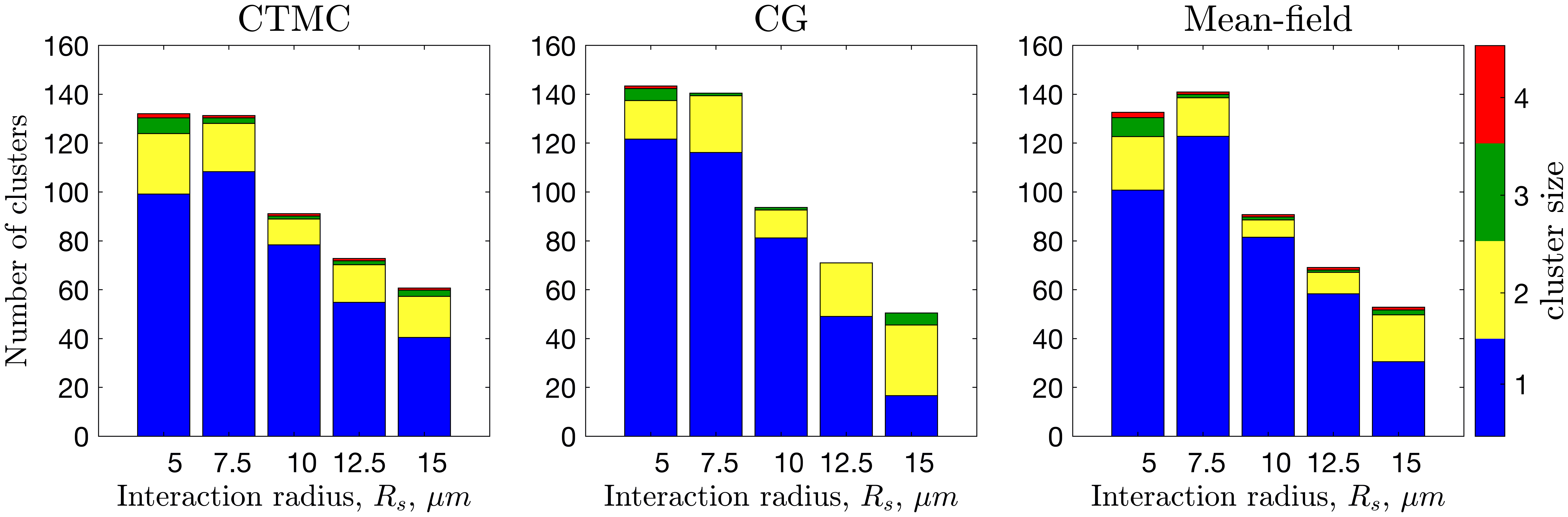}}
\caption{\textbf{Comparison of the dynamics of the multicellular VEGF-Delta-Notch model simulated on a cell monolayer using the full stochastic (CTMC), CG, and mean-field descriptions.} (a) The Delta-high cell proportion as a function of the cell-to-cell interaction radius, $R_s$, for varying noise amplitude, $\epsilon=1/ \Omega$ (the value of $\Omega$ is indicated in the title of each plot), for the full stochastic CTMC (black), CG (blue) and mean-field (red) descriptions. To explore different possible patterns in the deterministic mean-field system, we created a small initial perturbation to the initial configuration (\Cref{supp-InitialSetup_M}). (b) A series of barplots showing how the long-time distribution of Delta-high cell clusters changes as the interacton radius, $R_s$, varies for the full stochastic CTMC (\underline{left panel}), CG (\underline{middle panel}), and mean-field (\underline{right panel}) systems. The number of single Delta-high cells in the final pattern (i.e. at a fixed final simulation time) is shown in blue; the number of clusters with 2, 3, and 4 adjacent Delta-high cells is shown in yellow, green, and red, respectively. For these simulations, we fixed $\Omega=1000$ ($\epsilon = 0.001$). The results are averaged over 100 realisations. The remaining parameter values were fixed as indicated in \Cref{supp-Params}.} 
\label{ResultsMonolayer}
\end{figure}

We began by considering the dynamics of the Delta-high cell proportion for this spatial geometry (see \Cref{ResultsMonolayer_TipProportion}). Consistent with the previous results \cite{stepanova2021multiscale}, for all simulation frameworks (i.e. the full stochastic (CTMC), CG, and mean-field descriptions), the Delta-high cell proportion decreases as the cell interaction radius, $R_s$, increases. \Cref{ResultsMonolayer_TipProportion} confirms that, as expected, differences in this metric between the three systems decrease as the level of noise is reduced (i.e. as $\Omega$ increases). In particular, for high noise levels (i.e. lower values of $\Omega$), the patterns generated by the stochastic systems (full CTMC and CG frameworks) are more diverse, and the Delta-high cell proportions differ from those for the associated mean-field description. We note that the dynamics of the Delta-high cell proportion for the mean-field system (red lines) are identical in all subplots in \Cref{ResultsMonolayer_TipProportion} since noise is absent in deterministic systems (i.e. the system size parameter, $\Omega$, is irrelevant).

We also quantified the size distribution of the Delta-high cell clusters associated with the final patterns established on the cell monolayers. Since the dynamics of the three systems converge for larger values of the system size, $\Omega$ (as shown in \Cref{ResultsMonolayer_TipProportion}), \Cref{ResultsMonolayer_Clusters} shows results for this metric computed for simulations with $\Omega=1000$. The distributions are in good quantitative agreement for the three systems. The discrepancy for simulations with larger cell interaction radius (e.g. $R_s = 15 \mu m$) arises because (for this value of $\Omega$) the CG system is more likely to explore long timescale patterns which have a more `regular' structure and are more robust to noise (cells with Delta-high phenotype organised in similar clusters, see \Cref{subsec:PatterningCG}).

\paragraph*{\textbf{Setup 2: a branching network}} We next considered a more complex spatial geometry of a small branching network (see \Cref{supp-InitialSetup_N}) extracted from a simulation of a hybrid model of angiogenesis \cite{stepanova2021multiscale}. \Cref{supp-SnapshotsCGVascularNetwork} shows a series of patterns explored by the CG system at different time points during a typical simulation for this configuration (for the full simulation, see \hyperlink{MovieS3}{Movie S3}).

For this spatial configuration, we compared the three simulation frameworks using the same metrics as for the cell monolayer. The results for the Delta-high cell proportion are presented in \Cref{supp-ResultsVascularNetwork_TipProportion}. We find that the number of possible patterns generated by lateral inhibition is lower for the branching network geometry than for the cell monolayer (see \Cref{supp-SnapshotsCGVascularNetwork}). Consequently, the Delta-high cell proportions converge for smaller values of $\Omega$ (compare \Cref{ResultsMonolayer_TipProportion,supp-ResultsVascularNetwork_TipProportion}). We also note that, since, in the network configuration, cells have fewer neighbours, the values of this metric are higher than those computed for a cell monolayer 

\Cref{supp-ResultsVascularNetwork_Clusters} shows the size distribution of Delta-high cell clusters for simulations on the branching network. We note that, for this configuration, isolated Delta-high cells (i.e. cells not adjacent to another Delta-high cell) are predominant in the final spatial patterns and the patterns generated by the three frameworks are comparable.

Regarding the computational cost (see technical specifications of computers used in \hyperlink{FileSpecs}{File S1}), the CG method showed a great reduction in the average CPU time compared to the original stochastic system when performing a single realisation (see \Cref{supp-ResultsCPU}). Whereas the numerical cost of simulations of the full stochastic system (\Cref{supp-ResultsCPU}, left panels) increases exponentially as the system size, $\Omega$, grows, simulations of the CG system decrease in average computational time as $\Omega$ increases (\Cref{supp-ResultsCPU}, middle panels). This is because, as the noise level decreases (i.e. $\Omega$ increases), fewer transitions occur in a CG simulation for a fixed final simulation time. Interestingly, the CG simulations are also faster than the mean-field system (\Cref{supp-ResultsCPU}, right panels). The numerical integration of the mean-field system (we used the explicit scheme for the Euler-Lagrange method, although other schemes for numerical integration may show better performance) required evaluation of the non-linear right-hand-side of the equations of the mean-field description (see \Cref{supp-MulticellularMeanField} in \hyperlink{SuppMaterial}{Supplementary Material}) at each time step for every voxel in the lattice, whereas for the CG simulations only one voxel undergoes a change (i.e. a phenotype switch) at each iteration. 

To summarise, the CG method, while preserving stochasticity of transitions between cell phenotypes and producing spatial patterns comparable to those generated using the original stochastic and mean-field descriptions, significantly reduces computational time of simulations. 

\section{Discussion and conclusions}
\label{sec:Discussion}

Hybrid (multiscale) models of complex biological phenomena are often computationally inefficient, which hinders their potential utility. To address this issue, we have developed a coarse-graining (CG) method that reduces the numerical cost of simulations of multi-agent stochastic systems with multiple stable steady states. The CG technique is based on large deviation theory that allows the dynamics of a stochastic system to be reduced to a jump process (i.e. a continuous time Markov chain) on a discrete state space which comprises the stable steady states of all agents in the system. The CG system operates on a timescale on which transitions between these steady states take place. This allows the method to be applied to models whose dynamics act on timescales longer than the typical timescale for relaxation to an equilibrium (e.g., molecular or subcellular processes act on longer timescales when compared to higher spatial scales such as cell migration, dynamics of extracellular cues, etc.). Our results show good qualitative and quantitative agreement between CG simulations and other simulation methods (\Cref{ResultsMonolayer,supp-ResultsVascularNetwork}). Furthermore, the CG algorithm is numerically more efficient in terms of CPU time even when compared with the corresponding mean-field simulations (see \Cref{supp-ResultsCPU}). Likewise, the CG framework allows exploration of new emergent properties of the system, such as long timescale patterns in multicellular systems (\Cref{OptimalPattern}).

The implementation of the CG method requires pre-calculation of several look-up tables (for stable steady state solutions of the system that is being coarse-grained, quasipotential values for transitions between them and the corresponding prefactor of these transitions) which are used later in simulations. To do this, the values of model parameters must be fixed (except for the external variables). However, in order to perform sensitivity analysis with respect to any specific parameter, this parameter may be added to the set of external variables (thus, adding a new dimension to the look-up tables). Since the procedure of pre-calculating the look-up tables is done once, prior to model simulation, it does not increase the numerical cost of the algorithm. Likewise, the computational cost of computing the quasipotential via the geometric minimum action method (gMAM) is independent of the system size, $\Omega$, and an estimate for the required prefactor can be obtained from simulations of the full stochastic model for a single value of the system size parameter, $\Omega$, for which we provided an accurate estimate (see \Cref{OmegaEstimate} and \Cref{PrefactorFit}). Then the CG model can be efficiently simulated using the standard Gillespie algorithm for any value of $\Omega$ (or, equivalently, noise level, $\epsilon = 1/\Omega$).

After introducing the CG method (\cref{sec:CoarseGraining}), we applied it to a multi-agent model of phenotypic specification of cells via the VEGF-Delta-Notch signalling pathway. For this system, we demonstrated how the spatial patterning of cells with different phenotypes changes as CG transitions between these steady states (phenotypes) occur (\Cref{PatternConfigurations}). We then confirmed that the patterns generated by the CG system are quantitatively similar to steady state configurations of the original stochastic system and the associated mean-field limit for this model (see \Cref{ResultsMonolayer,supp-ResultsVascularNetwork}). We conclude that the CG method preserves the continuous cell phenotypes and stochasticity of the original system, while reducing the computational cost of simulations by several orders of magnitude (as compared to the numerical cost of simulations of the full stochastic system, see \Cref{supp-ResultsCPU}).

In this paper, we used the VEGF-Delta-Notch model to illustrate the benefits of the CG method. We note, however, that the CG method can be applied to a wider class of multi-agent models in which the behaviour of the agents is regulated by stochastic models with multiple stable attractors (e.g. steady states, limit cycles) and whose dynamics are controlled by external cues (e.g. morphogens, growth factors, levels of specific ligands/receptors in neighbouring cells, etc.). Examples of systems with subcellular dynamics which satisfy the requirements for application of the CG method include fate specification of cells in intestinal crypts \cite{kay2017role,buske2011comprehensive}, epithelial to mesenchymal phenotypic transition (and its reverse) in cancer invasion \cite{jolly2015implications} and development \cite{sha2019intermediate}, cell differentiation in neurogenesis \cite{formosa2013lateral}, and a general class of models describing cell decision switches \cite{guantes2008multistable}. These models are multistable and the timescale of simulations is longer than the timescale of the relevant subcellular signalling pathway. Nonetheless, the spectrum of models which are suitable for coarse-graining via the CG algorithm is not restricted to intracellular signalling pathways in animal cells; other examples include vegetation patterning in arid ecosystems \cite{kefi2010bistability} or plant morphogenesis mediated via the auxin hormone \cite{baldazzi2012towards,farcot2015modular}. The exact implementation of the CG system for the aforementioned models is beyond the scope of this paper. 

To conclude, the CG method developed in this paper paves the way for a systematic reduction of the dynamics of a wide class of multistable stochastic models. It allows for investigation of their behaviour on longer timescales than is possible with other frameworks (e.g. full stochastic simulations or deterministic equations). To our knowledge, this is the first example in which large deviation theory has been used to coarse-grain the dynamics of a multi-agent system. In future work we intend to further investigate the performance of the CG method by incorporating the CG system for the VEGF-Delta-Notch signalling into a multiscale model of angiogenesis \cite{stepanova2021multiscale}. 

%
%
\section*{Data management} All of the computational data output is included in the manuscript and/or in the supplementary material. The code of the numerical procedures used in this work is available upon request. 

\section*{Supplementary materials} 
\label{sec:SupMaterials}

\phantom{Text}
\vspace{10pt}

\paragraph*{\hypertarget{SuppMaterial}{\textbf{Supplementary Material}}} The file contains a more detailed description of the VEGF-Delta-Notch model, implementation of the CG method, and additional figures and tables. 


\paragraph*{\hypertarget{FileSpecs}{\textbf{File S1}}} 
Technical specifications of the computers used to perform simulations in this work.

\paragraph*{\hypertarget{MovieS1}{\textbf{Movie S1}}} \textbf{A simulation movie showing different pattern configurations explored by the CG system in a small 2D cell monolayer.} The colour bar indicates the levels of Delta. For this simulation, the interaction radius and system size were fixed at $R_s = 15 \mu m$ and $\Omega=100$, respectively; the values of the remaining parameters were fixed as indicated in \Cref{supp-Params}.

\paragraph*{\hypertarget{MovieS2}{\textbf{Movie S2}}} \textbf{A simulation movie showing emergence of robust pattern configurations in simulations of the CG system.} The colour bar indicates the levels of Delta. A single cell, whose dynamics is shown in \Cref{OptimalPattern_DeltaLevel,OptimalPattern_TransitionRate}, is outlined in cyan. For this simulation, the interaction radius and system size were fixed at $R_s = 15 \mu m$ and $\Omega=1000$, respectively; the values of all remaining parameters were fixed as indicated in \Cref{supp-Params}.

\paragraph*{\hypertarget{MovieS3}{\textbf{Movie S3}}} \textbf{A simulation movie showing different pattern configurations explored by the CG system in a branching network.} The colour bar indicates the levels of Delta. For this simulation, the interaction radius and system size were fixed at $R_s = 15 \mu m$ and $\Omega=100$, respectively; the values of the remaining parameters were fixed as indicated in \Cref{supp-Params}.

\bibliographystyle{siamplain}
\bibliography{references}

\makeatletter\@input{xxmain.tex}\makeatother

\end{document}


\maketitle
\makeatletter
\renewcommand{\thefootnote}{\fnsymbol{footnote}}
\makeatother
\footnotetext[2]{Centre de Recerca Matem\`atica, Bellaterra (Barcelona) 08193, Spain}
\footnotetext[3]{Departament de Matem\`atiques, Universitat Aut\`onoma de Barcelona, Bellaterra (Barcelona) 08193, Spain}
\footnotetext[4]{Wolfson Centre for Mathematical Biology, Mathematical Institute, University of Oxford, Oxford OX2 6GG, UK}
\footnotetext[5]{Instituci\'o Catalana de Recerca i Estudis Avan\c{c}ats (ICREA), Barcelona 08010, Spain}
\footnotetext[6]{\email{dstepanova@crm.cat}}

\renewcommand{\thefootnote}{\arabic{footnote}}

\section{VEGF-Delta-Notch system} 
\label{appendix:VEGFDeltaNotch}

The VEGF-Delta-Notch signalling pathway mediates phenotype specification of endothelial cells (ECs) during angiogenesis and vasculogenesis \cite{blanco2013vegf,gerhardt2003vegf}. In a two-dimensional geometry, lateral inhibition induced by Delta-Notch trans-interactions (i.e. binding of a Delta ligand on one cell to a Notch receptor on a neighbouring EC) produces an alternating pattern of cells with two distinct levels of gene expression which correspond to `Delta-high' and `Delta-low' phenotypes (in angiogenesis, the Delta-high (Delta-low) phenotype is referred to as tip (stalk) cell) \cite{jakobsson2010endothelial,ubezio2016synchronization}. Delta-high cells are characterised by elevated levels of Delta ligand and VEGFR2 (VEGF receptor 2), and low levels of Notch, whereas Delta-low cells are associated with low levels of Delta and VEGFR2, and high levels of Notch. When Delta-Notch signalling is combined with cell activation in response to extracellular VEGF stimulation, the robustness of the cell phenotypes increases \cite{boareto2015jagged}.

\begin{figure}[htbp]
\centering
\subfloat{\includegraphics[height = 6cm,valign=b]{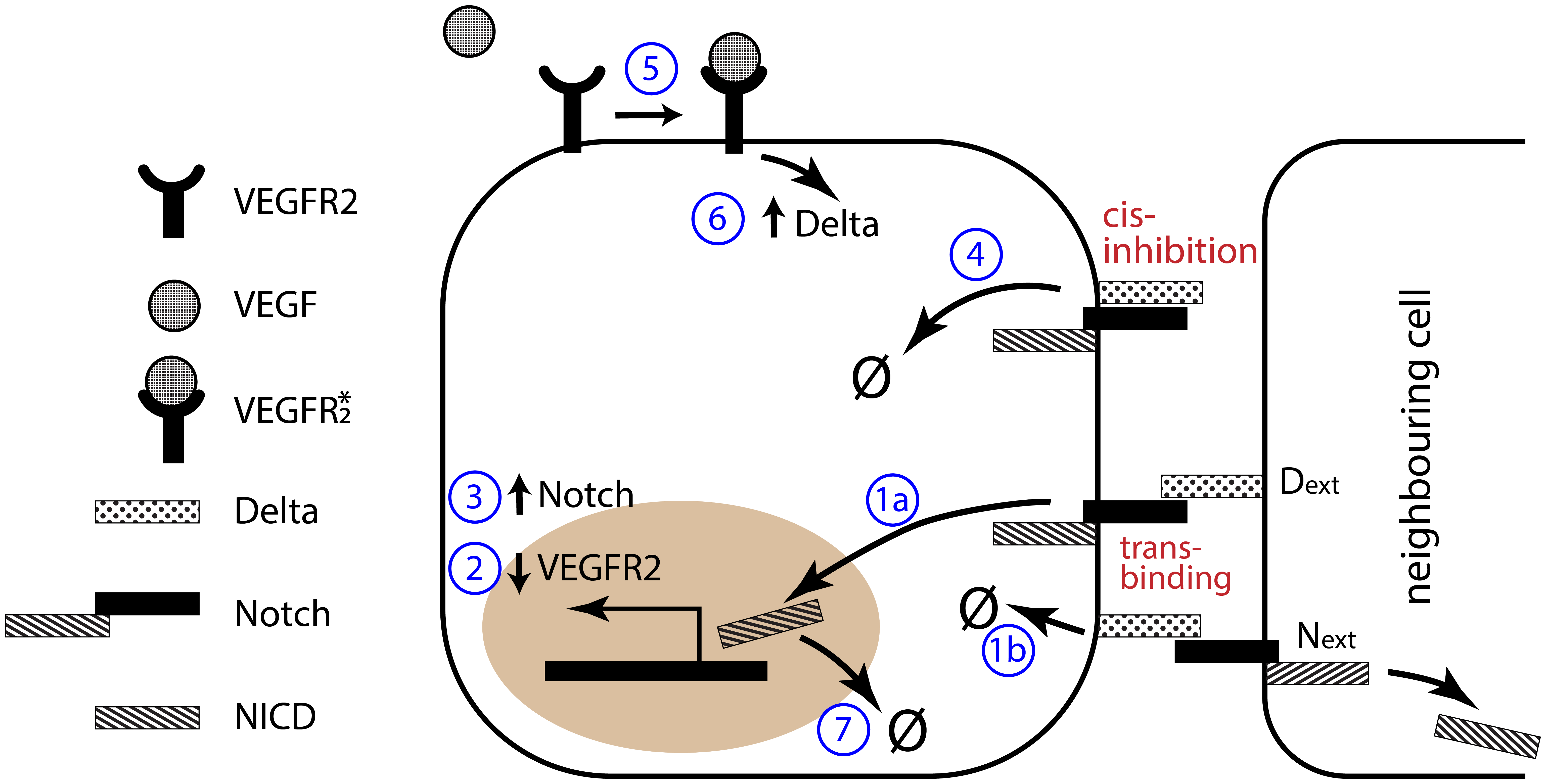}} 
\caption{\textbf{A schematic of the kinetic reactions incorporated in our mathematical model of the VEGF-Delta-Notch signalling pathway for the individual cell system \cite{stepanova2021multiscale}.} The reactions are labelled by blue circled numbers (see \Cref{FullStochasticSystem}). $N_{ext}$ and $D_{ext}$ represent extracellular levels of Notch and Delta, respectively. VEGFR${}_2^*$ denotes VEGFR2 receptor activated by extracellular VEGF.}
\label{DN_Reactions}
\end{figure}

\subsection{Individual cell system}
\label{subsection:IndividualSystem}

A schematic of the kinetic reactions involved in the VEGF-Delta-Notch signalling pathway for an individual cell system is shown in \Cref{DN_Reactions}. Briefly, a Notch receptor trans-binds a Delta ligand on a neighbouring cell (reaction \circled{1a} in \cref{DN_Reactions}). This leads to cleavage of the Notch Intracellular Domain (NICD) which, on translocation to the nucleus, inhibits (gene) expression of VEGFR2 (reaction \circled{\rlap{\hspace{2.6pt}2}\phantom{1a}}) and up-regulates (gene) expression of the Notch receptor (reaction \circled{\rlap{\hspace{2.6pt}3}\phantom{1a}}). After trans-binding, the Delta ligand is degraded or recycled by endocytosis (reaction \circled{1b}). Delta-Notch interactions can also occur on the membrane of the same cell, via a phenomenon termed cis-inhibition, which inhibits Delta and Notch from interacting with other proteins until they are eventually degraded (reaction \circled{\rlap{\hspace{2.6pt}4}\phantom{1a}}). At the same time, VEGF, on binding to its receptor, VEGFR2, activates this receptor (reaction \circled{\rlap{\hspace{2.6pt}5}\phantom{1a}}) which up-regulates production of Delta ligand (reaction \circled{\rlap{\hspace{2.6pt}6}\phantom{1a}}). The model also accounts for degradation of all proteins (see, for example, reaction \circled{\rlap{\hspace{2.6pt}7}\phantom{1a}} for NICD degradation).

Zooming out from the detailed description of each kinetic reaction, we note that trans-activation of Notch in a focal cell by Delta ligand from a neighbouring cell (production of an NICD) produces a positive feedback for Notch expression and negative feedback for Delta production. Thus, a focal cell with low levels of Delta expression can down-regulate Notch levels in fewer neighbouring cells while being further inhibited by its neighbours due to its up-regulated Notch level. This mechanism of crosstalk between cells (when ligand, here Delta, on one cell inhibits production of the same ligand in neighbours) is known as lateral inhibition. We refer to our previous work \cite{stepanova2021multiscale} in which we presented a detailed derivation and analysis of the kinetic reactions and the corresponding mean-field equations for this signalling pathway. Here we list the kinetic reactions and the corresponding transition rates of the full stochastic continuous time Markov chain (CTMC) model and its deterministic mean-field limit. The dependent variables and their non-dimensional concentrations are listed in \Cref{VariablesTable}. Given the size of the stochastic system, $\Omega$, we relate the number of molecules of a protein, $P$, with its non-dimensional concentration, $p$, via $p = {}^P/_{\Omega}$. 

\setlength{\doublerulesep}{2\arrayrulewidth}
\begin{table}[htbp]
\centering
\begin{tabular}{| p{6cm} | p{2.5cm} | p{2.8cm} |} \hline
\textbf{Protein} & \textbf{Dependent variable (dimensional)} & \textbf{Concentration (dimensionless)} \\ \hline \hline
Notch receptor & $N$ & $n$ \\ 
 Delta ligand & $D$ & $d$ \\ 
NICD & $I$ & $\iota$ \\
VEGFR2 & $R_2$ & $r_2$ \\
activated (bound to VEGF) VEGFR2 & $R_2^*$ & $r_2^*$ \\ \hline
\end{tabular}
\vspace{0.1cm}
\caption{\textbf{Dependent variables associated with our model of the VEGF-Delta-Notch signalling pathway.} Capital letters indicate numbers of protein molecules (used in the full stochastic system), while lowercase letters represent the corresponding dimensionless protein concentrations (used in the mean-field system of equations).}
\label{VariablesTable}
\end{table}

\renewcommand{\arraystretch}{1.4}
\setlength{\doublerulesep}{2\arrayrulewidth}
\begin{table}[htbp]
\small{
\begin{tabular}{| p{1.0cm} | p{2.8cm} | p{5.1cm} | p{2.4cm} |}
\hline 
\textbf{Reac-} \textbf{tion} \textbf{label, $r$} & \textbf{Kinetic reaction(s)} & \textbf{Transition rate(s), $\alpha_r$} &  \textbf{Stoichiometric} \textbf{vector(s), $\nu_r$} \\ \hline \hline
\circled{1a} & $N + D_{ext} \longrightarrow I + D_{ext} $ & $d_{ext}N$, $d_{ext} = {}^{D_{ext}}/_{\Omega}$ & $(-1,0,+1,0,0)^T$ \\[3pt] \hline
\circled{1b} & $D + N_{ext} \longrightarrow N_{ext} $ & $\eta n_{ext}D$, $n_{ext} = {}^{N_{ext}}/_{\Omega}$ & $(0,-1,0,0,0)^T$ \\[3pt] \hline
\multirow{2}{*}{\circled{\rlap{\hspace{2.6pt}2}\phantom{1a}}} & \multirow{2}{*}{ $\emptyset \rightleftarrows R_2$} & $\af{\rightarrow} \Omega \beta_{R_2}H^S(\rho_N I;\Omega,\lambda_{I,R_2},n_{R_2})$ & $(0,0,0,+1,0)^T$ \\
& & $\af{\leftarrow} R_2$ & $(0,0,0,-1,0)^T$ \\ \hline 
\multirow{2}{*}{\circled{\rlap{\hspace{2.6pt}3}\phantom{1a}}} & \multirow{2}{*}{ $\emptyset \rightleftarrows N$} & $\af{\rightarrow} \Omega \beta_{N}H^S(\rho_N I;\Omega,\lambda_{I,N},n_{N}) $ & $(+1,0,0,0,0)^T$ \\
& & $\af{\leftarrow} N$ & $(-1,0,0,0,0)^T$ \\ \hline
\circled{\rlap{\hspace{2.6pt}4}\phantom{1a}}  & $N + D \longrightarrow \emptyset$ & $\frac{\kappa}{\Omega}N D$ & $(-1,-1,0,0,0)^T$ \\[3pt] \hline
\multirow{2}{*}{\circled{\rlap{\hspace{2.6pt}5}\phantom{1a}}}  & $R_2 \longrightarrow R_2^*$ & $v_{ext}R_2$ & $(0,0,0,-1,+1)^T$ \\ 
& $R_2^*  \longrightarrow \emptyset $& $ \tau R_2^*$ & $(0,0,0,0,-1)^T$ \\ \hline
\multirow{2}{*}{\circled{\rlap{\hspace{2.6pt}6}\phantom{1a}}} & \multirow{2}{*}{ $\emptyset \rightleftarrows D$} & $\af{\rightarrow} \Omega \beta_{D}H^S(\rho_{R_2} R_2^*;\Omega,\lambda_{R_2^*,D},n_{D}) $ & $(0,+1,0,0,0)^T$ \\
& & $\af{\leftarrow} D$ & $(0,-1,0,0,0)^T$ \\ \hline
\circled{\rlap{\hspace{2.6pt}7}\phantom{1a}}  & $I \longrightarrow \emptyset$ & $\tau I$ & $(0,0,-1,0,0)^T$ \\[2pt] \hline
\end{tabular}
}
\vspace{0.1cm}
\caption{\textbf{Details of the kinetic reactions included in our full stochastic model of the VEGF-Delta-Notch signalling pathway within an individual cell.} Here $H^{S} \left(S; S_0, \lambda_{S,P}, n_P \right) = \frac{1+\lambda_{S,P} \left( {}^S/{}_{S_0}\right)^{n_P} }{ 1+\left( {}^S/{}_{S_0}\right)^{n_P} }$ is the so-called shifted Hill function. It represents the regulation of gene expression of protein, $P$, in response to the signalling variable, $S$. Production of $P$ is down-regulated (up-regulated) for $0 \leq \lambda_{S,P} < 1$ ($\lambda_{S,P} > 1$) as the level of signal, $S$, grows. The prefactor in front of $H^S()$ indicates the baseline production of the corresponding protein. A full list of parameter values is given in \Cref{Params}. $\Omega$ represents system size, which is used to scale the transition rates \cite{gillespie1976general}. The stoichiometric vectors, $\nu_r$, are indexed as $(N,D,I,R_2,R_2^*)^T$. Reaction labels are as in \Cref{DN_Reactions}.} 
\label{FullStochasticSystem}
\end{table}

\begin{figure}[htbp]
\centering 
\subfloat[][ cell monolayer\label{ResultsCPU_ML}]{\includegraphics[height = 4.3cm,valign=b]{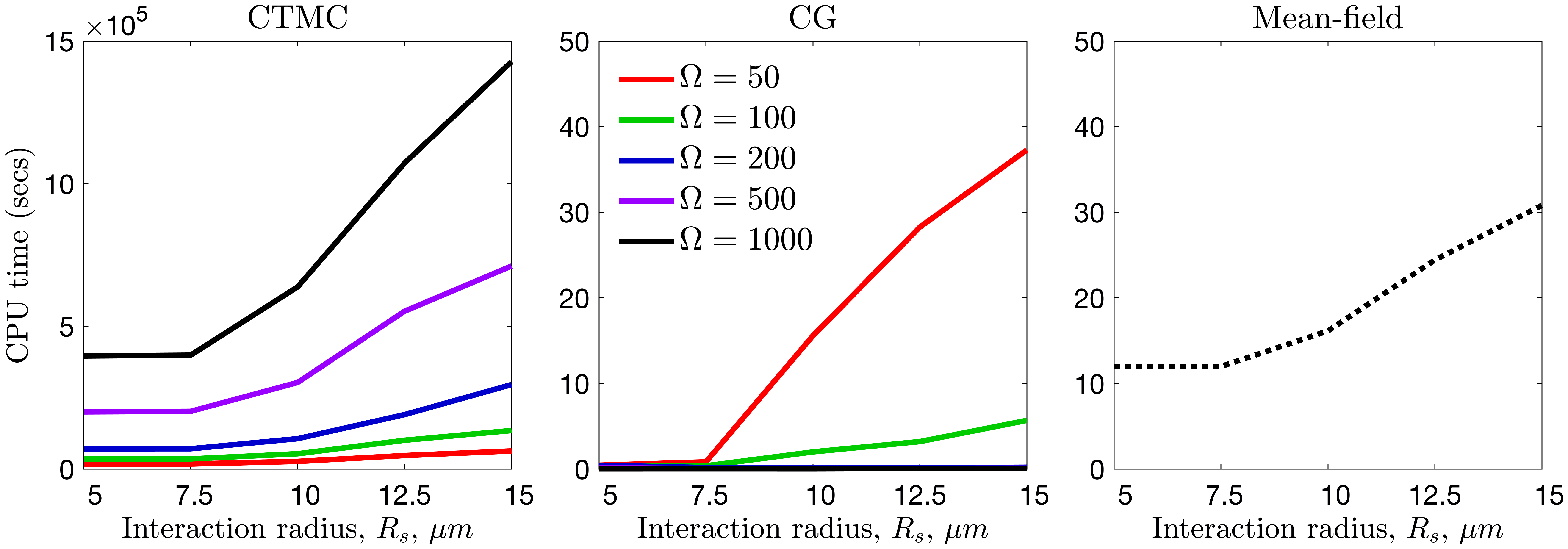}}

\subfloat[][branching network \label{ResultsCPU_N}]{\includegraphics[height = 4.3cm,valign=b]{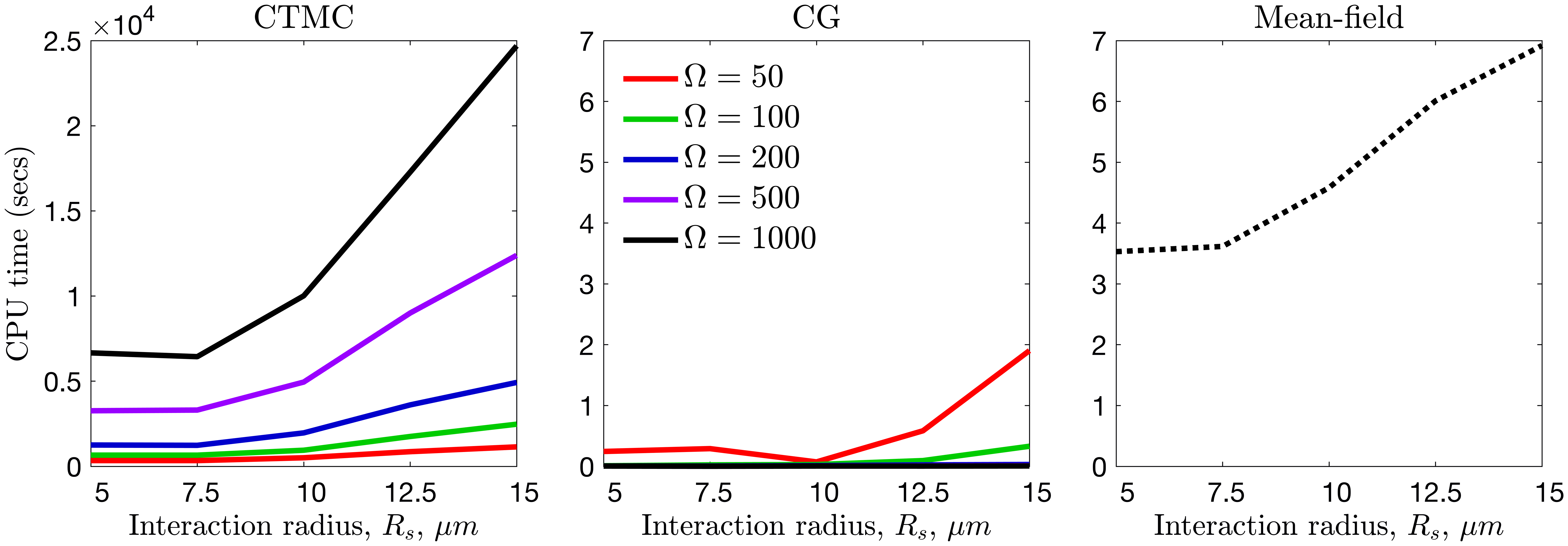}}
\caption{\textbf{Comparison of the mean CPU times to simulate the multicellular VEGF-Delta-Notch model.} The plots show how the average (100 realisations) CPU times (in seconds) to perform a single realization using the full stochastic CTMC (\underline{left panels}), CG (\underline{middle panels}), and mean-field (\underline{right panels}) descriptions changes as the cell-to-cell interaction radius, $R_s$, and the system size, $\Omega$, vary. The colour code indicates the system size, $\Omega$ (shown as insets in the middle panels); the mean-field description (dotted black line) corresponds to the limit $\Omega \rightarrow \infty$. The simulation setup was set to (a) a medium size cell monolayer, and (b) a small branching network. For both spatial geometries, the average CPU time for simulating the full stochastic CTMC is several orders of magnitude larger that those for CG and mean-field descriptions. For these simulations, the parameter values were fixed as indicated in \Cref{Params}.} 
\label{ResultsCPU}
\end{figure}

Given the transition rates and the corresponding stoichiometric vectors (listed in \cref{FullStochasticSystem}), we can formulate our individual cell stochastic model of the VEGF-Delta-Notch pathway in terms of the Chemical Master Equation (CME) 

\begin{equation} \label{CME}
\pdv{\prob(X,t)}{t} = \displaystyle \sum_r \of{ \alpha_r(X - \nu_r) \prob(X-\nu_r,t) - \alpha_r(X,t)\prob(X,t)},
\end{equation}
where $X = (N,D,I,R_2,R_2^*)^T$ and $\prob(X,t)$ is the probability of finding the system in state $X$ at time $t$. This CME cannot be solved analytically but can be numerically simulated via the Gillespie algorithm (GA) \cite{gillespie1976general} or the more efficient Next Subvolume (NSV) method \cite{elf2004spontaneous}. The computational cost of these simulations rapidly increases with the system size, $\Omega$. Furthermore, when extended to a multicellular system, the simulation time increases as the number of cells in the system increases (linearly for the GA or logarithmically for the NSV method). Thus, high computational power is required to perform numerical simulations for large numbers of cells ($O(10^2)-O(10^3)$) as in the angiogenesis model (see \Cref{ResultsCPU_ML,ResultsCPU_N}, left panels).

\renewcommand{\thesubfigure}{\alph{subfigure}}
\captionsetup{singlelinecheck=off}
\begin{figure}[htbp]
\centering 
\subfloat[][\label{BifurcationDiagram}]{\includegraphics[height = 6.0cm,valign=b]{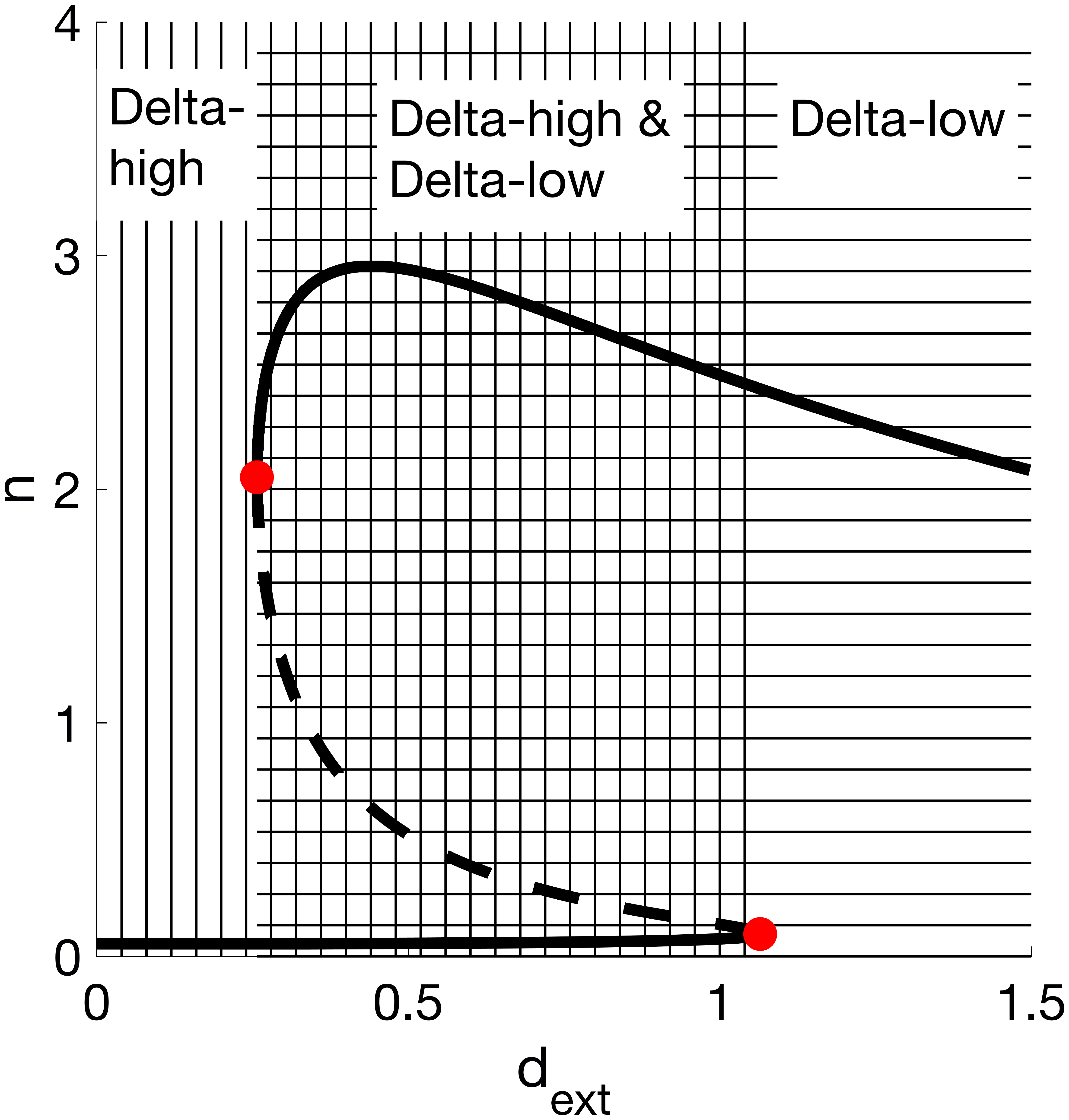}} \hfill
\subfloat[][\label{Tip2StalkTransition}]{\includegraphics[height = 6.2cm,valign=b]{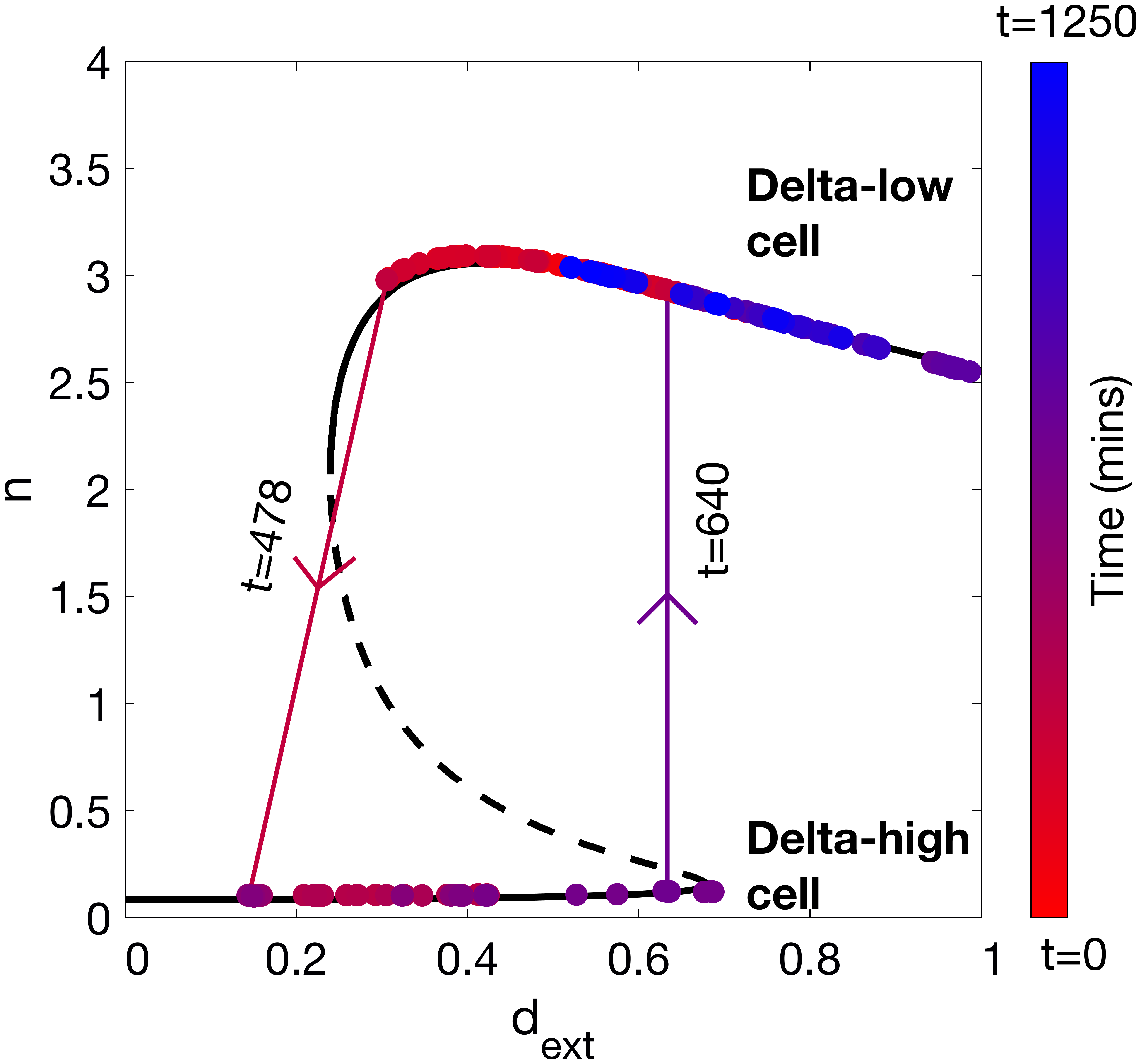}}
\caption{(a) A bifurcation diagram (corresponding to \Cref{eq:DN_single_nondimensional}) for the non-dimensional Notch concentration, $n$, as a function of extracellular Delta signalling, $d_{ext}$. The solid lines correspond to stable steady states; dashed line - to unstable saddle points. Saddle-node bifurcation points are indicated by red filled circles. Vertical (horizontal) hatching indicates a region in which the Delta-high (Delta-low) cell steady state exists. (b) Notch concentration, $n$, of a representative cell during simulation of a multiscale angiogenesis model \cite{stepanova2021multiscale} as the function of the cell's extracellular Delta signal, $d_{ext}$. The black line corresponds to a bifurcation curve (as in (a)) for a fixed value of $n_{ext}$ (although this value changes together with $d_{ext}$ in the simulations of our angiogenesis model, the bifurcation curve does not change significantly, so just one of them was plotted for simplicity). Arrows indicate phenotype switch transitions for the representative cell with the corresponding times. The colour bar represents simulation time.}
\label{DeltaNotchPlots}
\end{figure}

We can significantly decrease the computational cost of model simulations by deriving the mean-field limit equations, which read:

\begin{subequations}
\begin{align}
\dv{n}{t} &=\beta_N H^{S} (\rho_N \hspace{1pt} \iota;1.0,\lambda_{I,N},n_N) - n - d_{ext} \hspace{1pt} n- \kappa \hspace{1pt} n \hspace{1pt} d, \nonumber \\
\dv{d}{t} &=\beta_D H^{S} (\rho_{R_2} \hspace{1pt} r_2^*;1.0,\lambda_{R_2^*,D},n_D) - d - \eta \hspace{1pt} n_{ext} \hspace{1pt} d - \kappa \hspace{1pt} n \hspace{1pt} d, \nonumber \\
\dv{\iota}{t}&= d_{ext} \hspace{1pt} n - \tau \hspace{1pt} \iota, \tag{SM1.2}\label{eq:DN_single_nondimensional_subeq} \\
\dv{r_2}{t} &=\beta_{R_2} H^{S} (\rho_N \iota;1.0,\lambda_{I,R_2},n_{R_2}) - (1+v_{ext}) r_2, \nonumber \\
\dv{r_2^*}{t} &= v_{ext} \hspace{1pt} r_2 -\tau \hspace{1pt} r_2^*. \nonumber
\end{align} \label{eq:DN_single_nondimensional} 
\end{subequations} 

This system of equations is bistable for a range of values of the control (bifurcation) parameters, here the extracellular Delta and Notch concentrations, $d_{ext}$ and $n_{ext}$, respectively \cite{stepanova2021multiscale}. In \Cref{BifurcationDiagram}, we present a bifurcation diagram showing how the steady state non-dimensional Notch concentration, $n$, changes as $d_{ext}$ varies for a fixed value of $n_{ext}$. We note that for low (high) values of $d_{ext}$, the system is monostable, its only stable steady state corresponding to the Delta-high (Delta-low) cell phenotype. For intermediate values of $d_{ext}$, the system has two stable steady states (i.e. in this region both phenotypes coexist). 

The mean-field equations (\Cref{eq:DN_single_nondimensional}) are valid in the limit $\Omega \rightarrow \infty$. For finite values of $\Omega$ (as in any biological system), some level of noise is always present. This can affect the dynamics of the system, leading to behaviours which differ from the mean-field limit. In \Cref{Tip2StalkTransition}, we plot the non-dimensional Notch concentration of an individual cell during a single realisation of our multiscale model of angiogenesis \cite{stepanova2021multiscale}. Arrows on this plot indicate times at which the focal cell switched its phenotype. For example, a phenotype switch from Delta-low to Delta-high cell at $t=478$ minutes indicates that, during the simulation, the neighbourhood of the focal cell ($\uf{d_{ext},n_{ext}}$) changed and the cell was forced to adjust its phenotype accordingly. However, the phenotype switch, from Delta-high to Delta-low cell, at $t=640$ minutes is noise-induced since the cell's neighbourhood (i.e. $d_{ext}$) did not change at that time. Phenotype transitions of this type cannot be accounted for by the deterministic mean-field model. Likewise, \Cref{Tip2StalkTransition} confirms that fluctuations away from the mean-field steady state values are small since the simulated trajectory of the focal cell (circled markers) lies in a narrow neighbourhood around the deterministic bifurcation curve. This strengthens the case for the application of large deviation theory to coarse-grain the dynamics of this system.

\subsection{Multicellular system}
\label{subsection:MulticellularSystem}

In order to extend the individual cell CTMC model of the VEGF-Delta-Notch system given by the CME, \Cref{CME}, and the corresponding kinetic reactions listed in \Cref{FullStochasticSystem}, to a multicellular system in a 2D domain, we need only to specify the external levels of Delta, $\of{D_{ext}}_k$, and Notch, $\of{N_{ext}}_k$ (for the mean-field model, $\of{d_{ext}}_k$ and $\of{n_{ext}}_k$, respectively) for a cell situated in voxel $k$. A key parameter to determine cell cross-talk is the cell interaction radius, $R_s$, allowing for non-local (beyond immediate neighbours on a chosen lattice) interactions between cells. This is because in this model, cell position is known up to the position of its nucleus, and cell interactions are assumed to occur via membrane protrusions within a circular neighbourhood of its nucleus. Thus, $\of{D_{ext}}_k$ and $\of{N_{ext}}_k$ are computed as a normalised sum of the corresponding protein over all the neighbouring ECs within the interaction radius, $R_s$, from the focal cell's nucleus (for more details, see \cite{stepanova2021multiscale}):

\begin{subequations}
\begin{align}
\of{D_{ext}}_k &= \displaystyle \sum_{v_l ~\in~ neighbours(k) } \alpha_{k l} D_{l}, \\
\of{N_{ext}}_k & = \displaystyle \sum_{v_l ~\in~ neighbours(k) } \alpha_{k l} N_{l}.
\end{align} \label{ExtDN}
\end{subequations}
Here $k$ indicates the voxel index of the multicellular system. The weight $\alpha_{k l} = \frac{ \abs{v_{l} \cap \mathcal{B}_{R_s}(k)} }{A(k)}$, with $A(k) = \sum_{j} \abs{v_{j} \cap \mathcal{B}_{R_s}(k)}$, represents the proportion of the area of the circular neighbourhood of radius $R_s$ surrounding the focal cell in voxel $v_k$, $\mathcal{B}_{R_s}(k)$, that overlaps with voxel $v_l$.

When a bimolecular reaction of Delta-Notch trans-binding occurs, the neighbouring cell (in $neighbours(k)$) is chosen probabilistically (reaction \circled{1a}; reaction \circled{1b} is redundant for multicellular systems since it is accounted for as reaction \circled{1a} which takes place in neighbouring cells).  

The derivation of the mean-field system is straightforward. For each voxel, $v_k$, in the lattice we specify a system of ODEs:

\begin{subequations} 
\begin{align}
\dv{n_k}{t} &=\beta_N H^{S} (\rho_N \hspace{1pt} \iota_k;1.0,\lambda_{I,N},n_N) - n_k - \of{d_{ext}}_k \hspace{1pt} n_k- \kappa \hspace{1pt} n_k \hspace{1pt} d_k, \nonumber \\
\dv{d_k}{t} &=\beta_D H^{S} (\rho_{R_2} \hspace{1pt} \of{r_2^*}_k;1.0,\lambda_{R_2^*,D},n_D) - d_k - \eta \hspace{1pt} \of{n_{ext}}_k \hspace{1pt} d_k - \kappa \hspace{1pt} n_k \hspace{1pt} d_k, \nonumber \\
\dv{\iota_k}{t}&= \of{d_{ext}}_k \hspace{1pt} n_k - \tau \hspace{1pt} \iota_k, \tag{SM1.4} \\
\dv{\of{r_2}_k}{t} &=\beta_{R_2} H^{S} (\rho_N \iota_k;1.0,\lambda_{I,R_2},n_{R_2}) - (1+v_{ext}) \of{r_2}_k, \nonumber \\
\dv{\of{r_2^*}_k}{t} &= v_{ext} \hspace{1pt} \of{r_2}_k -\tau \hspace{1pt} \of{r_2^*}_k, \nonumber \\
\of{d_{ext}}_k &= \displaystyle \sum_{v_l ~\in~ neighbours(k) } \alpha_{k l} d_{l}, \nonumber \\
\of{n_{ext}}_k &= \displaystyle \sum_{v_l ~\in~ neighbours(k) } \alpha_{k l} n_{l}. \nonumber
\end{align} \label{MulticellularMeanField}
\end{subequations}
Here, again, cross-talk between neighbouring cells is accounted for via $\of{d_{ext}}_k$ and $\of{n_{ext}}_k$.

\vspace*{20pt}

\section{Geometric minimum action method (gMAM)} 
\label{appendix:gMAM}

The geometric minimum action method (gMAM) was developed in \cite{heymann2008geometric} as a technique for efficient numerical computation of the minimum action path (MAP) and the corresponding quasipotential (given as the minimum of the action functional) of a rare event. Briefly, starting with the geometric representation, $\widehat{S}(\phi)$ (\Cref{main-QuasipotentialGeometrical}), the Euler-Lagrange equation associated with the minimisation of $\widehat{S}(\phi)$ is derived, assuming $\hat{\theta}(\phi,\phi')$ is known. Then a (pre-conditioned) steepest descend algorithm can be used to solve this Euler-Lagrange equation, maintaining the standard arc length parametrisation of $\phi$. If no explicit formula for $\hat{\theta}(\phi,\phi')$ is available, then $\hat{\theta}(\phi,\phi')$ is computed in the inner loop of the algorithm, as a solution to the system

\begin{subequations}
\begin{align} 
&H(x, \hat{\theta}) = 0 \\
& \pdv{H(x, \theta)}{\theta} = \lambda \phi'
\end{align} \label{HamiltonianSystem}
\end{subequations}
for some $\lambda$.

For a diffusion process of type \Cref{main-SDEgeneral}, there are explicit expressions for the Lagrangian and the corresponding Hamiltonian

\begin{subequations}
\begin{align} 
L(x,y) & = \langle y - b(x), a^{-1}(x) (y- b(x)) = \norm{y-b(x)}^2_{a}, \\
H(x, \theta) & = \langle b(x), \theta \rangle + \frac{1}{2} \langle \theta, a(x) \theta \rangle,
\end{align} \label{HamiltonianDiffusionProcess}
\end{subequations}
\noindent where $\norm{p}^2_{a} = \langle p, a^{-1}(x) ~ p \rangle^{{}^1/_2}$ is a norm induced by the diffusion tensor, $a(x)$. This allows us to explicitly derive the action functional in its geometric reformulation \cite{heymann2008geometric},

\begin{equation*}
\widehat{S}(\phi) = \of{ \displaystyle \int_0^1 \norm{\phi ' }_{a} \norm{b(\phi)}_{a}  - \langle \phi ', a^{-1}(\phi) b(\phi) \rangle  }\mathop{d\alpha},
\end{equation*}
\noindent and the solution to the system of equations \Cref{HamiltonianSystem}:

\begin{subequations}
\begin{align} 
& \hat{\theta} (x,y) = a^{-1}(x) \of{\frac{\norm{b(x)}_{a}}{\norm{y}_{a}} - b(x)}, \\
& \lambda(x,y) = \frac{\norm{b(x)}_{a}}{\norm{y}_{a}}.
\end{align} \label{HamiltonianSystemSolution}
\end{subequations}

For a general birth-death CTMC of type \Cref{CME} with transition rates, $\alpha_r(x)$, and the corresponding stoichiometric vectors, $\nu_r$, the Hamiltonian reads:

\begin{equation} \label{HamiltonianCME}
H(x, \theta) = \displaystyle \sum_{r} \alpha_r (x) \of{\exp \of{\langle \theta, \nu_r \rangle} - 1}.
\end{equation}
\noindent In this case, there is no explicit solution to the system given by \Cref{HamiltonianSystem} but it can be computed in the inner loop of the gMAM \cite{heymann2008geometric}.

For the convergence of the gMAM for an SDE, $a$ and $b$ must be bounded and uniformly continuous, and $a$ has to be uniformly elliptic, whereas, for a CME, the rate functions, $\alpha_r$, must be uniformly bounded away from $0$ and $+\infty$ \cite{heymann2008geometric}.

\vspace*{20pt}

\section{System of stochastic differential equations of the VEGF-Delta-Notch model (individual cell)} 
\label{appendix:SDE}

Let $x^{\epsilon} = \of{n^{\epsilon}, d^{\epsilon}, \iota^{\epsilon}, r_2^{\epsilon}, r_2^{*\epsilon}}^T$. Then, the SDE for the stochastic VEGF-Delta-Notch system reads:

\begin{equation} \label{DN_SDE}
\mathrm{d} x^{\epsilon} (t) = b(x^{\epsilon}) \mathrm{d} t + \sqrt[]{\epsilon} \sigma (x^{\epsilon}) \mathrm{d} W.
\end{equation}
Here the drift vector, $b(x^{\epsilon}) \in \mathrm{R}^5$, is given by \Cref{DN_SDE_drift}, the diffusion tensor, $a(x^{\epsilon}) = (\sigma \sigma^T)(x^{\epsilon}) \in \mathrm{R}^{5 \times 5}$, and $\sigma^T(x^{\epsilon}) \in \mathrm{R}^{12 \times 5}$ is given by \Cref{DN_SDE_diffusion}. The level of noise is controlled by $\epsilon = \Omega^{-1}$. Finally, $W$ is a Wiener process in $\mathrm{R}^{12}$.

\begin{equation} \label{DN_SDE_drift}
b(x^{\epsilon}) = \begin{pmatrix}
\beta_N H^{S} (\rho_N \hspace{1pt} \iota^{\epsilon};1.0,\lambda_{I,N},n_N) - n^{\epsilon} - d_{ext} \hspace{1pt} n^{\epsilon}- \kappa \hspace{1pt} n^{\epsilon} \hspace{1pt} d^{\epsilon} \\
\beta_D H^{S} (\rho_{R_2} \hspace{1pt} r_2^{*\epsilon};1.0,\lambda_{R_2^*,D},n_D) - d^{\epsilon} - \eta \hspace{1pt} n_{ext} \hspace{1pt} d^{\epsilon} - \kappa \hspace{1pt} n^{\epsilon} \hspace{1pt} d^{\epsilon} \\
d_{ext} \hspace{1pt} n^{\epsilon} - \tau \hspace{1pt} \iota^{\epsilon} \\
\beta_{R_2} H^{S} (\rho_N \iota^{\epsilon};1.0,\lambda_{I,R_2},n_{R_2}) - (1+v_{ext}) r_2^{\epsilon} \\
v_{ext} \hspace{1pt} r_2^{\epsilon} -\tau \hspace{1pt} r_2^{*\epsilon}
\end{pmatrix}~.
\end{equation}
Here the shifted Hill function, $H^{S} (p; p_0, \lambda, n) = \frac{1+ \lambda \of{{}^p/_{p_0}}^n }{1+ \of{{}^p/_{p_0}}^n}$.


The matrix $\sigma^T(x^{\epsilon})$ takes the form:

\begin{equation} \label{DN_SDE_diffusion}
\sigma^T(x^{\epsilon}) = \begin{pmatrix}
S_1(x^{\epsilon}) & 0_{8 \times 2} \\
0_{4 \times 3} & S_2(x^{\epsilon}) 
\end{pmatrix},
\end{equation}
where  $0_{n \times m}$ is a zero block matrix of size $n \times m$, and block matrices $S_1(x^{\epsilon}) \in \mathrm{R}^{8 \times 5}$ and $S_2(x^{\epsilon}) \in \mathrm{R}^{4 \times 2}$ are defined as 

\small{
\begin{subequations}
\begin{align*}
S_1(x^{\epsilon}) &= \begin{pmatrix}
\sqrt[]{\beta_N H^{S} (\rho_N \hspace{1pt} \iota^{\epsilon};1.0,\lambda_{I,N},n_N)} & 0 & 0 \\
- ~ \sqrt[]{n^{\epsilon}} & 0 & 0 \\
0 & \sqrt[]{\beta_D H^{S} (\rho_{R_2} \hspace{1pt} r_2^{*\epsilon};1.0,\lambda_{R_2^*,D},n_D)} & 0\\
0 &  -~\sqrt[]{d^{\epsilon}} & 0 \\
 - ~ \sqrt[]{d_{ext}  \hspace{1pt} n^{\epsilon}} &  0 & \sqrt[]{ d_{ext} \hspace{1pt} n^{\epsilon}  } \\
 0 & -~\sqrt[]{ \eta \hspace{1pt} n_{ext} \hspace{1pt} d^{\epsilon} } & 0 \\
 0 & 0 & -~\sqrt[]{\tau \hspace{1pt} r_2^{*\epsilon}} \\
 - ~ \sqrt[]{\kappa \hspace{1pt} n^{\epsilon} \hspace{1pt} d^{\epsilon} } & -~\sqrt{\kappa \hspace{1pt} n^{\epsilon} \hspace{1pt} d^{\epsilon}} & 0 \\
 \end{pmatrix}, \nonumber \\
 S_2(x^{\epsilon}) &= \begin{pmatrix}
  \sqrt[]{\beta_{R_2} H^{S} (\rho_N \iota^{\epsilon};1.0,\lambda_{I,R_2},n_{R_2}) } & 0 \\
 -~\sqrt[]{v_{ext} \hspace{1pt} r_2^{\epsilon}} & -~\sqrt[]{v_{ext} \hspace{1pt} r_2^{\epsilon}} \\
  -~\sqrt[]{r_2^{\epsilon}} & 0 \\
 0 & -~\sqrt[]{r_2^{*\epsilon}} \\
\end{pmatrix}. \nonumber
 \end{align*}
 \end{subequations}
}

\normalsize

\vspace*{20pt}

\section{Minimum action path (MAP) for the VEGF-Delta-Notch system in an individual cell} 
\label{appendix:MAP}

\begin{figure}[htbp]
\centering 
\subfloat[][Delta-low cell $\rightarrow$ Delta-high cell \label{Map_s2t_3d}]{\includegraphics[height=4.8cm,valign=b]{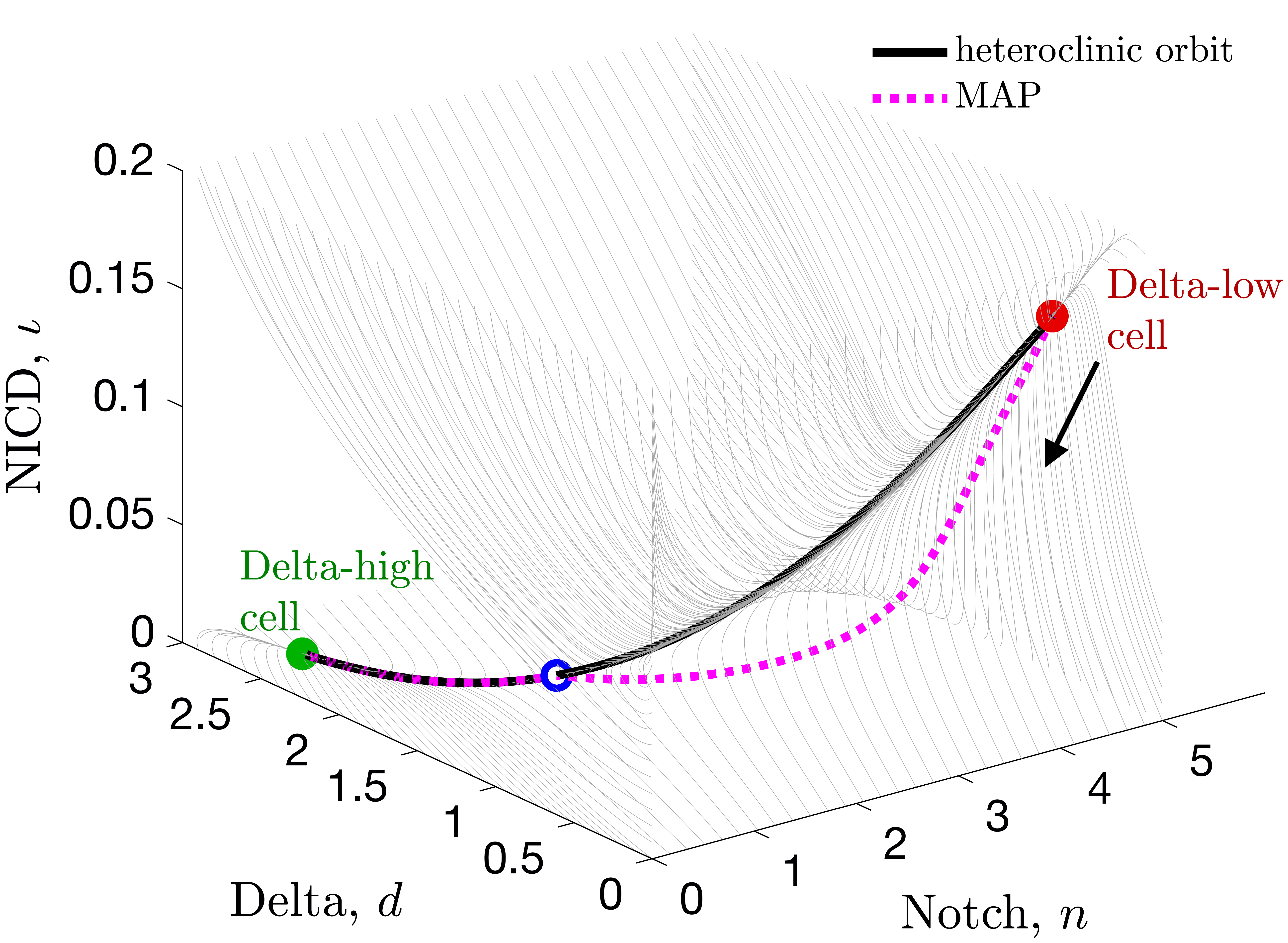}}
\hfill
\subfloat[][Delta-low cell $\rightarrow$ Delta-high cell \label{Map_s2t_2d}]{\includegraphics[height=4.8cm,valign=b]{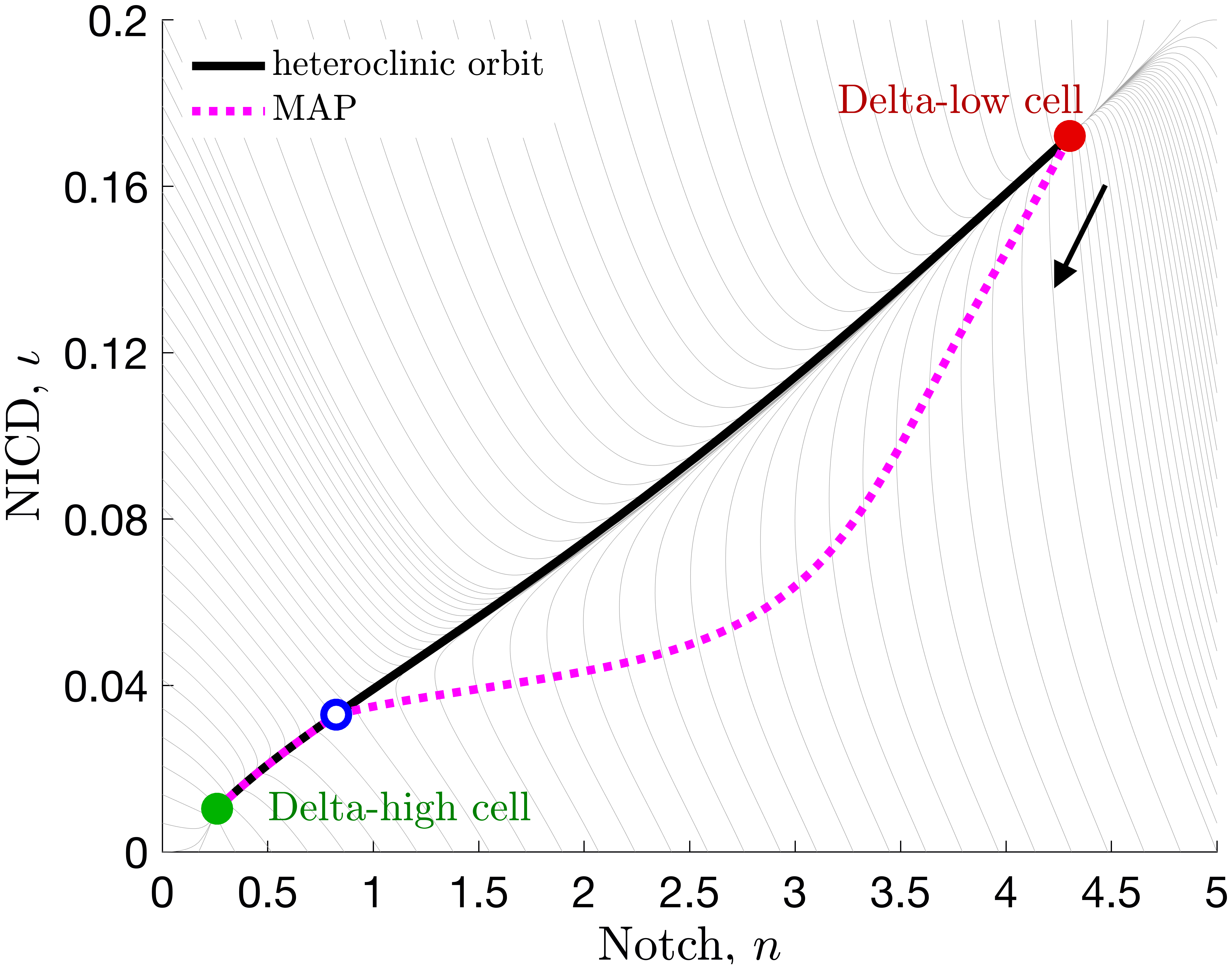}}

\subfloat[][Delta-high cell $\rightarrow$ Delta-low cell\label{Map_t2s_3d}]{\includegraphics[height=4.8cm,valign=b]{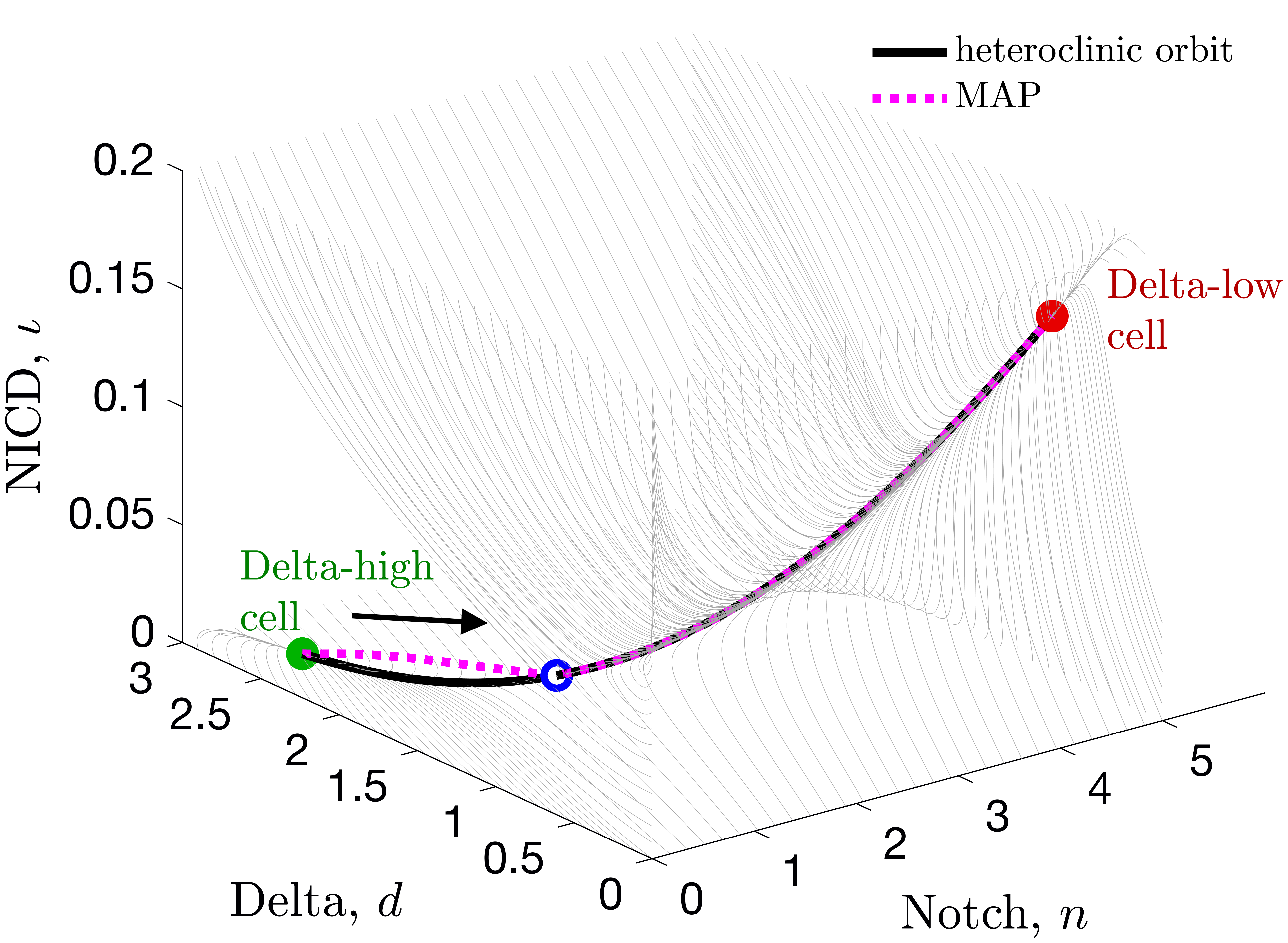}}
\hfill
\subfloat[][Delta-high cell $\rightarrow$ Delta-low cell\label{Map_t2s_2d}]{\includegraphics[height=4.8cm,valign=b]{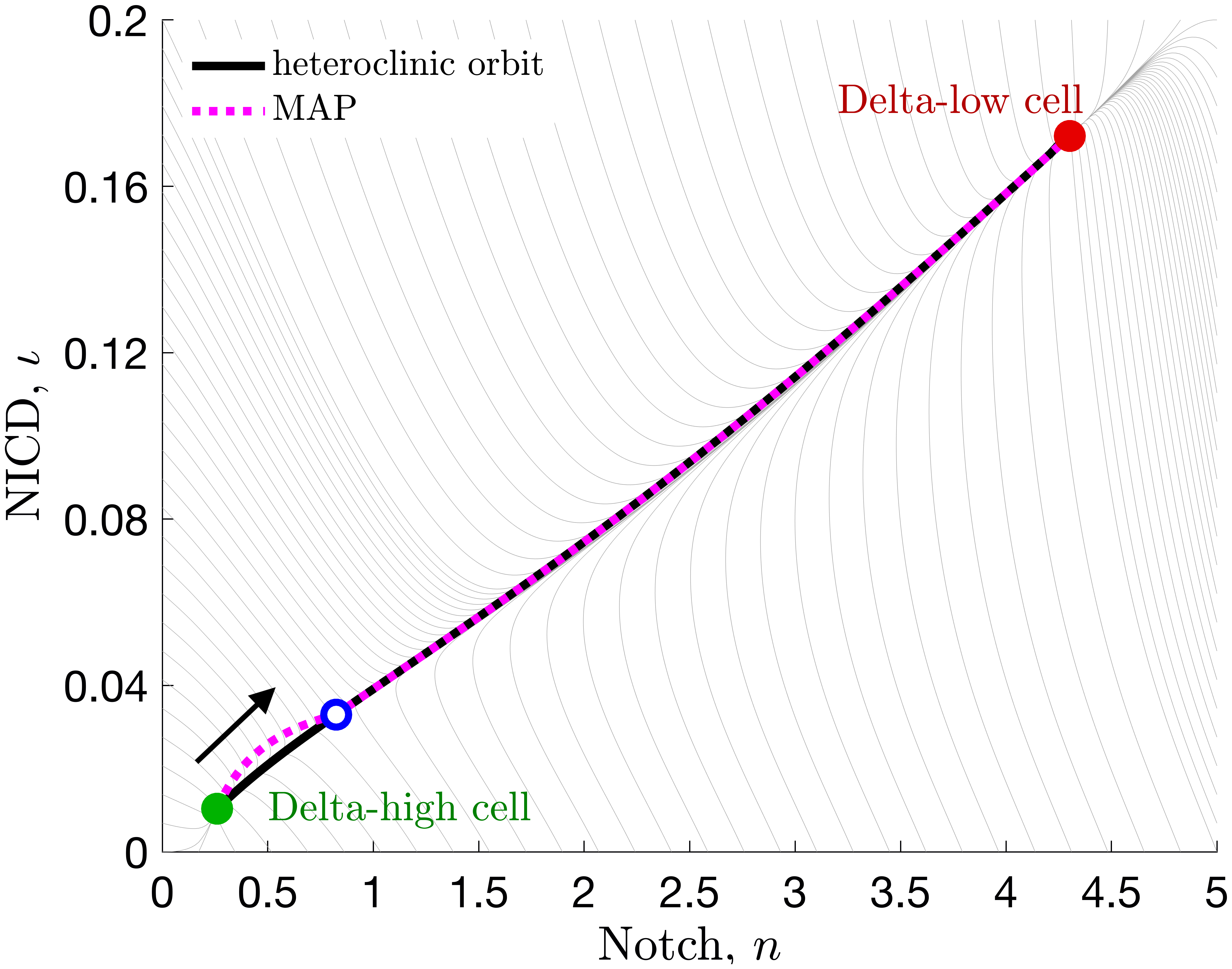}}
\vspace*{0.2cm}
\caption{\textbf{Minimum action paths.} Projections (3D, left panels; 2D, right panels) of the MAPs computed using the gMAM (dotted magenta lines) for transitions from (a)-(b) Delta-low cell $\rightarrow$ Delta-high cell; (c)-(d) Delta-high cell $\rightarrow$ Delta-low cell. Streamlines associated with the mean-field model are drawn in grey. Stable steady states corresponding to Delta-high (Delta-low) phenotype are indicated by filled green (red) circles; unstable saddle points are indicated by unfilled blue circles. Heteroclitic orbits, connecting steady states, are indicated by thick black lines.}
\label{MAPplot}
\end{figure}

Using the SDE for the VEGF-Delta-Notch system (\Cref{DN_SDE,DN_SDE_drift,DN_SDE_diffusion}), we implemented the gMAM to compute the minimum action path (MAP) and the corresponding quasipotential (using explicit expressions for the Hamiltonian and the momentum, \Cref{HamiltonianDiffusionProcess,HamiltonianSystemSolution}). An illustration of the MAPs for transitions between phenotypes for a fixed set of parameters (\Cref{Params}) is shown in \Cref{MAPplot}.

\renewcommand{\thesubfigure}{\alph{subfigure}}
\captionsetup{singlelinecheck=off}

\clearpage 

\section{Pseudocode algorithm for simulating the multi-agent CG model of a system with a region of multistability} 
\label{appendix:SimulationAlgorithm}
\phantom{Text}
\vspace{40pt}
\begin{algorithm}[tbh]
\caption{Pseudocode algorithm for simulating the multi-agent CG model of a system with a region of multistability.}
\label{SimulationAlgorithm}
\begin{algorithmic}[1]
\State Specify final simulation time, $T_{final}$, and the system size, $\Omega$.
\State Given a discretisation, $\{v_j\}_{j}$, of the external (bifurcation) variables, read look-up tables for steady states, quasipotential and prefactor values. 
\State For each $v_j$, compute CG transition rates, $k_{x_s \rightarrow x_l}$, $s,~l =1 \ldots S,~s \neq l$, defined by \Cref{main-CG_rates_DN} for the specified $\Omega$. Here $\{x_s\}_{s=1}^S$ is a set of stable steady states.
\State Initialise interpolation routines to establish an input-output relationship between an arbitrary $v \in \mathcal{V}$ and the CG transition rates, $k_{x_s \rightarrow x_l}$.
\State Initialise the system with a pre-pattern by using the original stochastic model or its mean-field limit (preferable).
\State Set the simulation time, $t=0$.
\While{$t<T_{final}$}
\State Set total propensity, $P=0$.
\For{each entity, $e$,}
\State Compute its external variables, $v^e$.
\For{$k,~l = 1 \ldots S,~k \neq l$,}
\State Interpolate $k^e_{x_s \rightarrow x_l}$ for the given $v^e$.
\State $P=P+k^e_{x_s \rightarrow x_l}$.
\EndFor
\EndFor
\State Sample the waiting time for the next transition to occur, $\bar{\tau} \sim \mathrm{Exp}(1/P)$, where 

\hspace{-8pt} $\mathrm{Exp}(\lambda)$ is an exponential distribution of intensity $\lambda$.
\State Probabilistically (as in the Gillespie algorithm), decide which transition occurs 

\hspace{-8pt} (in which entity, $\bar{e}$, and between which stable steady states). 
\State For this entity, compute again $v^{\bar{e}}$ and interpolate its new steady state after 

\hspace{-8pt} the transition.
\State Update the simulation time, $t=t+\bar{\tau}$.
\EndWhile
\State End of simulation.
\end{algorithmic}
\end{algorithm}


\clearpage

\section{Quantification metrics}
\label{appendix:Metrics}

We used the following metrics to compare our models.

\paragraph*{Delta-high cell proportion} In the angiogenesis model \cite{stepanova2021multiscale}, we used Delta levels as a proxy to determine cell phenotype. Thus, the number of Delta-high cells is given by the number of cells whose Delta level exceeds the threshold $d_{tip} = 0.1 \beta_D$. Then the Delta-high cell proportion at time, $t$, can be computed as the ratio of the number of Delta-high cells to the total number of cells. Here, $\beta_D$ is the characteristic expression of Delta in a cell (see \Cref{Params}).

\begin{figure}[h]
\subfloat[][\label{ClustersR10}]{\includegraphics[height = 6cm,valign=b]{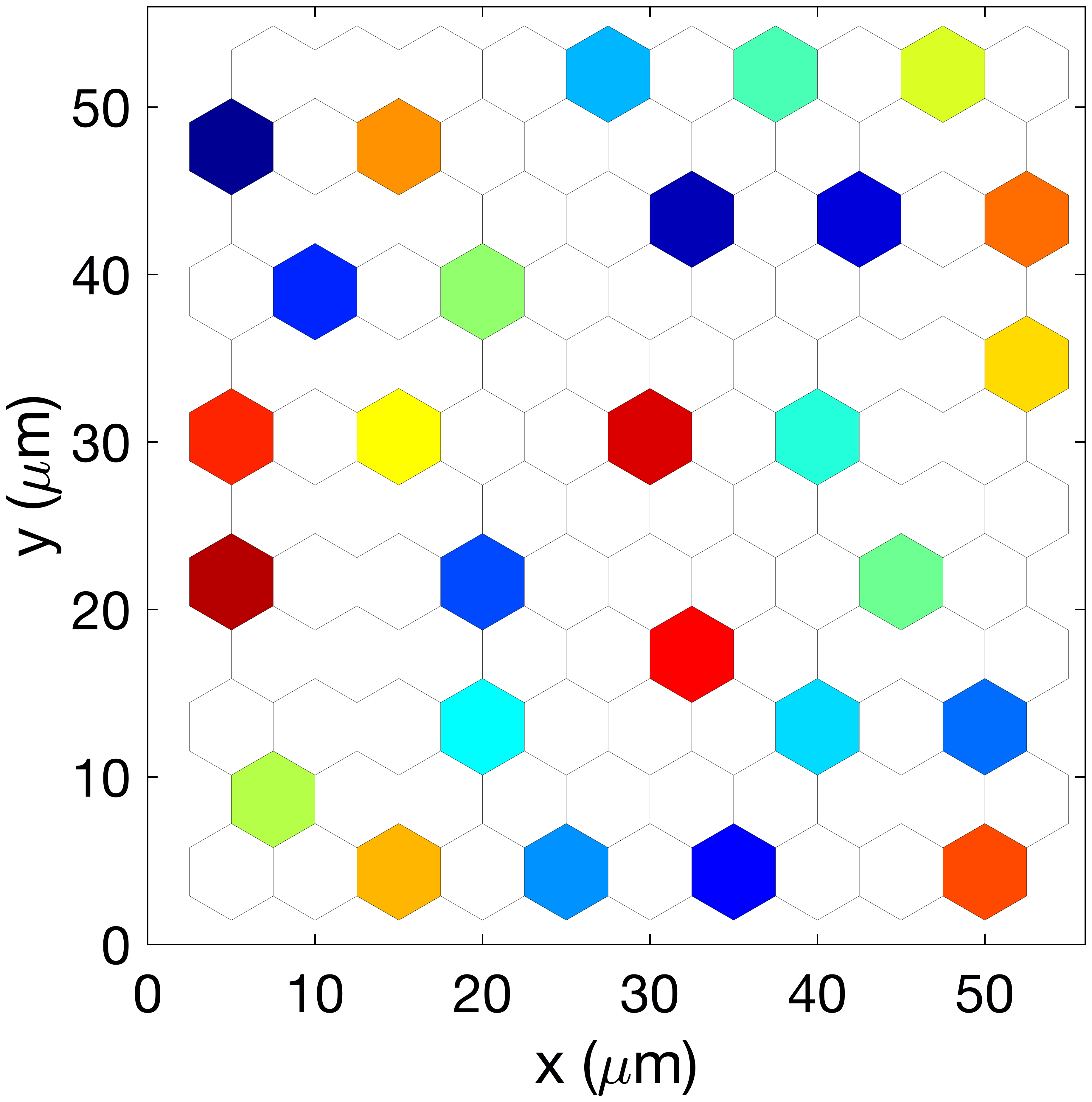}}
\hfill
\subfloat[][\label{ClustersR30}]{\includegraphics[height = 6.1cm,valign=b]{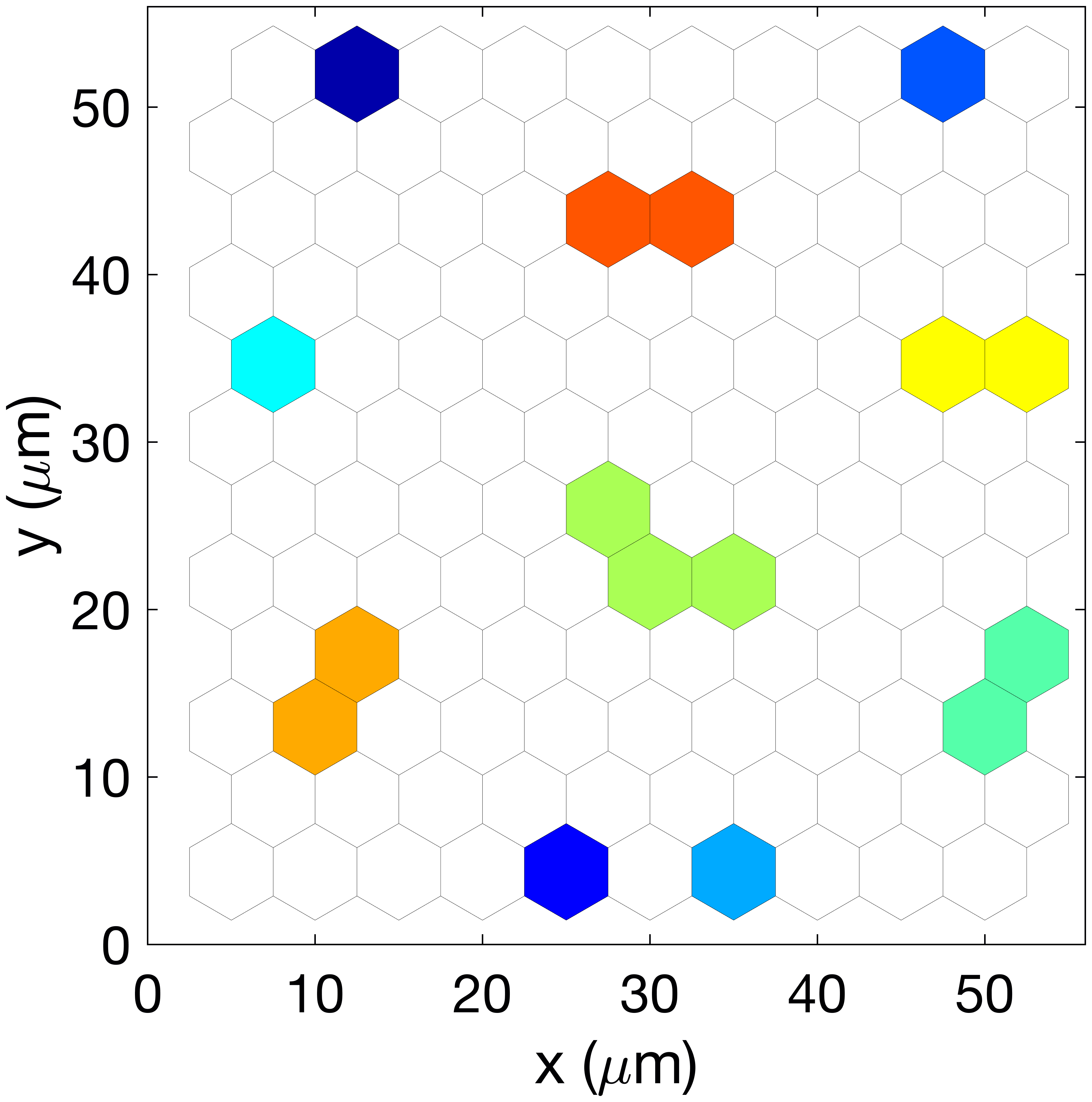}}
\caption{\textbf{Simulation results showing how the long time distribution of Delta-high cell clusters in a small monolayer of cells changes as the cell-to-cell interaction radius varies.} Cell interaction radius (a) $R_s = 5 \mu m$ (b) $R_s = 15 \mu m$. Each group of Delta-high cells (a cluster) is coloured by a distinct colour (randomly chosen). Delta-low cells are left white. }
\label{ClustersIllustration}
\end{figure}

\paragraph*{Delta-high cell cluster distribution} Depending on the parameter values, in the final spatial pattern (at a fixed final simulation time), Delta-high cells can form small clusters, i.e. be adjacent (see \Cref{ClustersIllustration}). We extracted the distribution of the cluster sizes for final configurations of the spatial phenotype pattern.  

\paragraph*{Computational cost} Computational cost is defined to be the mean CPU times (in seconds) required to perform a single realisation of a model simulation. Technical specifications of computers used to perform the simulations are indicated in \hyperlink{main-FileSpecs}{File S1}.

\renewcommand{\thesubfigure}{\alph{subfigure}}
\captionsetup{singlelinecheck=off}
\clearpage

\section{Supplementary figures and tables}
\label{appendix:SupFigs}

\phantom{Text}

\renewcommand{\thesubfigure}{\alph{subfigure}}
\captionsetup{singlelinecheck=off}
\begin{figure}[htbp]
\centering 
\subfloat[][$R_s=5 \mu m$ \label{PatternVaryingR10}]{\includegraphics[height=6cm,valign=b]{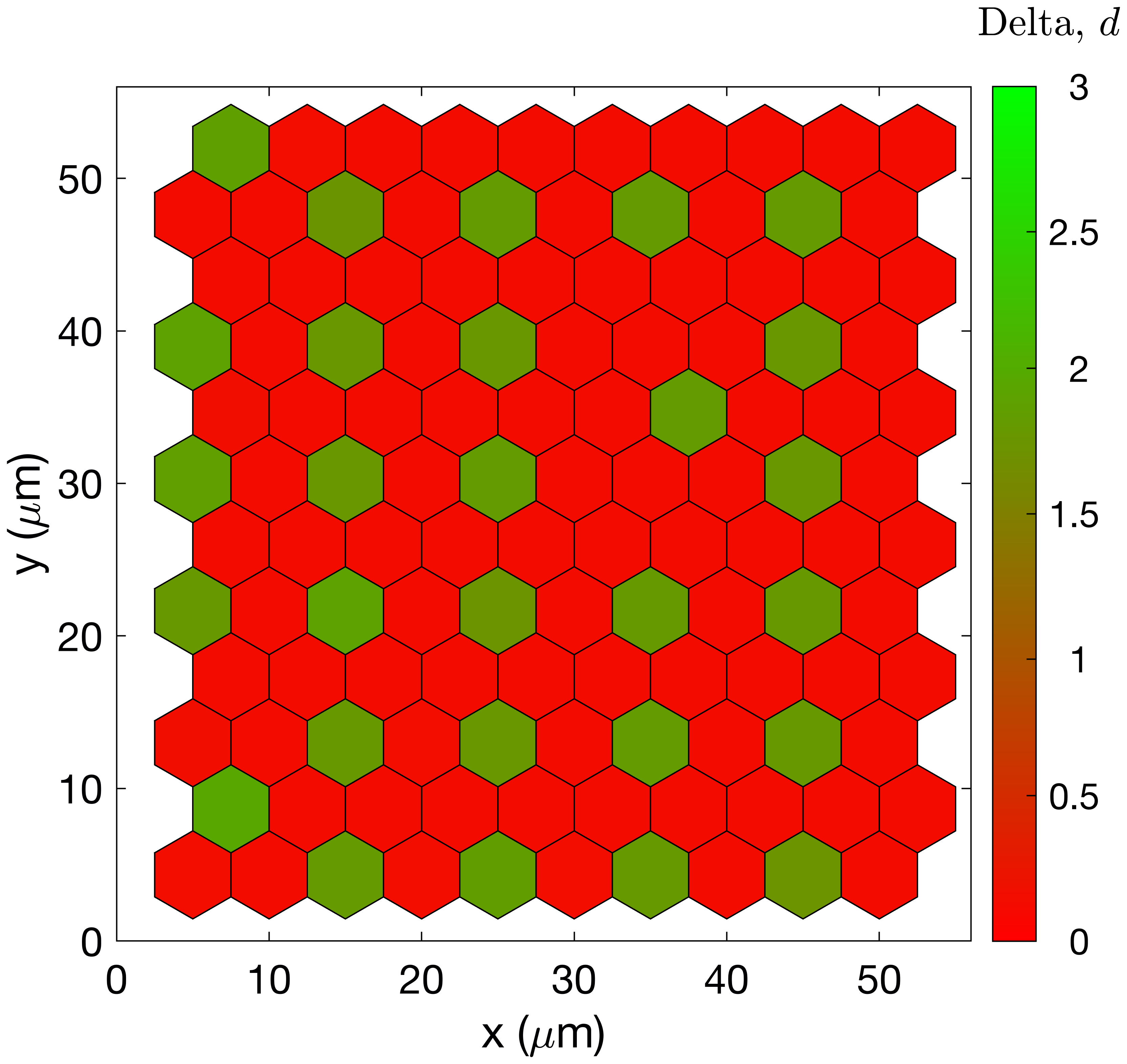}}
\hfill
\subfloat[][$R_s=10 \mu m$\label{PatternVaryingR20}]{\includegraphics[height=6cm,valign=b]{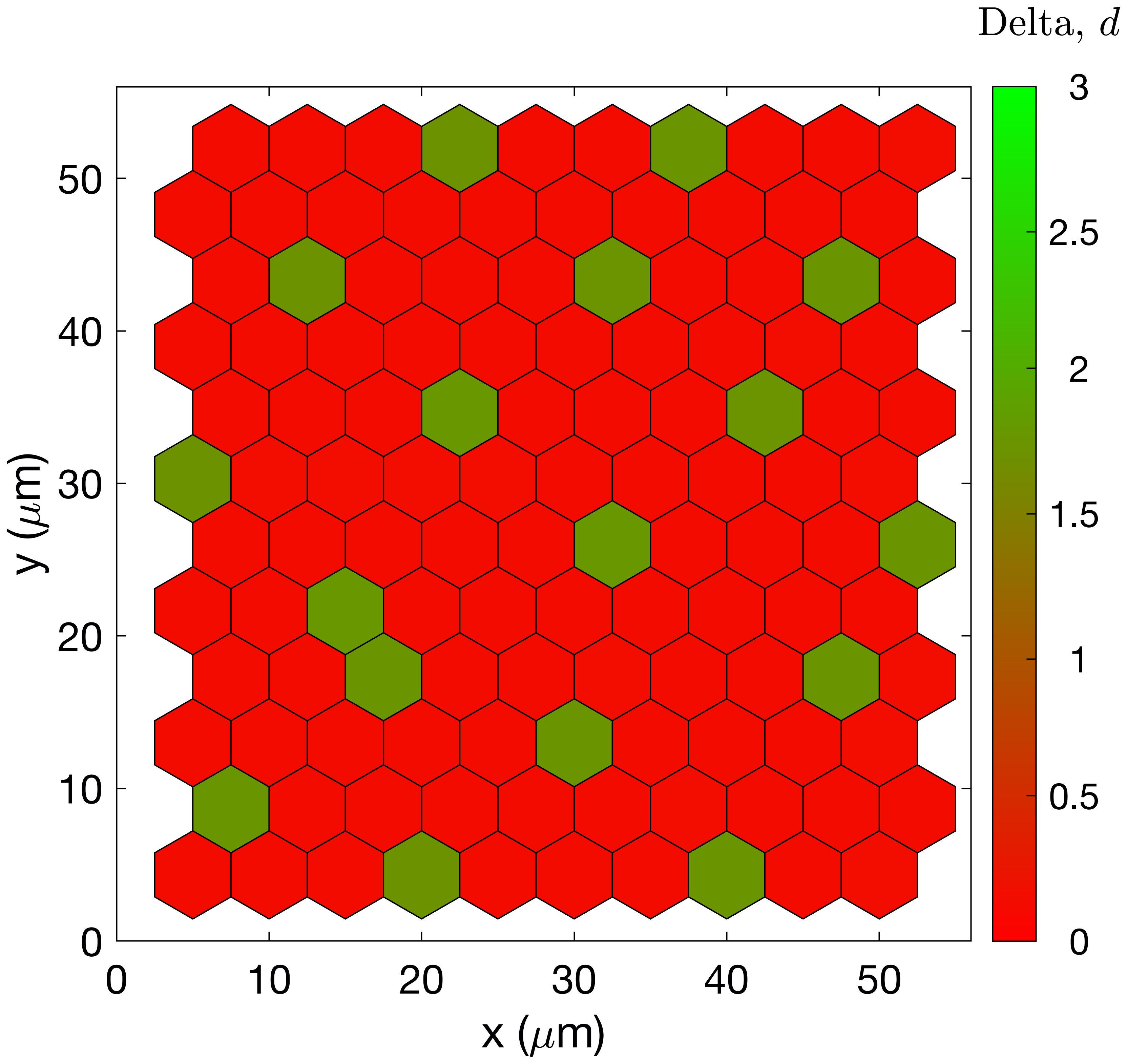}}

\subfloat[][$R_s=15 \mu m$ \label{PatternVaryingR30}]{\includegraphics[height=6cm,valign=b]{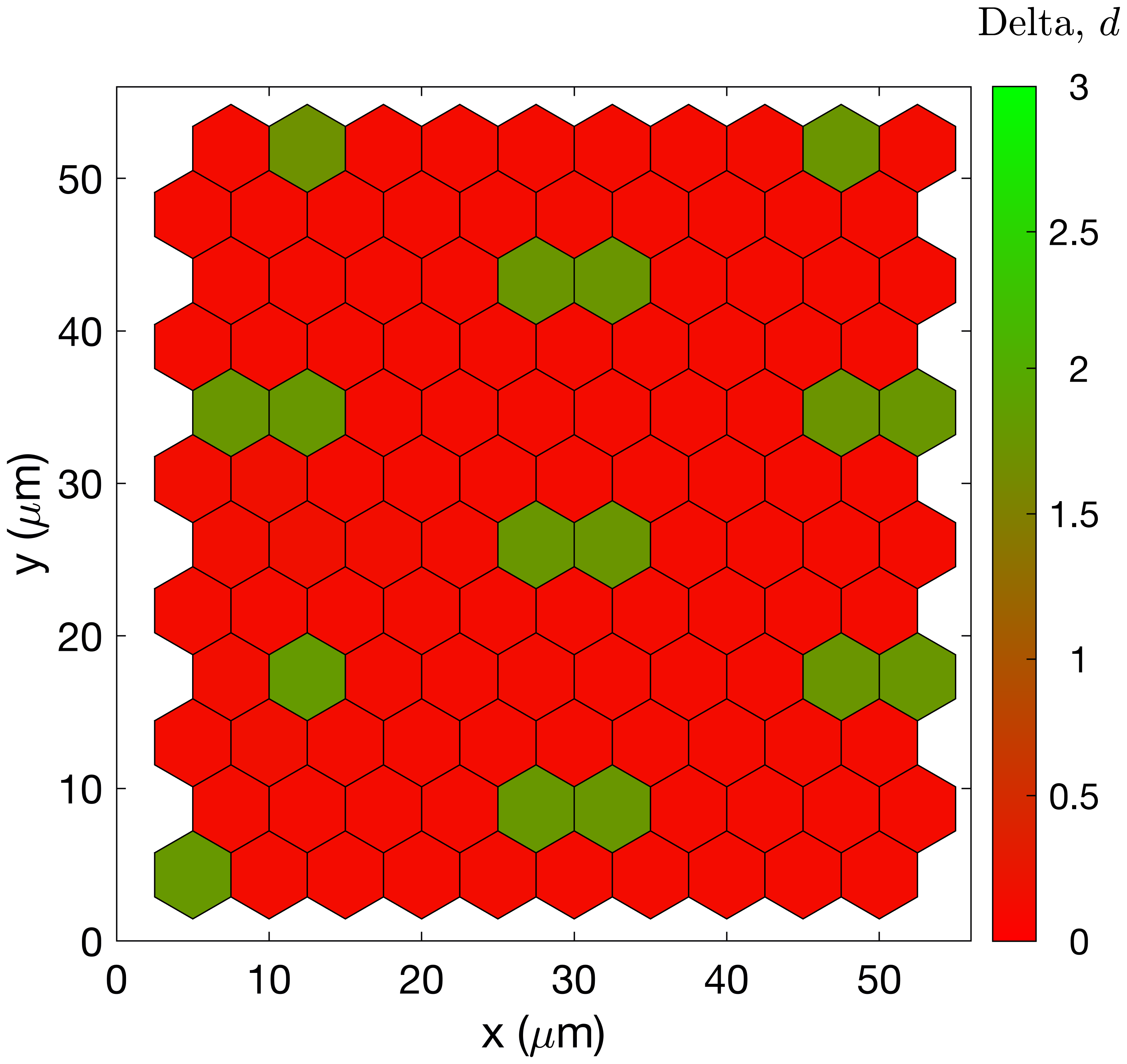}}
\hfill
\subfloat[][$R_s=20 \mu m$ \label{PatternVaryingR40}]{\includegraphics[height=6cm,valign=b]{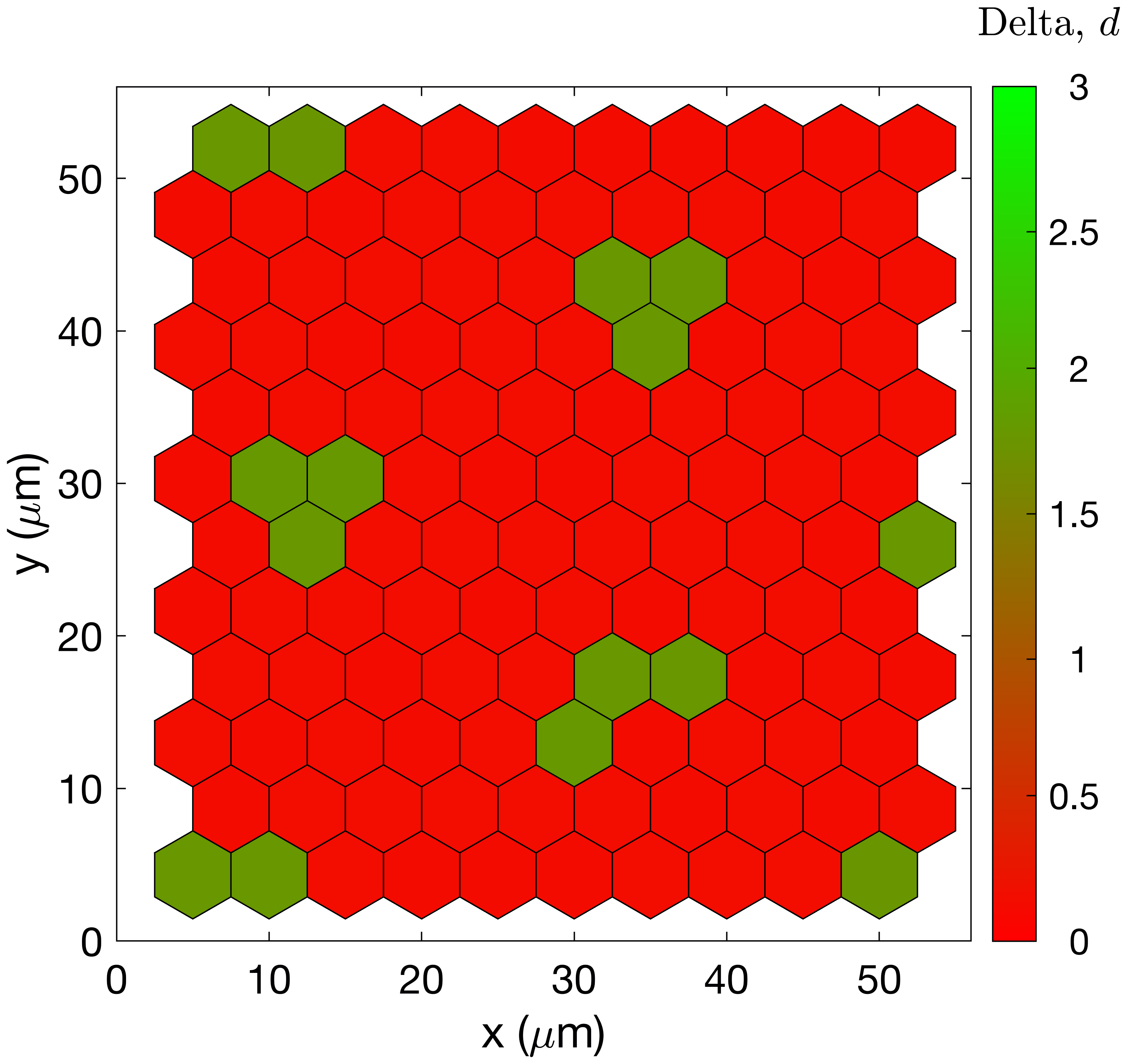}}
\caption{\textbf{A series of plots showing how the spatial patterns, generated by the VEGF-Delta-Notch signalling in a cell monolayer, become more clustered as the interaction radius, $R_s$, increases.} For these simulations, the interaction radius is fixed at (a) $R_s=5 \mu m$; (b) $R_s=10 \mu m$; (c) $R_s=15 \mu m$; (d) $R_s=20 \mu m$. The system size is fixed at, $\Omega=100$; the rest of the parameter values were fixed as indicated in \Cref{Params}. The colour bar indicates the levels of Delta, $d$, in each cell.}
\label{PatternVaryingR}
\end{figure}

\renewcommand{\thesubfigure}{\alph{subfigure}}
\captionsetup{singlelinecheck=off}
\begin{figure}[htbp]
\centering 
\includegraphics[height=7cm,valign=b]{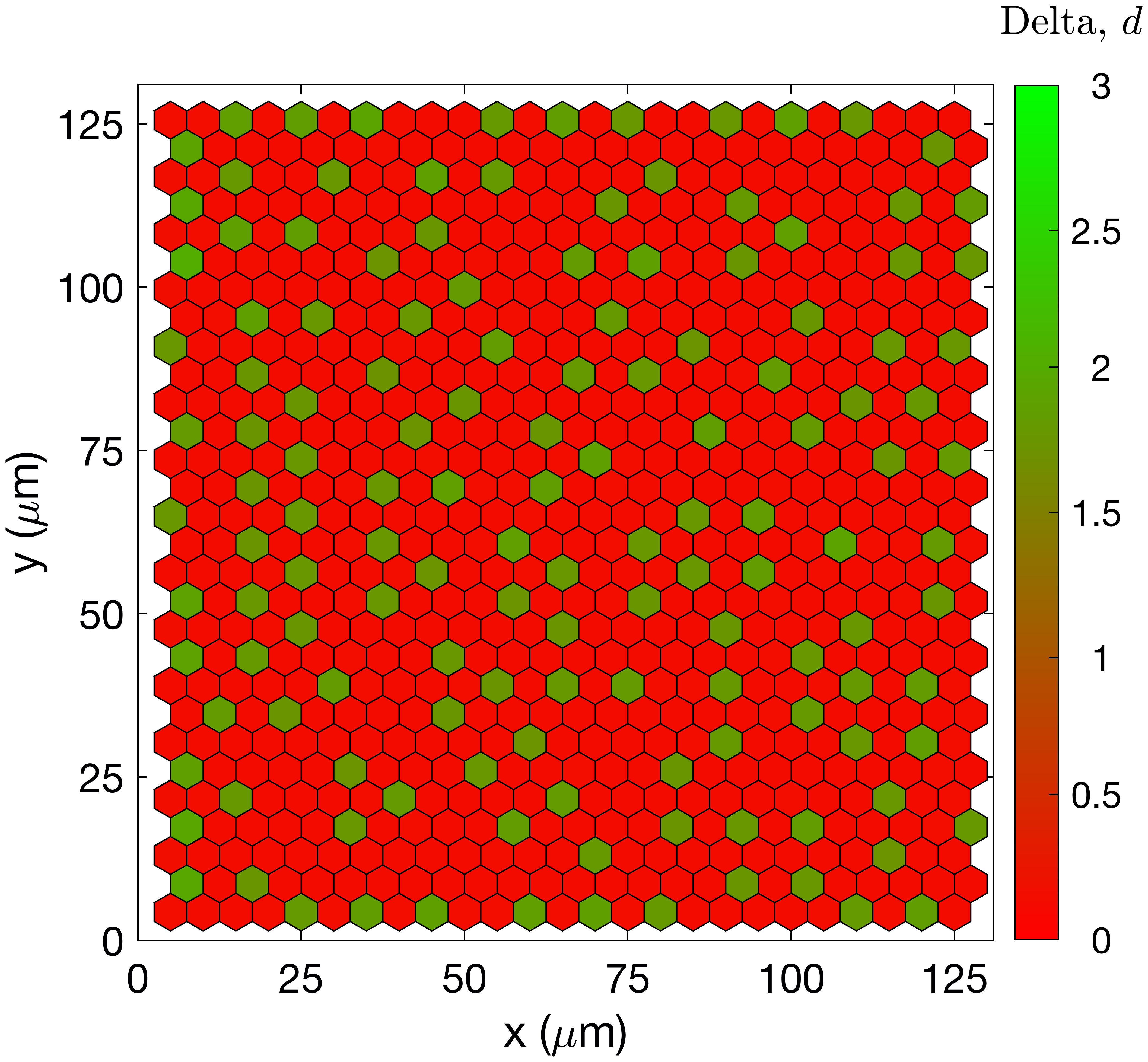}
\caption{\textbf{Initial configuration of a cell monolayer for numerical simulation.} The size of the monolayer is $25 \times 29$ voxels. The colour bar indicates the initial levels of Delta, $d$, in each cell.}
\label{InitialSetup_M}
\end{figure}

\renewcommand{\thesubfigure}{\alph{subfigure}}
\captionsetup{singlelinecheck=off}
\begin{figure}[htbp]
\centering 
\includegraphics[height=7cm,valign=b]{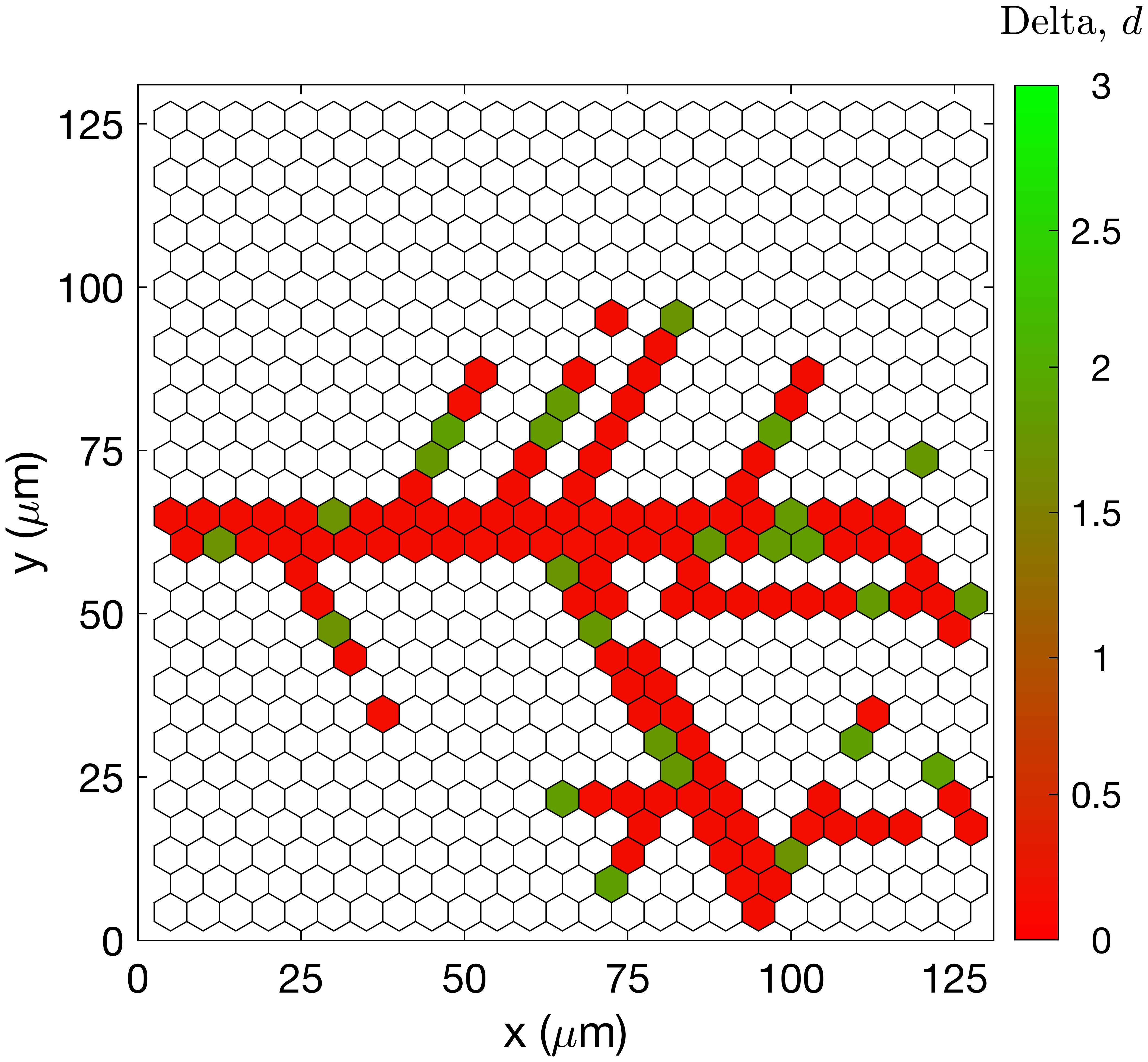}
\caption{\textbf{Initial setup configuration of a branching network for numerical simulation.} We extracted this configuration from a simulation of the angiogenesis model \cite{stepanova2021multiscale}. The colour bar indicates the level of Delta, $d$. Voxels without cells are left white in the plot.}
\label{InitialSetup_N}
\end{figure}

\renewcommand{\thesubfigure}{\alph{subfigure}}
\captionsetup{singlelinecheck=off}
\begin{figure}[htbp]
\centering 
\subfloat[][\label{SnapshotsCGVascularNetwork1}]{\includegraphics[height=6cm,valign=b]{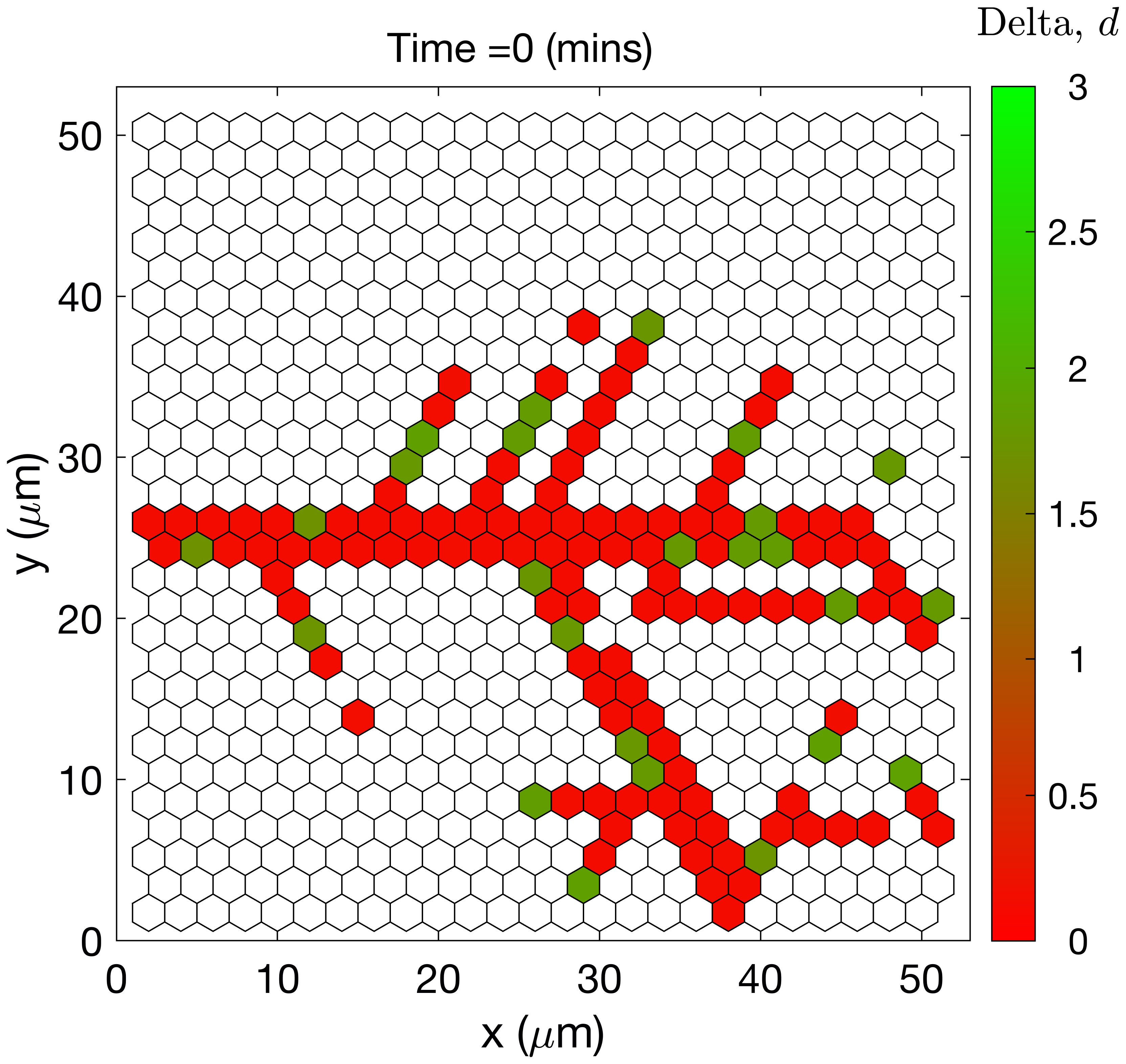}}
\hfill
\subfloat[][\label{SnapshotsCGVascularNetwork2}]{\includegraphics[height=6cm,valign=b]{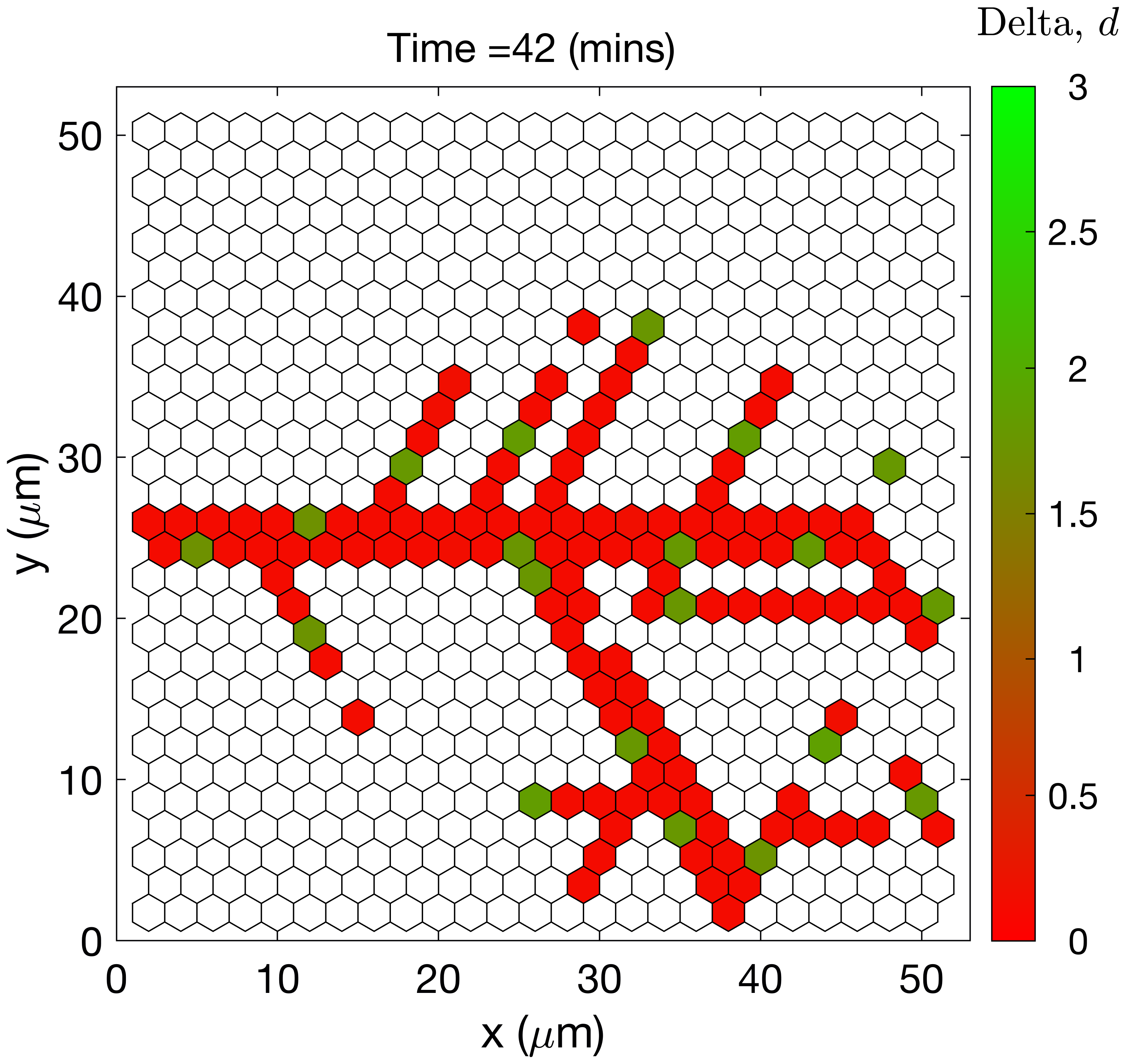}}

\subfloat[][\label{SnapshotsCGVascularNetwork3}]{\includegraphics[height=6cm,valign=b]{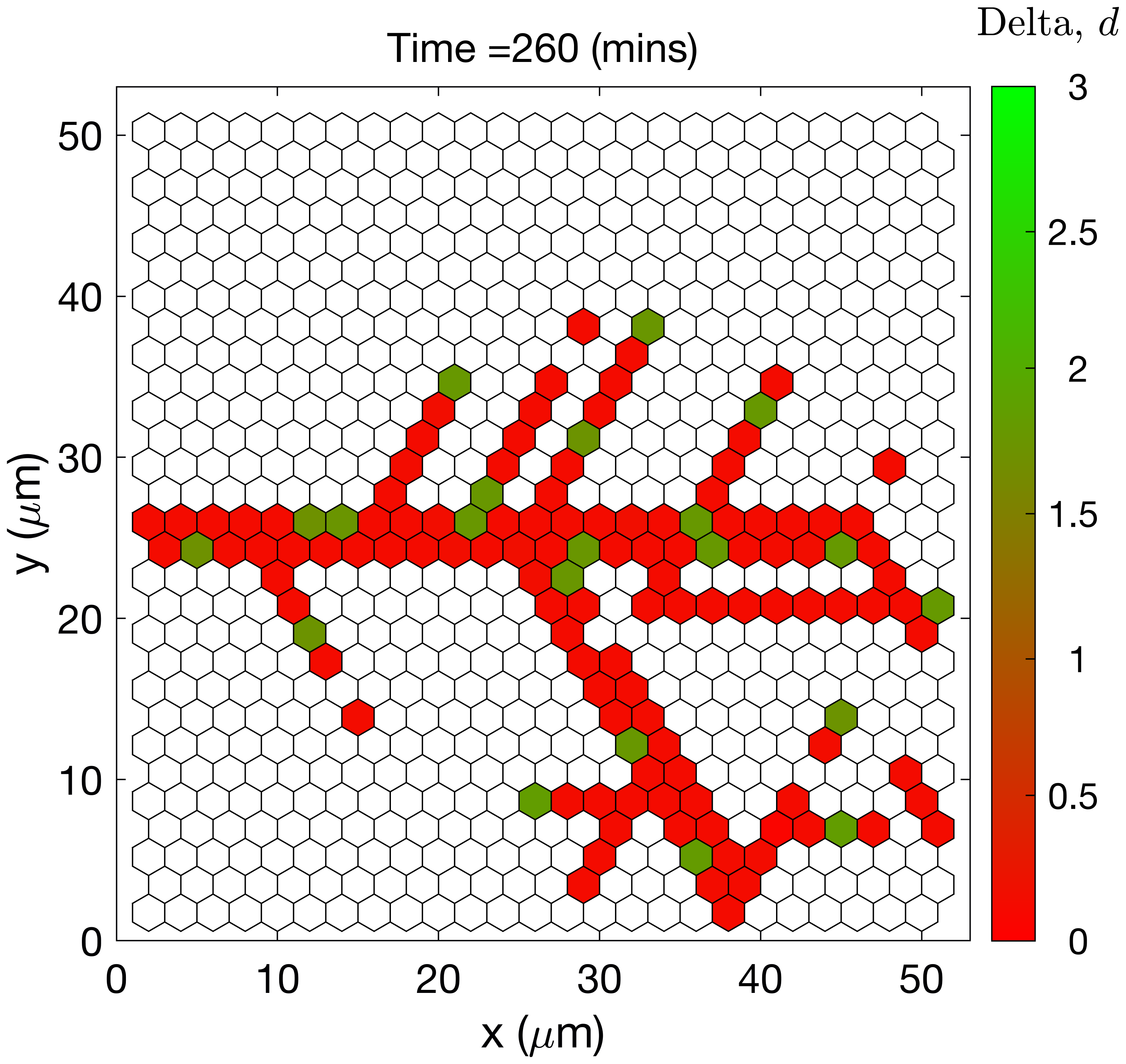}}
\hfill
\subfloat[][\label{SnapshotsCGVascularNetwork4}]{\includegraphics[height=6cm,valign=b]{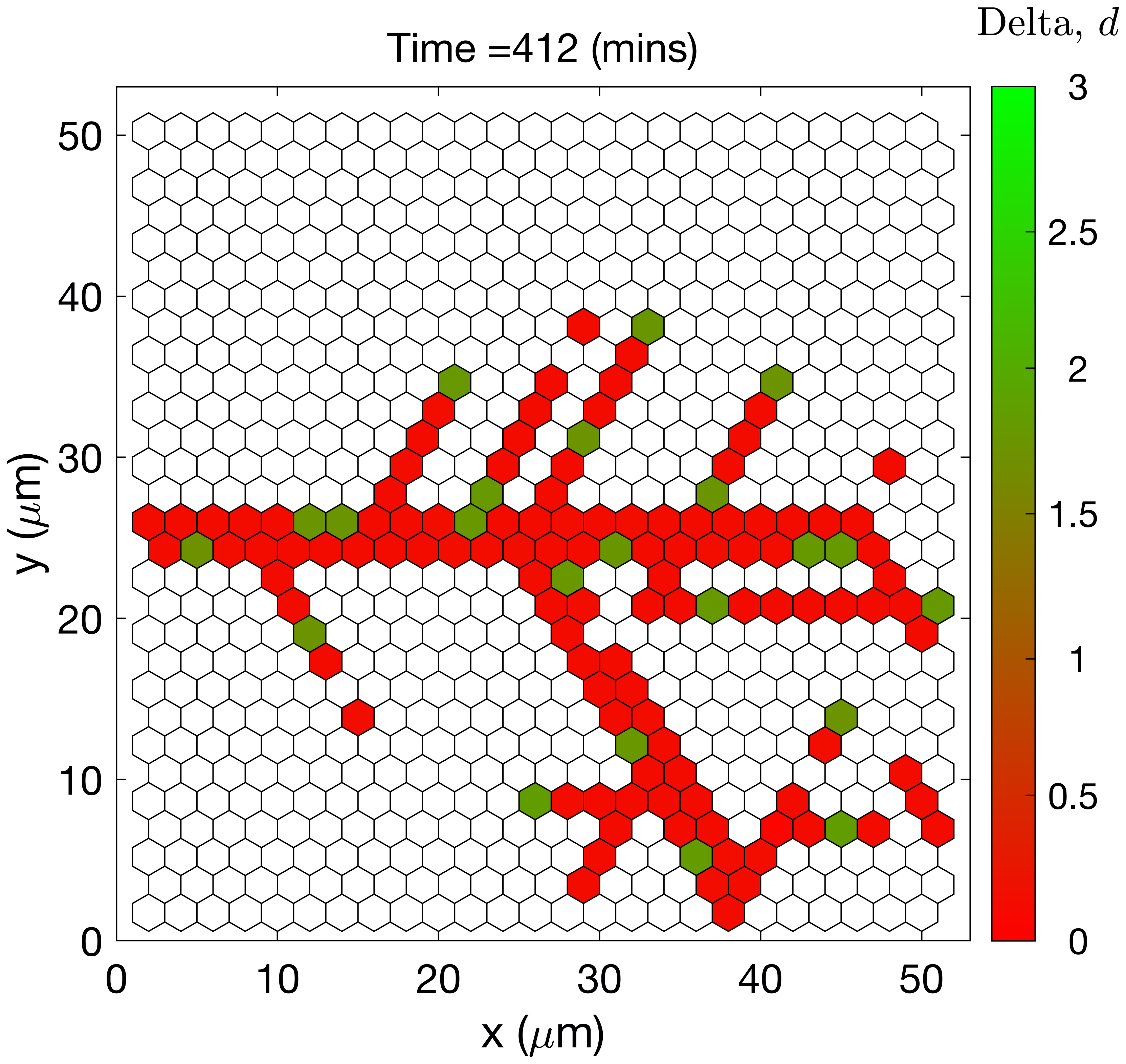}}
\caption{\textbf{A series of plots illustrating the time evolution of the phenotype patterning in a branching network in a typical simulation of the CG system for the multicellular VEGF-Delta-Notch signalling pathway.} Time points (indicated in the title of each plot) are (a) $t = 0$, (b) $t=42$, (c) $t=260$, (d) $t=412$ minutes. The colour bar indicates the levels of Delta, $d$, in each cell. For these simulations, the interaction radius and system size were fixed at $R_s=15 \mu m$ and $\Omega=100$, respectively. The remaining parameter values were fixed as indicated in \Cref{Params}.}
\label{SnapshotsCGVascularNetwork}
\end{figure}

\renewcommand{\thesubfigure}{\alph{subfigure}}
\captionsetup{singlelinecheck=off}
\begin{figure}[htbp]
\centering 
\subfloat[][\label{ResultsVascularNetwork_TipProportion}]{\includegraphics[height = 4.3cm,valign=b]{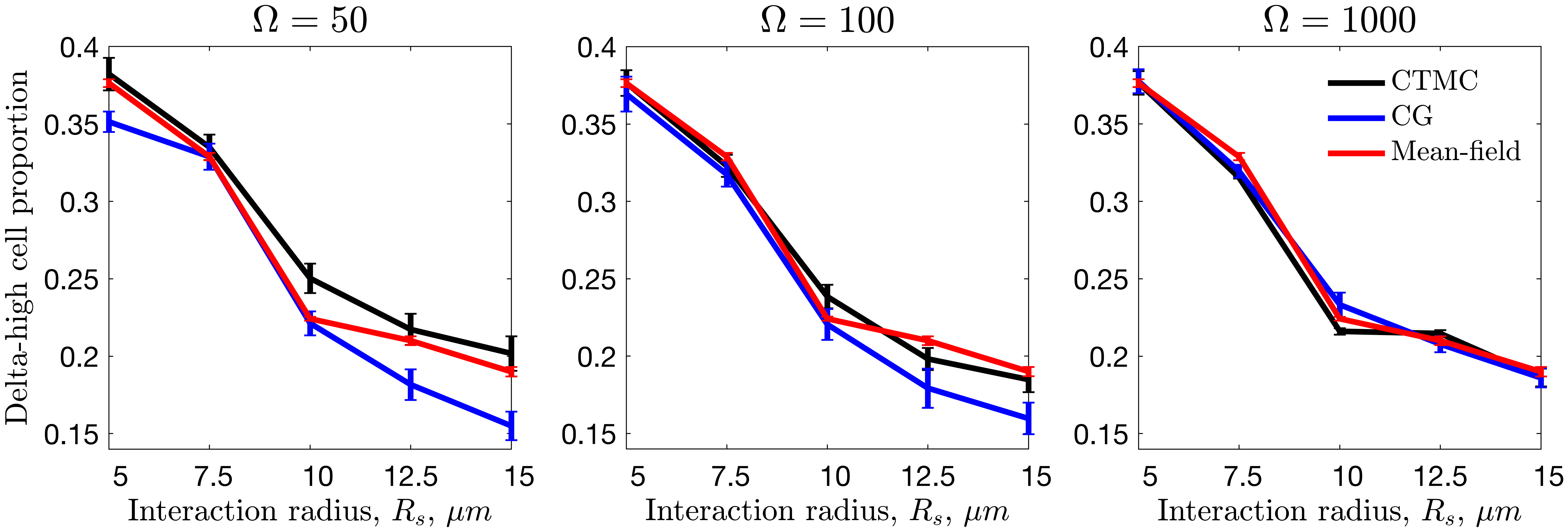}}

\subfloat[][\label{ResultsVascularNetwork_Clusters}]{\includegraphics[height = 4.2cm,valign=b]{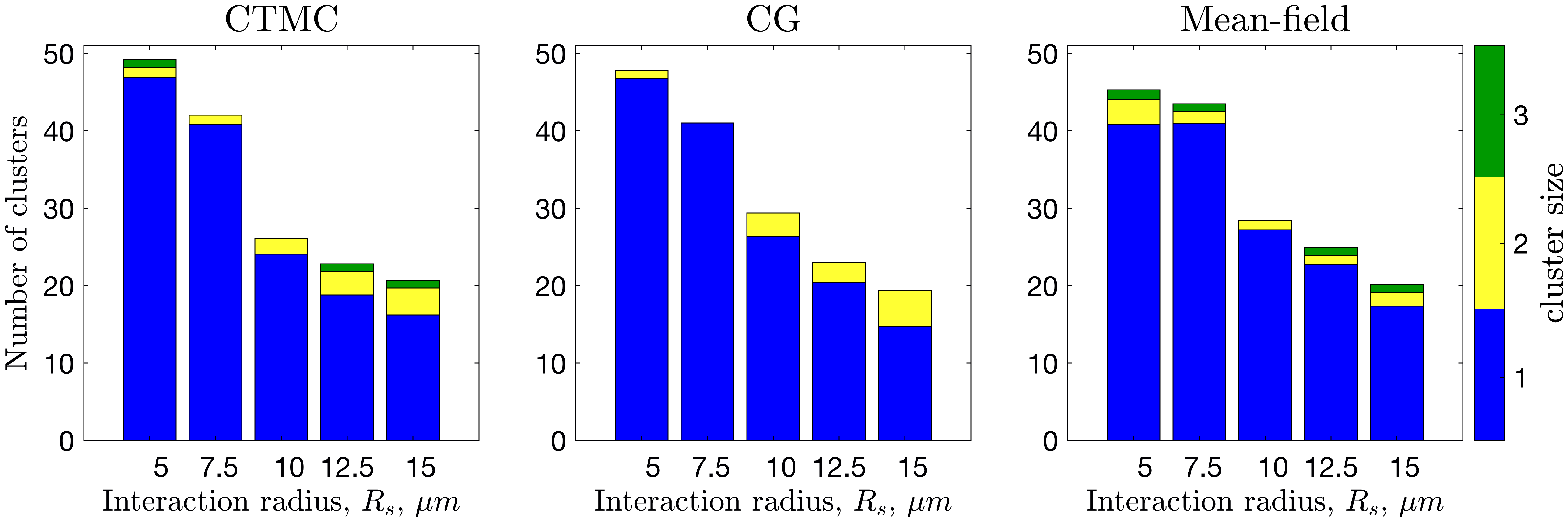}}
\caption{\textbf{Comparison of the dynamics of the multicellular VEGF-Delta-Notch model simulated on a branching network using the full stochastic (CTMC), CG, and mean-field descriptions.} (a) The Delta-high cell proportion as a function of the cell-to-cell interaction radius, $R_s$, for varying noise amplitude, $\epsilon=1/ \Omega$ (the value of $\Omega$ is indicated in the title of each plot), for the full stochastic CTMC (black), CG (blue) and mean-field (red) descriptions. (b) A series of barplots showing how the long-time distribution of Delta-high cell clusters changes as the interacton radius, $R_s$, varies for the full stochastic CTMC (\underline{left panel}), CG (\underline{middle panel}), and mean-field (\underline{right panel}) systems. The number of single Delta-high cells in the final pattern (i.e. at a fixed final simulation time) is shown in blue; the number of clusters with 2, and 3 adjacent Delta-high cells is shown in yellow and green, respectively. For these simulations, we fixed $\Omega=1000$ ($\epsilon = 0.001$). The results are averaged over 100 realisations. The remaining parameter values were fixed as indicated in \Cref{Params}.} 
\label{ResultsVascularNetwork}
\end{figure}

\renewcommand{\thesubfigure}{\alph{subfigure}}
\captionsetup{singlelinecheck=off}

\clearpage

\renewcommand{\arraystretch}{2.0}
\setlength{\doublerulesep}{2\arrayrulewidth}
\begin{table}[htbp]
\begin{tabular}{| p{2.2cm} | p{9.9cm} |}
\hline
\textbf{Parameter} & \textbf{Value}  \\ \hline \hline
$\beta_N $ & $2.5$, fixed in all figures.  \\
$\beta_D$ & $4.0$, fixed in all figures. \\
$\beta_{R_2}$ & $4.0$, fixed in all figures. \\ 
$\rho_N$ & $20.0$, fixed in all figures.  \\ 
$\rho_{R_2}$ & $10.0$, fixed in all figures.  \\ 
$\lambda_{I,N}$ & $4.0$, fixed in all figures.  \\ 
$\lambda_{I,R_2}$ & $0.0$, fixed in all figures.  \\ 
$\lambda_{R_2^*,D}$ & $2.0$, fixed in all figures.  \\ 
$n_N$ & $2$, fixed in all figures.  \\ 
$n_D$ & $1$, fixed in all figures.  \\
$n_{R_2}$ & $1$, fixed in all figures.  \\
$\eta$ & $0.5$, fixed in all figures.  \\ 
$\tau$ & $5.0$, fixed in all figures.  \\ \hline
$\kappa$ & $4.0$ for \Cref{main-Map_with_SamplePath,main-OmegaConvergence,main-PrefactorFit,MAPplot};  $12.0$ for all other figures and simulations. \\
$v_{ext}$ & $0.1$ for \Cref{main-Map_with_SamplePath,main-OmegaConvergence,main-PrefactorFit,MAPplot};  $1.25$ for all other figures and simulations. \\
$d_{ext}$ & $0.2$ for \Cref{main-Map_with_SamplePath,main-OmegaConvergence,main-PrefactorFit,MAPplot};  $d_{ext} \in [0.0, 4.0]$ and the exact value is determined during the simulations (depending on the neighbourhood of each cell) for all other figures. \\
$n_{ext}$ & $0.5$ for \Cref{main-Map_with_SamplePath,BifurcationDiagram,main-OmegaConvergence,main-PrefactorFit,MAPplot};  $d_{ext} \in [0.0, 4.0]$ and the exact value is determined during the simulations (depending on the neighbourhood of each cell) for all other figures. \\ \hline
voxel width, $h$ & $5~\mu m$ in all multicellular simulations (hexagon width). \\
$R_s$ & $15~\mu m$ in \Cref{main-PatternConfigurations,main-OptimalPattern,SnapshotsCGVascularNetwork}; for all other multicellular simulations $R_s$ is indicated in the text and/or figure captions. \\ \hline
final simulation time, $T_{final}$ & $12000$ mins for \Cref{main-PatternConfigurations,PatternVaryingR}; $5 \cdot 10^{23}$ mins for \Cref{main-OptimalPattern}; $1500$ mins for \Cref{main-ResultsMonolayer,ResultsVascularNetwork,ResultsCPU}; $9000$ mins for \Cref{SnapshotsCGVascularNetwork}. \\ \hline
\end{tabular}
\vspace{0.1cm}
\caption{\textbf{Non-dimensional parameters of the VEGF-Delta-Notch system.} } \label{Params}
\end{table}

\clearpage 

\bibliographystyle{siamplain}
\bibliography{references}

\makeatletter\@input{xxsupp.tex}\makeatother